\documentclass[traditabstract]{aa} % for the abstract without structuration 
\usepackage{graphicx}
\usepackage{txfonts}
\usepackage{color}
\begin{document}

\title{Molecular gas properties of a lensed star-forming galaxy at $z\sim 3.6$: a case study\thanks{Based on observations carried out with the IRAM Plateau de Bure Interferometer, the IRAM 30~m telescope, and the NRAO Karl G. Jansky Very Large Array. The Institut de Radioastronomie Millim\'etrique (IRAM) is supported by CNRS/INSU (France), the MPG (Germany), and the IGN (Spain). The National Radio Astronomy Observatory (NRAO) is a facility of the National Science Foundation operated under cooperative agreement by Associated Universities, Inc.}}

%\subtitle{}

\author{M. Dessauges-Zavadsky\inst{1},
        M. Zamojski\inst{2,1},
        W. Rujopakarn\inst{3,4},
        J. Richard\inst{5},
        P. Sklias\inst{1},
        D. Schaerer\inst{1,6},
                F. Combes\inst{7},\\                  
                H. Ebeling\inst{8},        
        T.~D. Rawle\inst{9}, 
        E. Egami\inst{10},
        F. Boone\inst{6},
        B. Cl\'ement\inst{5},
        J.-P. Kneib\inst{11,12},
        K. Nyland\inst{13},
        \and 
        G. Walth\inst{14}
        }

\offprints{miroslava.dessauges@unige.ch}

\institute{
Observatoire de Gen\`eve, Universit\'e de Gen\`eve, 51 Ch. des Maillettes, 1290 Versoix, Switzerland
\and
Independent researcher, New York, NY, USA
\and
Department of Physics, Faculty of Science, Chulalongkorn University, 254 Phayathai Road, Pathumwan, Bangkok 10330, Thailand
\and
Kavli Institute for the Physics and Mathematics of the Universe (WPI), University of Tokyo Institutes for Advanced Study, Kashiwa, Chiba 277-8583, Japan
\and
Universit\'e Lyon, Universit\'e Lyon 1, Ens de Lyon, CNRS, Centre de Recherche Astrophysique de Lyon UMR5574, 69230 Saint-Genis-Laval, France
\and 
CNRS, IRAP, 14 Avenue E. Belin, 31400 Toulouse, France
\and
Observatoire de Paris, LERMA, 61 Avenue de l'Observatoire, 75014 Paris, France
\and
Institute for Astronomy, University of Hawaii, 2680 Woodlawn Drive, Honolulu, HI 96822, USA  
\and  
ESA / Space Telescope Science Institute (STScI), 3700 San Martin Drive, Baltimore, MD 21218, USA
\and               
Steward Observatory, University of Arizona, 933 North Cherry Avenue, Tucson, AZ 85721, USA
\and
Laboratoire d'Astrophysique, Ecole Polytechnique F\'ed\'erale de Lausanne (EPFL), 1290 Versoix, Switzerland
\and
Aix Marseille Universit\'e, CNRS, LAM, UMR 7326, 13388, Marseille, France  
\and
National Radio Astronomy Observatory, 520 Edgemont Rd, Charlottesville, VA 22903, USA
\and 
University of California, Center for Astrophysics and Space Sciences, 9500 Gilman Drive, San Diego, CA 92093, USA
}

\date{}

\authorrunning{Dessauges-Zavadsky et~al.}

\titlerunning{Molecular gas properties of a lensed star-forming galaxy at $z\sim 3.6$}

% \abstract{}{}{}{}{} 
% 5 {} token are mandatory
 
\abstract%{a}{b}{c}{d}{e}
{We report on the galaxy MACSJ0032-arc at $z_{\rm CO} = 3.6314$ discovered during the {\it Herschel} Lensing snapshot Survey of massive galaxy clusters, and strongly lensed by the cluster MACS\,J0032.1+1808. The successful detections of its rest-frame ultraviolet (UV), optical, far-infrared (FIR), millimeter, and radio continua, and of its CO emission enable us to characterize, for the first time at such a high redshift, the stellar, dust, and molecular gas properties of a compact star-forming galaxy with a size smaller than 2.5~kpc, a fairly low stellar mass of $4.8^{+0.5}_{-1.0}\times 10^9~M_{\sun}$, and a moderate IR luminosity of $4.8^{+1.2}_{-0.6}\times 10^{11}~L_{\sun}$.
%placing MACSJ0032-arc in the moderately luminous infrared galaxy category.
%representative of the more numerous galaxies below the turnover of the stellar mass and IR luminosity functions at $3<z<4$. 
%within less than a factor of 2.5 of typical main sequence parametrisations. 
By combining the stretching effect of the lens with the high angular resolution imaging of the CO(1--0) line emission and the radio continuum at 5~GHz, we find that the bulk of the molecular gas mass and star formation seems to be spatially decoupled from the rest-frame UV emission. About 90\% of the total star formation rate is undetected at rest-frame UV wavelengths because of severe obscuration by dust, but is seen through the thermal FIR dust emission and the radio synchrotron radiation.
%is seen through the thermal FIR dust emission and the radio synchrotron radiation, and is undetected at rest-frame UV wavelengths because of severe obscuration by dust. 
The observed CO(4--3) and CO(6--5) lines demonstrate that high-$J$ transitions, at least up to $J=6$, remain excited in this galaxy, whose CO spectral line energy distribution resembles that of high-redshift submm galaxies, even though the IR luminosity of MACSJ0032-arc is ten times lower. This high CO excitation is possibly due to the compactness of the galaxy. We find evidence that this high CO excitation has to be considered in the balance when estimating the CO-to-H$_2$ conversion factor. Indeed, the respective CO-to-H$_2$ conversion factors as derived from the correlation with metallicity and the FIR dust continuum can only be reconciled if excitation is accounted for. The inferred depletion time of the molecular gas in MACSJ0032-arc supports the decrease in the gas depletion timescale of galaxies with redshift, although to a lesser degree than predicted by galaxy evolution models. Instead, the measured molecular gas fraction as high as 60--79\% in MACSJ0032-arc favors the continued increase in the gas fraction of galaxies with redshift as expected, despite the plateau observed between $z\sim 1.5$ and $z\sim 2.5$.
%%hk \LEt{ long sentence, ambiguous commas. Perhaps: ... galaxy evolution models; instead, its measured molecular gas fraction alone (which is as high as 60-79\%) favors the continued increase in the gas fraction of galaxies with ... }
}
%The extent of the molecular gas depletion timescale and the molecular gas fraction studies toward the high redshift of the MACSJ0032-arc first confirms the decrease of the former with cosmic time, although shallower than predicted by models, and second sustains the expected steady increase of the latter with cosmic time, despite the plateau observed between $z\sim 1.5$ and $z\sim 2.5$.


\maketitle
%
%________________________________________________________________

\section{Introduction}

In order to track the star formation in galaxies and understand how gas is 
converted into stars, it is essential to study the prevalence and distribution 
of molecular gas across cosmic time. In recent years, molecular gas mass 
measurements have finally become feasible for normal galaxies, that is 
star-forming galaxies (SFGs) lying on the main sequence (MS) at redshifts 
$z<2.5$ and featuring stellar masses of $M_* > 2.5\times 10^{10}~M_{\sun}$ 
\citep[e.g.,][]{daddi10,genzel10,geach11,tacconi10,tacconi13}. These galaxies, 
following the fairly tight relation between $M_*$ and star formation rate 
($\mathit{SFR}$) with a dispersion of $\pm 0.3~\rm dex$, contribute 
$\sim 80-90$\% of the cosmic $\mathit{SFR}$ density at redshifts $0<z<3$ 
\citep{daddi07,rodighiero11,wuyts11,rodighiero15,schreiber15}.

%
%__________________________________________________________________

\begin{figure*}
\centering
\includegraphics[width=8.2cm,clip]{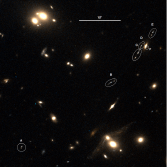}
\raisebox{4.15cm}[0pt][0pt]{\includegraphics[width=4.05cm,clip]{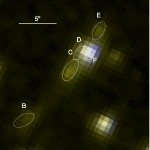}}
\raisebox{0cm}[0pt][0pt]{\hspace{-4.22cm} \includegraphics[width=4.05cm,clip]{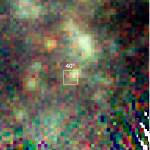}}
\caption{Color-composite image of MACSJ0032-arc obtained using the HST/ACS {\it F606W} and {\it F814W} filters (left panel). The studied galaxy at $z_{\rm CO} = 3.6314$, strongly lensed by the galaxy cluster MACS\,J0032.1+1808, is composed of six multiple images well resolved in the HST image, labeled A, B, C, D, E, and F and encircled in white. They form a giant arc extending over $42.4''$. 
%The counter-image F is extremely faint and located right next to image D, very close to a galaxy in the lensing cluster. 
The right panels show, respectively, the zoom-in color-composite image obtained using the {\it Spitzer}/IRAC 3.6 and $4.5~\mu$m bands over the multiple images B, C, D, E, and F (top) and the zoom-out color-composite image obtained using the {\it Herschel}/SPIRE 250, 350, and $500~\mu$m bands; the white box represents a $40''\times 40''$  area.}
\label{fig:thumbnail}
\end{figure*}
%
%__________________________________________________________________

With the increasing number of CO measurements obtained for MS SFGs, we are 
starting to highlight the significant role that the molecular gas plays in 
galaxies in general and in the implementation of the MS itself.  Indeed, in the 
local Universe it has now been established that the location of a galaxy in the 
$\mathit{SFR}$--$M_*$ plane is primarily governed by its supply of molecular 
gas, whereas variations in the star formation efficiency ($\mathit{SFE}$) only 
play a secondary role \citep{saintonge12}. Similar conclusions have been 
reached for high-redshift galaxies \citep{tacconi13,dessauges15}, although 
\citet{genzel15} find equal contributions from the $\mathit{SFE}$ and the 
molecular gas content for the offset of galaxies from the MS. Likewise, the 
observed evolution of the MS with redshift that results in the increase in the 
specific star formation rate ($\mathit{sSFR}$) with cosmic time 
\citep{schaerer09,rodighiero10,schreiber15} is closely linked to the 
behavior of the molecular gas. Current observations convincingly show that the 
rapid rise in the $\mathit{sSFR}$ of galaxies up to $z\sim 2$ can be explained 
by the comparable rise in their molecular gas fraction ($f_{\rm molgas}$) with 
cosmic time and their slowly varying molecular gas depletion timescale 
($t_{\rm depl}$) \citep{geach09,geach11,genzel10,genzel15,daddi10,
bauermeister13,tacconi10,tacconi13}. Beyond $z\gtrsim 2$, it remains a subject 
of debate whether $f_{\rm molgas}$ continues to increase or whether, instead, 
the decrease in $t_{\rm depl}$ is steepening 
\citep{saintonge13,tan13,dessauges15}, primarily due to the lack of molecular 
gas mass measurements available so far for MS SFGs at $z>2.5$. 

While determining the molecular gas content in individual MS SFGs at $z>2.5$ is 
challenging,
%, and even more via CO emission line measurements. 
it is feasible with the help of strong gravitational lensing, which enables us 
to push beyond current instrumental sensitivity thresholds. Indeed, of the five 
CO measurements performed to date for $z>2.5$ MS SFGs, four were obtained in 
strongly lensed SFGs, and the fifth was obtained in a very massive 
($M_* = 10^{11}~M_{\sun}$) unlensed SFG \citep{riechers10,johansson12,tan13,
saintonge13}. Indirect estimates of the molecular gas masses in some 53 
%another eight MS SFGs at $2.7<z<4$ with $M_* = (2-40)\times 10^{10}~M_{\sun}$
massive MS galaxies at $2.7<z<4$ with $M_* > 2\times 10^{10}~M_{\sun}$ were 
recently reported by \citet{scoville16} and \citet{schinnerer16} based on 1~mm 
dust continuum emissions detected with the Atacama Large Millimeter Array 
(ALMA). 
%and calibrations of rest-frame $\approx 300~\mu$m fluxes to cold gas masses from \citet{scoville14} and \citet{groves15}.
%their own calibration of the dust-to-gas mass ratio \citep{scoville14}.

We report here on the CO(6--5), CO(4--3), and CO(1--0) line detections for a 
newly discovered star-forming galaxy at $z_{\rm CO} = 3.6314$, strongly lensed 
by the galaxy cluster MACS\,J0032.1+1808 \citep{ebeling01}, hereafter denoted 
MACSJ0032-arc. Featuring the highest redshift at which an estimate of the 
molecular gas mass has been obtained from CO in a normal low-mass 
%(non-ULIRG\footnote{By ``non-ULIRG'' (non-Ultra-Luminous IR Galaxy), we mean a star-forming galaxy with an infrared luminosity lower than $10^{12}~L_{\sun}$.}) 
SFG, this galaxy is of particular interest for a detailed study aimed at 
addressing key questions: What are the $f_{\rm molgas}$ and $t_{\rm depl}$ 
evolutionary trends at very high redshift? Is the CO-to-H$_2$ conversion factor 
significantly different at $z\sim 3.6$? Does the CO spectral line energy 
distribution (SLED) vary with respect to the SLEDs of lower redshift MS SFGs? 
Is the cold gas CO emission spatially correlated with the stellar UV emission?

Our paper is organized as follows. In Sect.~\ref{sect:target} we introduce our 
target, and in Sect.~\ref{sect:observations} we describe the multiwavelength 
observations performed from the optical through the infrared, far-infrared, 
and millimeter to the radio regime. The analysis of the data including the 
gravitational lens modeling and the inferred stellar, dust, and CO molecular 
gas properties of MACSJ0032-arc can be found in 
Sect.~\ref{sect:analysis-results}. We discuss these results in the context of MS 
SFGs with CO measurements and explore the redshift evolution of both the 
molecular gas depletion timescale and the molecular gas fraction in 
Sect.~\ref{sect:discussion}. We also estimate the CO-to-H$_2$ conversion factor 
using two different methods and highlight their respective strengths and 
weaknesses. A summary and conclusions follow in Sect.~\ref{sect:conclusions}. 
Throughout the paper, we assume a $\Lambda$CDM cosmology with 
$H_0 = 70~\rm km~s^{-1}~Mpc^{-1}$, $\Omega_{\rm M} = 0.29$, and 
$\Omega_{\Lambda} = 0.71$, and the \citet{chabrier03} initial mass function.
%\footnote{The values from the literature are scaled by a factor of 1.7 when the \citet{salpeter55} initial mass function is used.}.

%
%__________________________________________________________________

\section{The target: MACSJ0032-arc}
\label{sect:target}

We report on a newly discovered galaxy, MACSJ0032-arc, identified as a bright 
FIR emitter in the {\it Herschel} Lensing snapshot Survey (HLS-snapshot; Egami 
et~al., in prep.). Designed to find strongly lensed sources at high redshift, 
the survey used {\it Herschel}/SPIRE to image the fields around more than 300 
massive galaxy clusters.
%(even in shallow images). 
Located at $\rm RA = 00$:32:07.776, $\rm DEC = +18$:06:47.80, MACSJ0032-arc is 
strongly lensed by the galaxy cluster MACS\,J0032.1+1808 at $z=0.377$ (Ebeling 
\& Repp, in prep.). The redshift of the lensed system was first  accessed 
from weak rest-frame UV interstellar medium (ISM) lines and the Ly$\alpha$ break 
detected in absorption in the spectrum we obtained with LRIS on the Keck\,I 
telescope, $z_{\rm ISM} = 3.626\pm 0.001$ (Richard et~al., in prep.). It was 
then  confirmed with our firm detections of the CO(4--3) and CO(6--5) 
emission lines at $z_{\rm CO} = 3.6314\pm 0.0005$ with the IRAM 30~m telescope. 
Very recently, we  also detected the nebular 
[O\,{\sc iii}]\,$\lambda$5007 and H$\beta$ emission lines at $z_{\rm neb} = 
3.633\pm 0.003$, redshifted in the near-IR, with LUCI on the LBT (Walth et al., 
in prep.). The systemic redshift of MACSJ0032-arc is hence well constrained by 
both $z_{\rm CO}$ and $z_{\rm neb}$, which agree within their $1\,\sigma$ 
errors. The ISM absorption lines at $z_{\rm ISM}$ appear blueshifted with 
respect to the systemic redshift of the galaxy by $-350\pm 65~\rm km~s^{-1}$.
%is observed between the ISM absorption lines at $z_{\rm ISM} = 3.626\pm 0.008$ and the systemic redshift of the galaxy, $z_{\rm CO} = 3.6314\pm 0.0005$, as traced by the CO lines.

MACSJ0032-arc was also observed in the infrared (IR) as part of the 
{\it Spitzer}/IRAC Lensing Survey, and in the optical regime in ongoing 
efforts to image strong-lensing clusters with the {\it Hubble} Space Telescope 
(HST). It has been detected at all these wavelengths, as is shown in 
Fig.~\ref{fig:thumbnail}. These observations combined with millimeter (mm) and 
radio data obtained with the Plateau de Bure Interferometer (PdBI) and the 
Jansky Very Large Array (JVLA), respectively, enable us to derive the stellar, 
dust, and molecular gas properties of this $z\sim 3.6$ galaxy.

The high-quality, high-resolution rest-frame UV images obtained with HST 
%(see Sect.~\ref{sect:photometry} for observational details) 
reveal six images of this high-redshift galaxy, labeled A, B, C, D, E, and F 
throughout the paper, as illustrated in Fig.~\ref{fig:thumbnail}. The 
counter-image F is very faint and barely visible, located very close to a 
lensing cluster galaxy member. A seventh counter-image exists, but is 
de-amplified and hence invisible. 
%(see Sect.~\ref{sect:lensmodel}). 
All of these images originate from the same background galaxy, amplified by 
different magnification factors, and form a giant arc extending over $42.4''$. 
Moreover, the lensed galaxy is resolved into two compact, UV-bright knots, 
separated by $\sim 0.8''$ in the most amplified multiple images B, C, D, and E, 
plus a tail attached to the brightest knot. 
%With the help of the gravitational lens stretching, we expect to resolve the CO(6--5) and CO(1--0) emissions too, obtained, respectively, with the PdBI and
The high-resolution JVLA observations enable us to resolve the CO(1--0) line 
emission and the 5~GHz radio continuum as well within the multiple images B, C, 
D, and E, and to compare the location
%Expecting to spatially resolve the lensed galaxy's JVLA CO(1--0) line emission and 5~GHz radio continuum emission too, we plan to determine the spatial location 
of the molecular gas and dust-obscured star formation with respect to the 
rest-frame UV emission.

In what follows, we focus our analysis only on the brightest multiple 
images B, C, D, and E, covered by all our data and detected in most of them, as 
shown and discussed in Sect.~\ref{sect:observations}.

%
%__________________________________________________________________

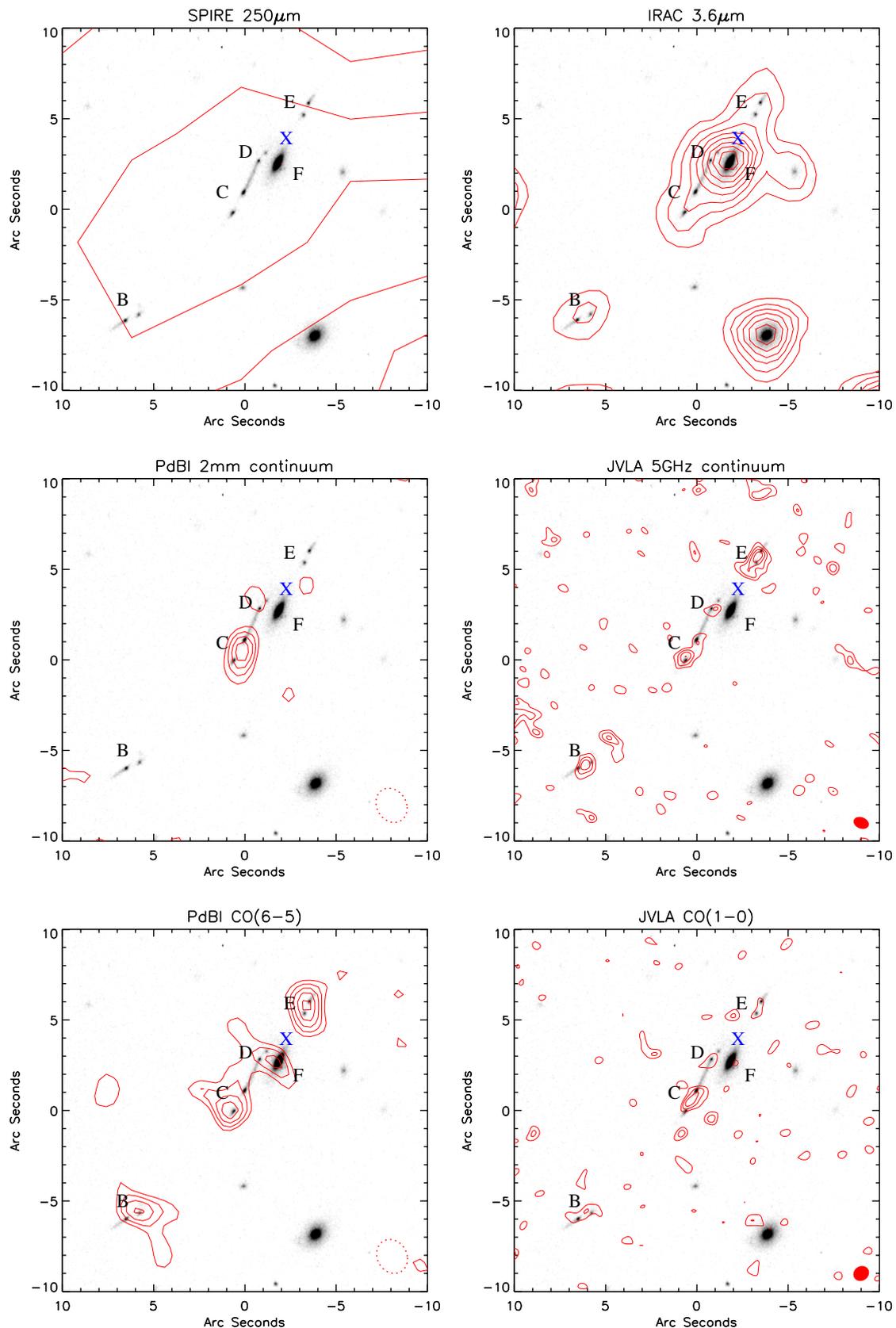
\begin{figure*}\setlength{\unitlength}{7.5cm}
{\centering
\includegraphics[width=7.5cm,clip]{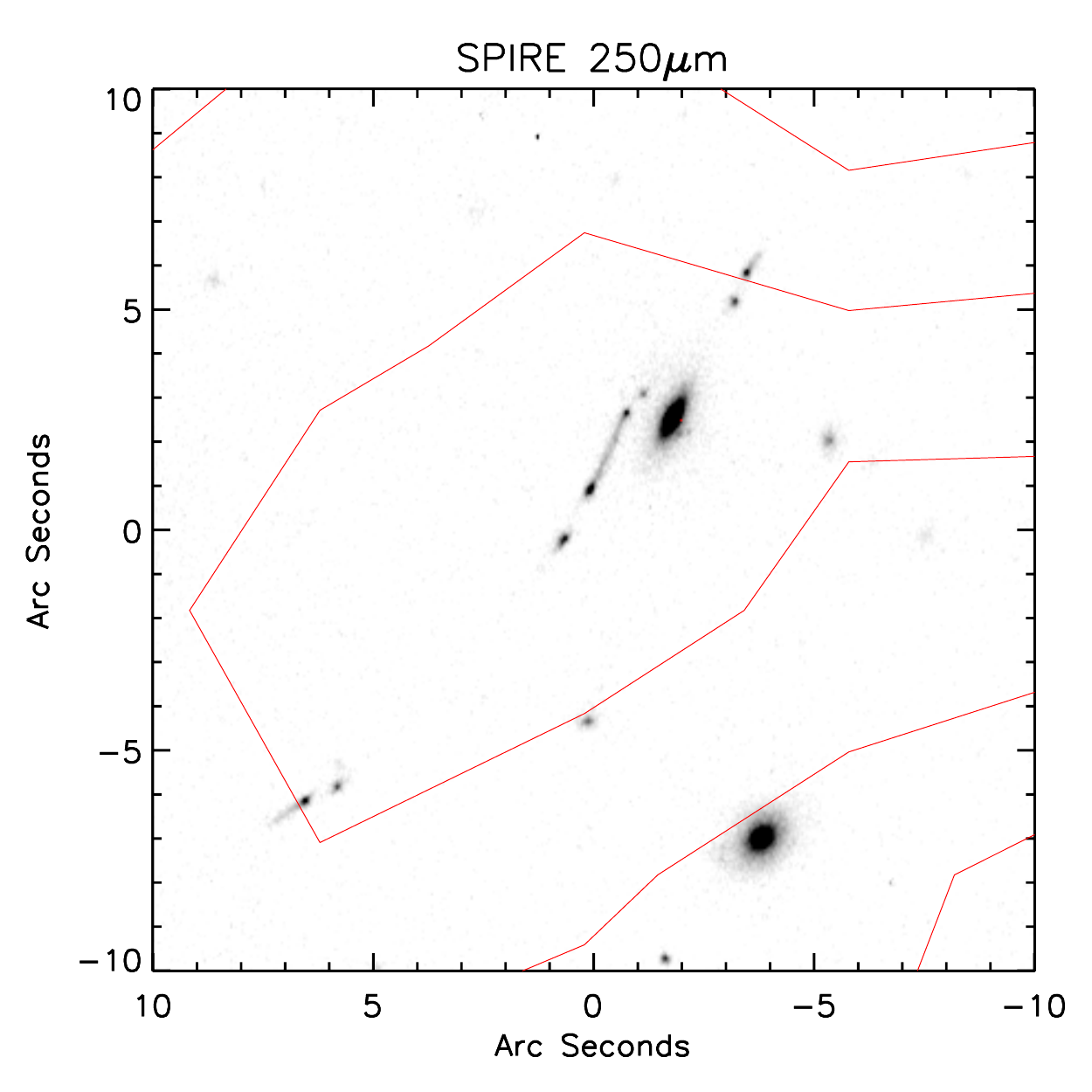}
%%\multiput(0,0)(-0.1,0){11}{\line(0,1){1}}
%%\multiput(-1,0)(0,0.1){11}{\line(1,0){1}}
\put(-0.74,0.3){\small B}
\put(-0.52,0.54){\small C}
\put(-0.47,0.63){\small D}
\put(-0.37,0.74){\small E}
\put(-0.35,0.58){\small F}
\put(-0.38,0.66){\color{blue}{\small X}}
\includegraphics[width=7.5cm,clip]{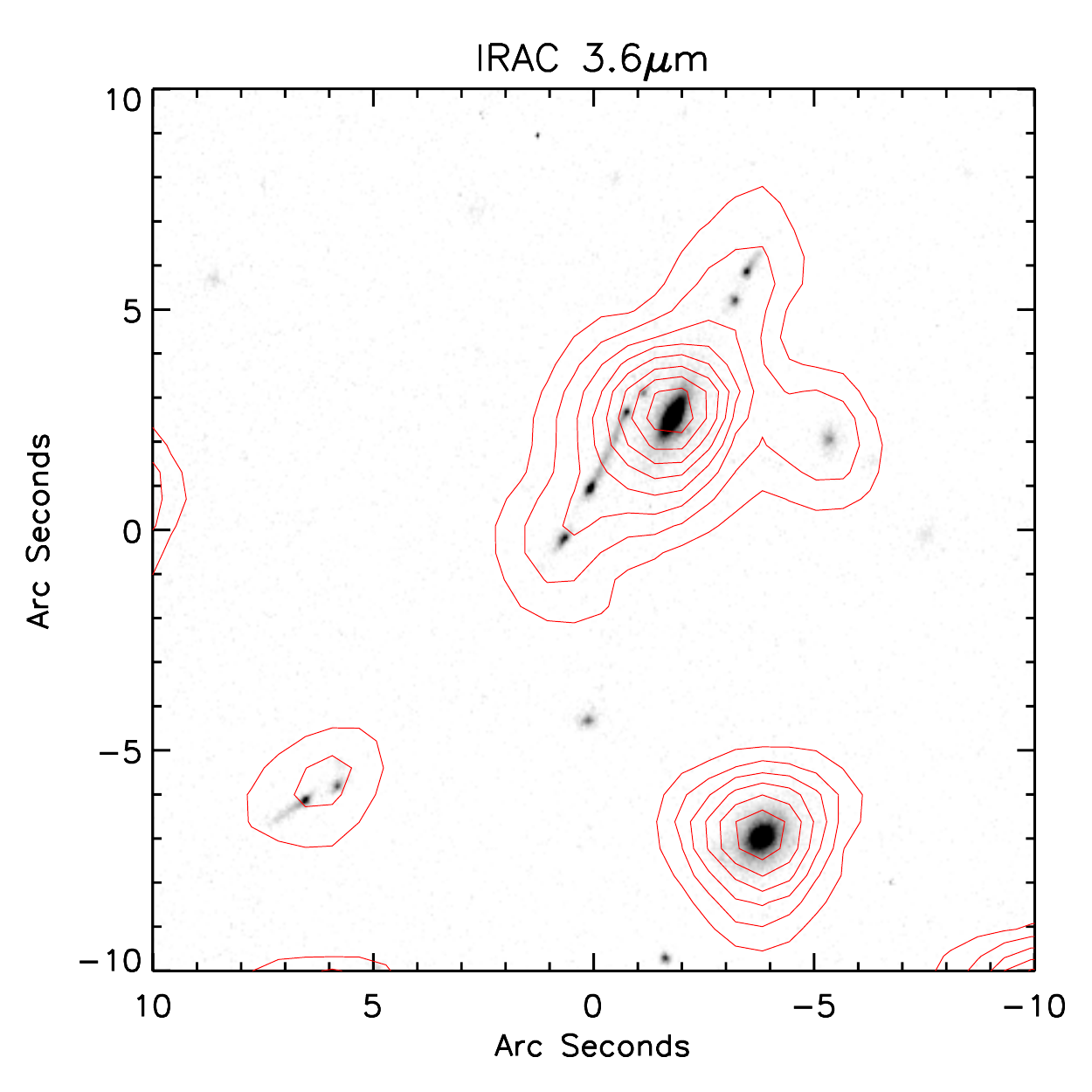} 
\put(-0.74,0.3){\small B}
\put(-0.52,0.54){\small C}
\put(-0.47,0.63){\small D}
\put(-0.37,0.74){\small E}
\put(-0.35,0.58){\small F}
\put(-0.38,0.66){\color{blue}{\small X}}

\includegraphics[width=7.5cm,clip]{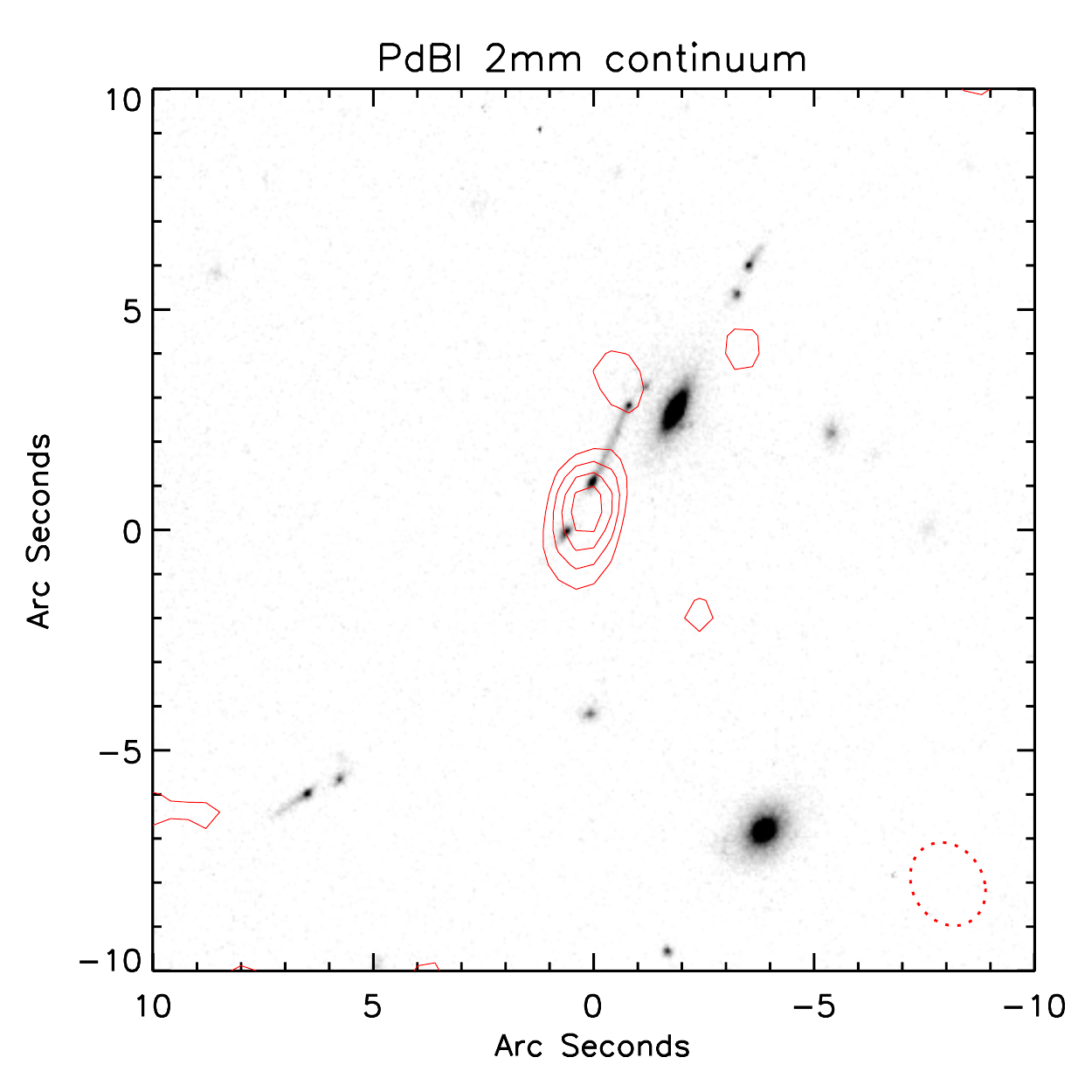}
\put(-0.74,0.3){\small B}
\put(-0.52,0.54){\small C}
\put(-0.47,0.63){\small D}
\put(-0.37,0.74){\small E}
\put(-0.35,0.58){\small F}
\put(-0.38,0.66){\color{blue}{\small X}}
\includegraphics[width=7.5cm,clip]{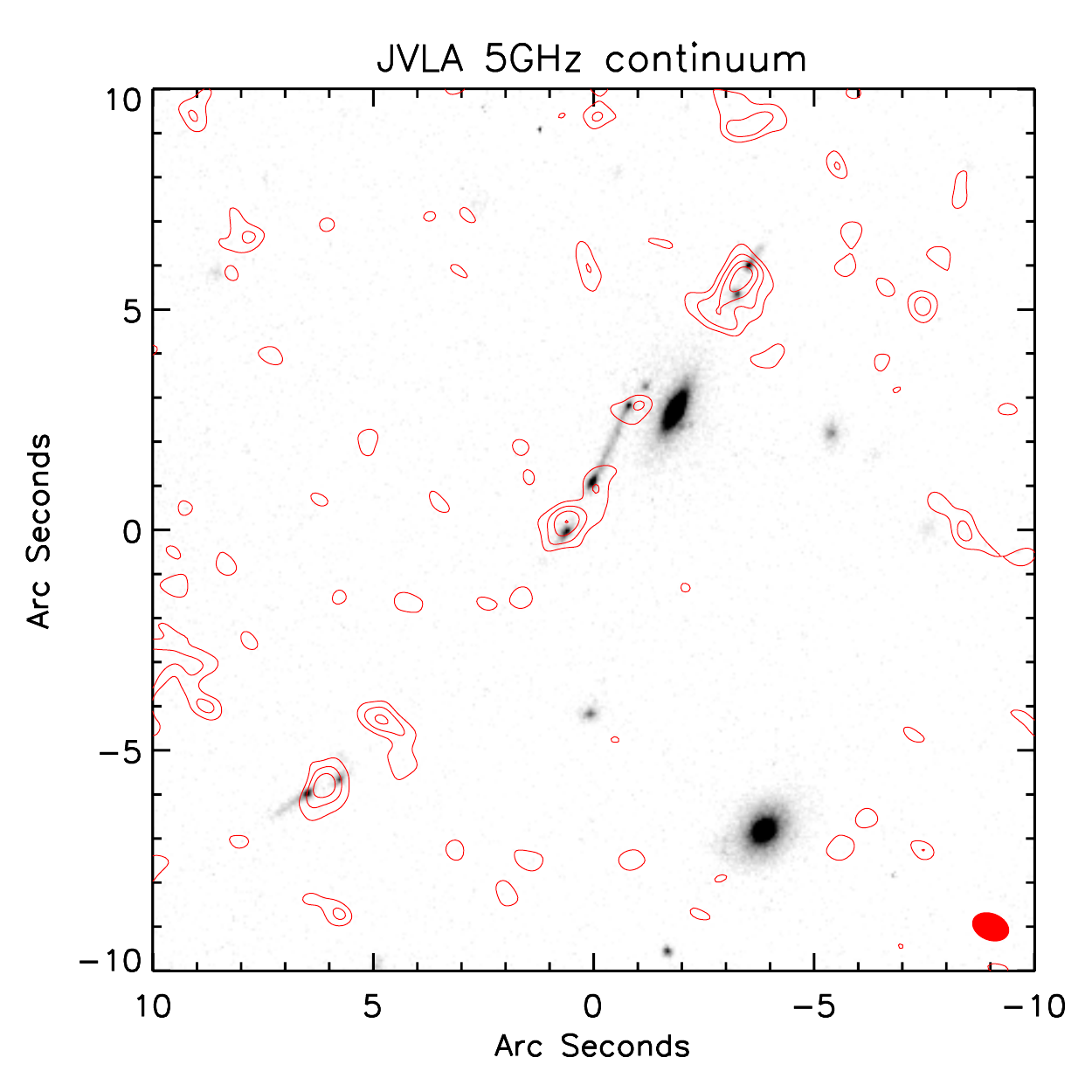}
\put(-0.74,0.3){\small B}
\put(-0.52,0.54){\small C}
\put(-0.47,0.63){\small D}
\put(-0.37,0.74){\small E}
\put(-0.35,0.58){\small F}
\put(-0.38,0.66){\color{blue}{\small X}}

\includegraphics[width=7.5cm,clip]{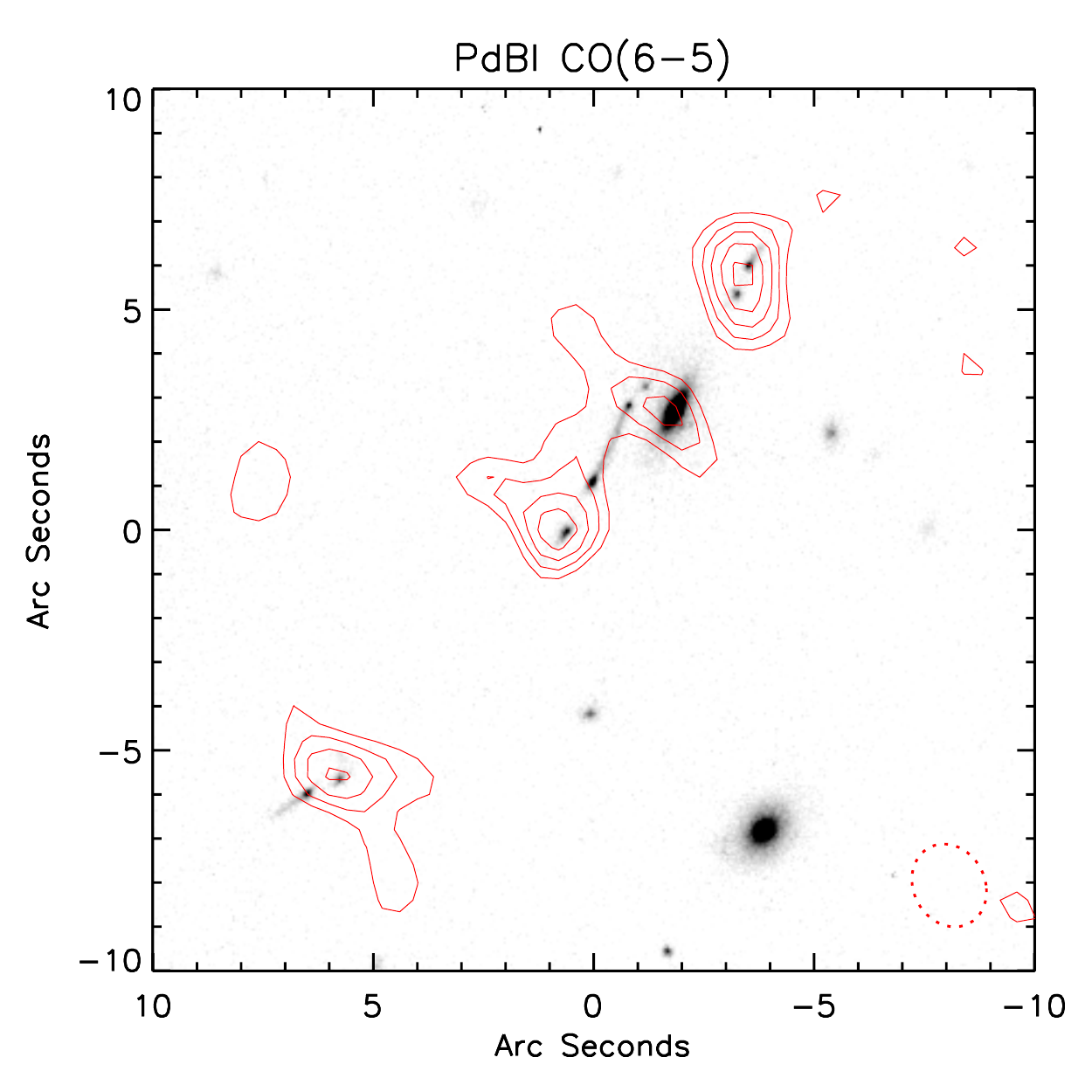}
\put(-0.74,0.3){\small B}
\put(-0.52,0.54){\small C}
\put(-0.47,0.63){\small D}
\put(-0.37,0.74){\small E}
\put(-0.35,0.58){\small F}
\put(-0.38,0.66){\color{blue}{\small X}}
\includegraphics[width=7.5cm,clip]{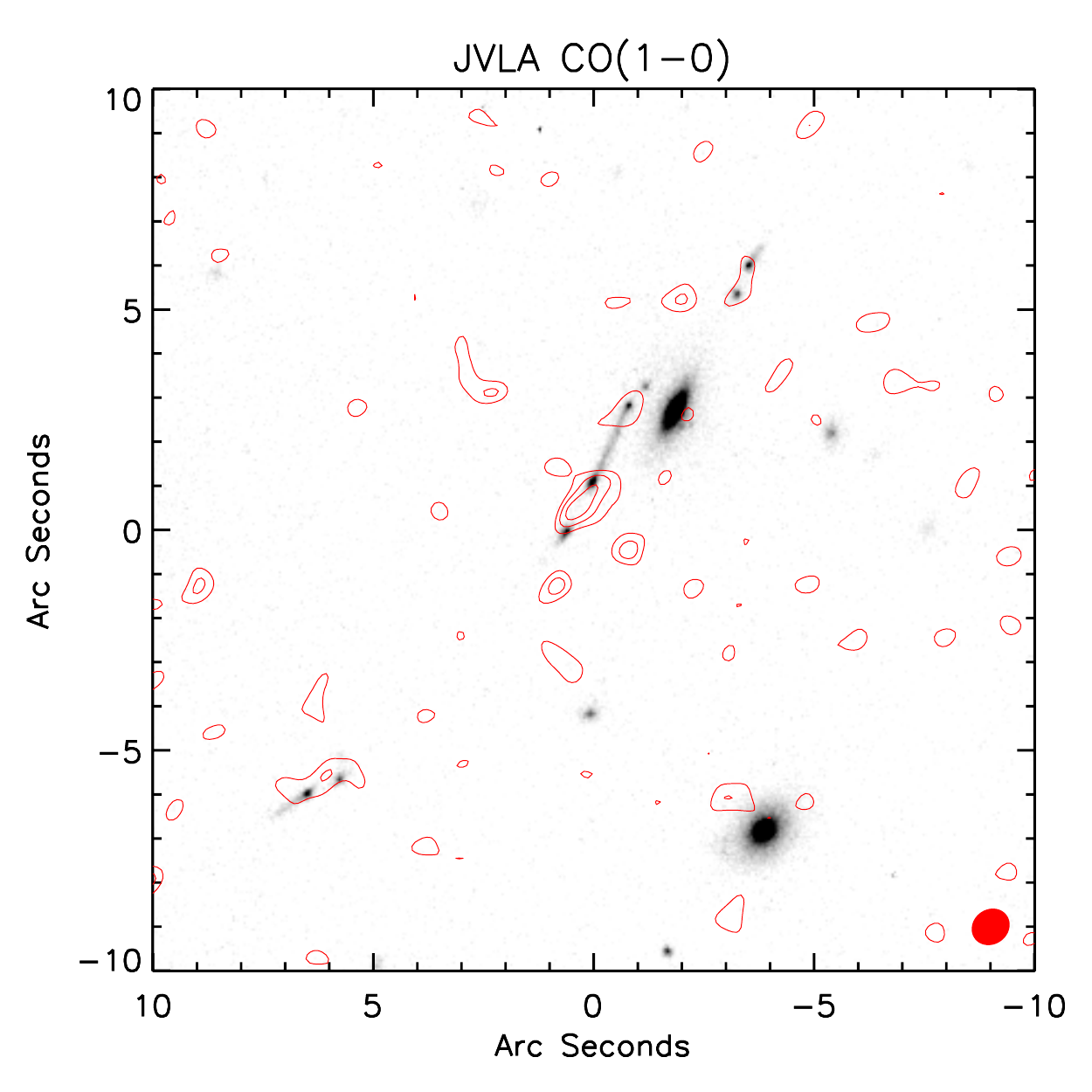}
\put(-0.74,0.3){\small B}
\put(-0.52,0.54){\small C}
\put(-0.47,0.63){\small D}
\put(-0.37,0.74){\small E}
\put(-0.35,0.58){\small F}
\put(-0.38,0.66){\color{blue}{\small X}}

}

\caption{{\it Herschel}/SPIRE $250~\mu$m, {\it Spitzer}/IRAC $3.6~\mu$m, PdBI 2~mm continuum, JVLA 5~GHz continuum, PdBI CO(6--5) line emission, and JVLA CO(1--0) line emission contours (in red) overlaid, from left to right and from top to bottom, on the HST/ACS {\it F814W} multiple images B, C, D, and E of MACSJ0032-arc. The contribution of the counter-image F is negligible. X refers to a galaxy member of the lensing cluster.
%(a member of the lensing cluster) located between the counter-images D and F. 
Contour levels start at $2\,\sigma$ and are spaced in steps of $1\,\sigma$, expect for IRAC $3.6~\mu$m where the contours are in steps of 9, 16, 25, 36, $49\,\sigma$, etc. The SPIRE detection extends beyond the multiple images B, C, D, and E, as shown in Fig.~\ref{fig:thumbnail}, but peaks at $4\,\sigma$ over B, C, D, and E.
%reaches a $4\,\sigma$ detection. 
The RMS levels of the 2~mm, 5~GHz, CO(6--5), and  CO(1--0) data are 57~$\mu$Jy, 1.7~$\mu$Jy, 93~$\rm mJy~km~s^{-1}$, and 6.2~$\rm mJy~km~s^{-1}$, respectively. The size and orientation of the beam are indicated by the dotted (PdBI) or filled (JVLA) red ellipse in the bottom right corner.
%rms 2mm = 57 muJy / rms CO65 = 93 mJy km/s / rms CO10 = 6.2 mJy km/s / rms 5GHZ = 1.7 muJy
}
\label{fig:contours}
\end{figure*}
%
%__________________________________________________________________

\begin{table*}
\caption{From optical to radio
%hk \LEt{ ratio? (leave hyphens) or range? (i.e., from optical to radio, take out hyphens) }
photometry and CO line integrated fluxes of MACSJ0032-arc.}             
\label{tab:photometry}      
\centering          
\begin{tabular}{l r@{ }l r@{ }l r@{ }l r@{ }l r@{ }l}     
\hline\hline    
Band/Line & \multicolumn{8}{c}{Multiple images} & \multicolumn{2}{c}{Total} \\
          & \multicolumn{2}{c}{B} & \multicolumn{2}{c}{C} & \multicolumn{2}{c}{D} & \multicolumn{2}{c}{E} & \multicolumn{2}{c}{B+C+D+E} \\
\hline
Magnification & 12.7 $\pm$&3 & 24.7 $\pm$&2 & 12.6 $\pm$&3 & 10.8 $\pm$&4 & 62 $\pm$&6 \\
\hline 
%ACS {\it F606W} & 23.73 $\pm$&0.05 & 23.31 $\pm$&0.05 & 24.32 $\pm$&0.10 & 24.15 $\pm$&0.05 & 22.30 $\pm$&0.05 \\
ACS {\it F606W} ($\mu$Jy) & 1.169 $\pm$&0.002 & 1.722 $\pm$&0.004 & 0.679 $\pm$&0.003 & 0.794 $\pm$&0.002 & 4.365 $\pm$&0.010 \\
%ACS {\it F814W} & 23.09 $\pm$&0.05 & 22.66 $\pm$&0.05 & 23.72 $\pm$&0.10 & 23.40 $\pm$&0.05 & 21.64 $\pm$&0.05 \\
ACS {\it F814W} ($\mu$Jy) & 2.109 $\pm$&0.005 & 3.133 $\pm$& 0.007 & 1.180 $\pm$& 0.005 & 1.585 $\pm$&0.003 & 8.017 $\pm$&0.019 \\
%IRAC $3.6~\mu$m  & 20.92 $\pm$&0.09 & \multicolumn{2}{c}{blended} & \multicolumn{2}{c}{blended} & \multicolumn{2}{c}{blended} & 19.49 $\pm$&0.09\tablefootmark{a} \\
IRAC $3.6~\mu$m ($\mu$Jy)  & 15.56 $\pm$&0.07 & \multicolumn{2}{c}{blended} & \multicolumn{2}{c}{blended} & \multicolumn{2}{c}{blended} & 58.08 $\pm$&0.27\tablefootmark{a} \\
%IRAC $4.5~\mu$m  & 20.79 $\pm$&0.08 &\multicolumn{2}{c}{blended} & \multicolumn{2}{c}{blended} & \multicolumn{2}{c}{blended} & 19.35 $\pm$&0.08\tablefootmark{a} \\
IRAC $4.5~\mu$m ($\mu$Jy) & 17.54 $\pm$&0.07 &\multicolumn{2}{c}{blended} & \multicolumn{2}{c}{blended} & \multicolumn{2}{c}{blended} & 66.07 $\pm$&0.27\tablefootmark{a} \\
SPIRE $250~\mu$m (mJy) & \multicolumn{2}{c}{blended} &\multicolumn{2}{c}{blended} & \multicolumn{2}{c}{blended} & \multicolumn{2}{c}{blended} & 58 $\pm$&8 \\
SPIRE $350~\mu$m (mJy) & \multicolumn{2}{c}{blended} &\multicolumn{2}{c}{blended} & \multicolumn{2}{c}{blended} & \multicolumn{2}{c}{blended} & 72 $\pm$&10 \\
SPIRE $500~\mu$m (mJy) & \multicolumn{2}{c}{blended} &\multicolumn{2}{c}{blended} & \multicolumn{2}{c}{blended} & \multicolumn{2}{c}{blended} &  52 $\pm$&9 \\
PdBI 2~mm (mJy) & \multicolumn{2}{c}{$<0.16$\tablefootmark{$\dag$}} & 0.37 $\pm$&0.07 & \multicolumn{2}{c}{$<0.12$\tablefootmark{$\dag$}} & \multicolumn{2}{c}{$<0.14$\tablefootmark{$\dag$}} & 0.65 $\pm$&0.25\tablefootmark{b\,c} \\
JVLA 5~GHz ($\mu$Jy) & 18.9 $\pm$&3.0 & 25.9 $\pm$&4.4 & 7.8 $\pm$&2.9 & 40.8 $\pm$&4.1\tablefootmark{d} & 93.4 $\pm$&14.4 \\
PdBI CO(6--5) (Jy km~s$^{-1}$) & 0.68 $\pm$&0.15 & 0.86 $\pm$&0.15 & 0.53 $\pm$&0.20 & 0.71 $\pm$&0.10\tablefootmark{d} & 2.78 $\pm$&0.60 \\
30~m CO(4--3) (Jy km~s$^{-1}$) &--& &--& &--& &--& & 2.30 $\pm$&0.55\tablefootmark{e} \\
JVLA CO(1--0) (Jy km~s$^{-1}$) & 0.047 $\pm$&0.020\tablefootmark{$\dag$} & 0.08 $\pm$&0.02 & 0.030 $\pm$&0.015\tablefootmark{$\dag$} & 0.035 $\pm$&0.018\tablefootmark{$\dag$} & 0.21 $\pm$&0.06\tablefootmark{b\,f} \\
\hline                  
\end{tabular}
\tablefoot{
Observed (not lensing-corrected) continuum fluxes
%AB magnitudes up to, and including, the IRAC bands, then continuum fluxes, 
and CO line integrated fluxes, all with $1\,\sigma$ uncertainties. The ``total'' refers to the sum of the signal coming from the multiple images B, C, D, and E, the contribution of the counter-image F being negligible. 
\tablefoottext{$\dag$}{Tentative detections below $3\,\sigma$.}
\tablefoottext{a}{Estimated by propagating the IRAC--{\it F814W} colors properly measured in the counter-image B over the sum of multiple images B, C, D, and E (``total'').}
%by adopting the flux scaling between the single multiple image B and the multiple images B, C, D, and E altogether (total) of the HST photometry.
\tablefoottext{b}{Obtained by integrating the flux over the same apertures as those used for the CO(6--5) emission detected over all multiple images B, C, D, and E.}
\tablefoottext{c}{By applying the 2~mm--{\it F814W} color conservation over the multiple images, we get a higher total 2~mm continuum flux of 0.9~mJy.}
\tablefoottext{d}{The color conservation is not satisfied, the measured flux is in excess.}
\tablefoottext{e}{All the multiple images B, C, D, and E contribute to the observed CO(4--3) line integrated flux given the IRAM 30~m telescope half power beam width of $24''$ at the redshifted frequency of the CO(4--3) line.}
\tablefoottext{f}{By applying the CO(1--0)--{\it F814W} color conservation over the multiple images, we get a similar total CO(1--0) line integrated flux of 0.20 Jy~km~s$^{-1}$.}
}
\end{table*}
%
%__________________________________________________________________

\section{Observations, data reduction, and flux estimates}
\label{sect:observations}

\subsection{HST, Spitzer, and Herschel data}
\label{sect:photometry}

%Our study of MACSJ0032-arc benefits from multi-wavelength imaging, covering the optical to FIR regime. Specifically, the lensing cluster was observed with HST/ACS in the {\it F606W} and {\it F814W} passbands, %(GO-12166, PI Ebeling), {\it Spitzer}/IRAC in the 3.6 and 4.5~$\mu$m passbands, and {\it Herschel}/SPIRE in the 250, 350 and 500~$\mu$m passbands. We focus on the multiple images B, C, D, and E, because of PdBI CO(6--5) detections associated with them. 

%The HST/ACS {\it F606W} and {\it F814W} rest-frame UV images reveal six multiple images of the $z=3.6314$ galaxy (see Fig.~\ref{fig:thumbnail}). 
%All the multiple images are composed of two compact knots, separated by $\sim 0.8''$ in the brightest multiple images B, C, D, and E, plus a tail attached to the brightest knot. 
The lensing cluster has been observed in two HST/ACS passbands, {\it F606W} and 
{\it F814W}. The astrometry of the two HST images has been adjusted to the 
astrometry of the interferometric JVLA 5~GHz continuum image by cross-matching 
positions of seven bright point-like sources detected in both the HST/ACS 
{\it F606W+F814W} images and in the JVLA image at more than $10\,\sigma$ and 
$8\,\sigma$, respectively. Median corrections of $+0.06''$ in right ascension 
and $-0.17''$ in declination were applied to the HST astrometry. The resulting 
absolute positional error is about 45~mas for the two HST passbands. 

In these high-resolution HST optical images, the multiple images of 
MACSJ0032-arc and the associated substructure are all resolved, and are 
generally unblended with foreground galaxy light in the cluster field. The sole 
exception is the elliptical galaxy, a cluster member labeled X, which affects 
the photometry of the very faint counter-image F and marginally affects image D 
(see Figs.~\ref{fig:thumbnail} and \ref{fig:contours}). We fit and subtracted 
the light profile of galaxy X using {\tt GALFIT} \citep{peng02}. To measure the 
flux of each counter-image, we first created segmentation maps by selecting, in 
the {\it F814W} image convolved with a 3-pixel full width half maximum (FWHM) 
Gaussian, all pixels above $1\,\sigma$. We separate counter-images C and D, that 
form a continuous arc, at the location of the critical line (see 
Fig.~\ref{fig:thumbnail}). We use these segments to integrate the flux in the 
{\it F814W} band and in the properly aligned {\it F606W} image. We hence obtain 
clean rest-frame UV-photometry individually for each of the multiple images B, 
C, D, and E.

%Things become more complicated in the {\it Spitzer}/IRAC 3.6 and 4.5~$\mu$m rest-frame optical images. 
Because of the much coarser resolution of the {\it Spitzer}/IRAC IR images 
available in the 3.6 and 4.5~$\mu$m passbands, the multiple images C, D, E, and 
F are blended with each other, as well as with the cluster member X mentioned 
above. This can be clearly seen in Fig.~\ref{fig:contours}, where we overlay the 
IRAC 3.6~$\mu$m contours on the HST {\it F814W} image. The IRAC flux even seems 
to peak at the position of galaxy X. Therefore, we derive the rest-frame optical 
%IRAC 3.6 and 4.5~$\mu$m 
photometry only for the well-resolved and unblended counter-image B.
For this purpose, we performed prior-based photometry, as in \citet{sklias14}. 
We used the reddest HST band to produce the so-called stamp image of the 
counter-image B. This stamp image includes only pixels within the SExtractor 
aperture. The pixels define the prior shape of the image B that is then 
convolved with the IRAC point-spread function (PSF) and scaled in flux.
%using aperture photometry with an appropriate aperture correction. 
The photometry of the other images C, D, and E, and of the multiple images B, C, 
D, and E taken together, are then recovered by equally applying the 
IRAC~$3.6~\mu$m--{\it F814W} and IRAC~$4.5~\mu$m--{\it F814W} colors measured 
in B. This is fully justified, since color is conserved from one counter-image 
to another, which is a well-established property of gravitational lensing.

In the three {\it Herschel}/SPIRE FIR images at 250, 350, and 500~$\mu$m, 
the multiple images B, C, D, E, and F are all blended within a single 
{\it Herschel} beam, with the exception of the counter-image A which falls 
outside the beam\footnote{The {\it Herschel} beam  varies in size from $18''$  
at 250~$\mu$m to $37''$ at 500~$\mu$m.}.
%with no hope of deblending 
Since MACSJ0032-arc is otherwise unblended with other SPIRE-detected sources 
(see Fig.~\ref{fig:thumbnail}, bottom right panel), simple aperture photometry 
with the standard SPIRE color and aperture 
corrections\footnote{http://herschel.esac.esa.int/hcss-doc-16.0/load/spire\_drg/html/ch06s09.html} yields a robust FIR flux measurement for the sum of the 
multiple images B, C, D, E, and F. We are confident that the contribution of 
the cluster member X to the {\it Herschel} flux is negligible. Indeed, at the 
redshift of the MACS\,J0032.1+1808 cluster, very few cluster galaxies besides 
the brightest cluster galaxies (BCGs) are likely to be detected in the SPIRE 
bands \citep{rawle12} and the galaxy X is not a BCG. Moreover, if the 
contribution of the galaxy X were significant, then one would expect the 
combined dust peak to be wider than observed (see Fig.~\ref{fig:SED}). To get 
the observed 250/350~$\mu$m color, any cluster galaxy contribution at 
250~$\mu$m has to be small and thus its contribution to the total FIR flux 
negligible. Other cluster members are more than $5-10''$ away from the 
MACSJ0032-arc; their contribution, if any, to the {\it Herschel} flux would 
yield some elongation in the SPIRE point source, which is not observed.

%The determination of the corresponding total HST/ACS {\it F606W} and {\it F814W} magnitudes is straightforward from the HST detections of the individual counter-images B, C, D, and E. In the case of the {\it Spitzer}/IRAC images, where accurate photometry is obtained only for the counter-image B, we take the IRAC~$3.6~\mu$m--{\it F814W} and IRAC~$4.5~\mu$m--{\it F814W} colours measured in B and apply them equally to the multiple images C, D, and E, as well as the sum of B, C, D, and E to derive the total B+C+D+E IRAC photometry. This is fully justified since colour is conserved from one counter-image to the other, which is a well-established property of gravitational lensing. 

The resulting multiwavelength photometry for the individual counter-images 
(when available) and the B, C, D, and E images taken together is summarized in 
Table~\ref{tab:photometry}. Because the {\it Herschel}/SPIRE photometry for the 
multiple images B, C, D, and E cannot be deblended (the very faint counter-image 
F can be neglected), we note that we are forced to work on the total B+C+D+E 
emission in all available bands, including the CO data, in order to get a 
coherent global picture.

%
%__________________________________________________________________

\subsection{30~m telescope data}

The IRAM 30~m telescope observations of MACSJ0032-arc were conducted on March 28 
and 29, 2012, as part of a blind CO line search program in the 3~mm band aimed 
at identifying the redshifts of bright HLS-snapshot FIR emitters. 
%(Dessauges-Zavadsky et~al., in prep.). 
For this purpose, we used the Eight MIxer Receiver (EMIR) combined with the 
32~GHz IF system, which includes 24 fast Fourier Transform Spectrometers (FTS) 
working at a spectral resolution of 200~kHz. A first redshift solution of 
$z_{\rm ISM} = 3.626\pm 0.008$ was obtained for MACSJ0032-arc from ISM 
absorption lines detected in the KeckI/LRIS spectrum (see 
Sect.~\ref{sect:target}).
%based on weak rest-frame UV ISM absorption lines and the Ly$\alpha$ break detected with the Keck\,I/LRIS spectrograph. 
Therefore, the 32~GHz bandwidth of our EMIR observation was distributed in two 
8~GHz wide IF outputs  centered on the E090 band (3~mm) and the E150 band (2~mm) 
each, in dual polarization. The two bands were tuned to the redshifted 
frequencies of the CO(4--3) and CO(6--5) lines at 99.663~GHz and 149.475~GHz, 
respectively. The corresponding half-power beam widths are $24''$ and $16''$, 
respectively. All multiple images B, C, D, E, and F contribute to the observed 
CO(4--3) and CO(6--5) fluxes, with the exception of the counter-image A which 
falls outside the 30~m beam (see Fig.~\ref{fig:thumbnail}). The observations 
were conducted in wobbler-switching mode with a frequency of 0.5~Hz and a 
symmetrical azimuthal wobbler throw of $50''$ to maximize the baseline 
stability. 
%Series of 12 ON/OFF subscans of 30 seconds each were performed and calibrations were repeated every 6 minutes. 
The total on-source integration time was 1.4~hours.

Data reduction was performed with the IRAM {\tt GILDAS} software package CLASS. 
Since the FTS suffers from severe platforming effects, corrections were applied 
using a dedicated script provided by the IRAM 30~m observatory. This correction 
was applied individually to each scan and led to baseline-subtracted spectra. 
All the scans obtained with the FTS backends tuned on the CO(4--3) line and 
those 
%obtained with the FTS backends 
tuned on the CO(6--5) line were averaged independently, using the temporal scan 
length as weight. The resulting CO(4--3) and CO(6--5) spectra were then Hanning 
smoothed to a resolution of $40.16~\rm km~s^{-1}$. They reach an RMS noise level 
of 2.4~mJy in 13.335~MHz channels and 2.2~mJy in 20~MHz channels, respectively, 
and reveal firm CO(4--3) and CO(6--5) emission line detections at 
$z_{\rm CO} = 3.6314\pm 0.0005$ (see Fig.~\ref{fig:COspectra}). The CO(4--3) 
line integrated flux listed in Table~\ref{tab:physicalparameters} is derived 
from a double Gaussian fit applied to the observed CO(4--3) line profile.

%
%__________________________________________________________________

\subsection{PdBI data}
\label{sect:PdBI}

The IRAM PdBI observations of MACSJ0032-arc were carried out on November 11 and 
16, 2013, in the C-configuration and in the 2~mm band, tuned to the redshifted 
frequency of the CO(6--5) transition at 149.300~GHz, as computed from the arc 
CO redshift $z_{\rm CO} = 3.6314\pm 0.0005$. 
%measured from the CO(4--3) and CO(6--5) lines detected with the IRAM 30~m telescope. 
%Six antennas in the C-configuration were used. The C-configuration still offers a sufficiently high sensitivity, while at the same time providing enough angular resolution at the tuned frequency to resolve individually the multiple images of MACSJ0032-arc. 
We used the WideX correlator that provides a continuous frequency coverage of 
3.6~GHz in dual polarization with a fixed channel spacing of 1.95~MHz 
resolution. A total of 3840 visibilities were obtained during 3.2~hours of 
on-source integration time.

Standard data reduction was performed with the IRAM {\tt GILDAS} software 
packages CLIC and MAPPING, with flux, bandpass, and phase calibration performed 
using the calibrators most suitable for our target. The data were mapped with 
the CLEAN procedure using the HOGBOM deconvolution algorithm and combined with 
``natural'' weighting, thus giving priority to mapping sensitivity rather than 
angular resolution. The resulting synthesized beam size is $1.94''\times 1.62''$ 
($\rm PA = +27.3\degr$). We reach an RMS noise per beam of 57~$\mu$Jy in the 
continuum after averaging the PdBI data over the full 3.6~GHz spectral range and 
having excluded the channels where CO emission is detected. The resulting 2~mm 
continuum is then subtracted from the UV table in order to obtain the UV table 
for the CO(6--5) line only. 

To improve the mapping accuracy for the CO(6--5) line emission, we added the 
short spacings from the IRAM 30~m single-dish data over the CO(6--5) detection 
channel interval using the UVSHORT procedure in MAPPING. This yielded an 
improvement of approximately 10\% and an RMS noise per beam of 0.7~mJy when 
resampling the PdBI+30~m data to a bandwidth of 20~MHz ($40.16~\rm km~s^{-1}$).
Using the GO MOMENT procedure, we determine the CO(6--5) first moment map 
(velocity-integrated map) of the cleaned, weighted images over the nine velocity 
channels where the CO(6--5) emission is detected. 
%The corresponding CO(6--5) spectrum (Fig.~\ref{fig:COspectra}) is extracted by spatially integrating the datacube over the total B+C+D+E CO(6--5) detection. 

The resulting CO(6--5) velocity-integrated contours overlaid on the HST/ACS 
{\it F814W} image are shown in Fig.~\ref{fig:contours}. The CO(6--5) emission is 
successfully detected in the four most strongly amplified multiple images of 
MACSJ0032-arc, namely images B, C, D, and E, with a signal-to-noise ratio (S/N) 
varying between $4\,\sigma$ and $6\,\sigma$. Counter-images A and F remain 
undetected in CO(6--5) because of their much lower magnification factors (see 
Sect.~\ref{sect:lensmodel}). 
%In all four images B, C, D, and E, the CO(6--5) emission is observed over 9 velocity channels. 
We derive the CO(6--5) line integrated fluxes of the individual images B, C, D, 
and E, and of their sum, using custom apertures over the velocity-integrated 
map, large enough to include, for each counter-image, all the signal above the 
background. The inferred CO(6--5) line integrated fluxes are listed in 
Table~\ref{tab:photometry}.

The 2~mm continuum emission is detected at $5.8\,\sigma$ in the most strongly 
amplified counter-image C only, as shown in Fig.~\ref{fig:contours} where the 
2~mm contours are overlaid on the HST/ACS {\it F814W} image. The other images B, 
D, and E remain undetected. Although this is surprising, we note that their 
fluxes are expected to reach at most the $3\,\sigma$ level when scaling the 
observed flux in image C by the respective B, D, and E magnification factors 
(see Fig.~\ref{fig:simulatedimageplane}). Emissions at this low confidence level 
may stochastically happen to be cancelled out, in large part by noise.
%At this detection level, continuum emission is usually hardly accessible. 
%By integrating {\bf over an aperture large enough},
Therefore, we derive a direct 2~mm continuum flux for the counter-image C, and 
estimate the total B+C+D+E flux  
%needed for comparison with the {\it Herschel} FIR flux as discussed in Sect.~\ref{sect:photometry}, 
by integrating the 2~mm continuum map over the same apertures as those used for 
the CO(6--5) emission detected over all multiple images B, C, D, and E. We find 
this estimate to be smaller by 38\% than
%consistent, within $1\,\sigma$ uncertainty, with 
the total flux derived when applying the 2~mm--{\it F814W} color conservation 
over the multiple images of MACSJ0032-arc (see Table~\ref{tab:photometry}). We 
thus set the error on the total 2~mm continuum flux in order to match the two 
flux estimates within $1\,\sigma$ uncertainty.

%
%__________________________________________________________________

\subsection{JVLA data}
\label{sect:JVLA}

The NRAO JVLA 5~GHz continuum observations of MACSJ0032-arc were conducted on 
February 9, 2014, and August 27, 2015, in the most extended A and BnA 
configurations (program IDs VLA/13B-358 and VLA/15A-215). 
%with a half-power beam width of $260''$.
%as single-pointing observations centred at the counter-image C. 
The half-power beam width is $260''$, ensuring that all the multiple images are 
within the central area of the primary beam. 
%with $\gtrsim 99$\% beam power. 
We used the WIDAR correlator and the C-band receiver with the 8-bit samplers 
tuned to $4.20-6.20~\rm GHz$ ($1024\times 2~\rm MHz$ channel, 2.0~GHz total 
bandwidth) to maximize the bandwidth unaffected by strong radio frequency 
interferences (RFI) at $4.0-4.2~\rm GHz$. 
%We observed 3C48 for the flux calibration and J0010+1724 for the phase calibration every 15 minutes.  
The total on-source time was 4.5~hours. 

The raw data were Hanning smoothed, calibrated by the {\tt CASA} pipeline 
provided by NRAO with appropriate calibrators used for the flux, bandpass, and 
phase calibration, 
%(Chandler et~al., in prep.), 
and manually flagged to exclude any remaining strong RFI. We imaged the data 
using the CLEAN procedure opting for the Multi-Scale Multi-Frequency Synthesis 
algorithm (MS-MFS) deconvolution algorithm \citep{rau11}, the Cottom-Schwab PSF 
mode, the Briggs weighting with a robust parameter of 0.5, and the tapering of 
baselines beyond 100~k$\lambda$ to facilitate the detection of any extended 
components. The resulting map has a synthesized beam size of 
$0.87''\times 0.62''$ ($\rm PA = +68.0\degr$) and an RMS noise per beam of 
$1.7~\mu$Jy.

The detected 5~GHz continuum contours overlaid on the HST/ACS {\it F814W} image 
are shown in Fig.~\ref{fig:contours}. All  four individual counter-images B, 
C, D, and E are detected at $3\,\sigma$ and more. The corresponding 5~GHz 
fluxes, and that of their sum are listed in Table~\ref{tab:photometry}. 
%we list the 5~GHz fluxes of the individual counter-images B, C, D, and E, as well as of their sum, derived 
They were derived using the \texttt{PyBDSM} software, performing multi 
two-dimensional Gaussian fits. 

We observed the CO(1--0) line emission of MACSJ0032-arc on June 6, 12, 23, and 
25, 2015, in the C configuration with a half-power beam width of $99''$ (program 
ID VLA/13A-385).
%again as single-pointing observations centred at the counter-image C. 
%The half-power beam width is $99''$, which encompasses all the multiple images. 
We used the WIDAR correlator and the K-band receiver with the 8-bit samplers 
tuned to two frequencies at 24.962~GHz and 24.986~GHz ($1024\times 2~\rm MHz$ 
channel, 2.0~GHz total bandwidth; the 24~MHz gap helps mitigate the effect of 
the lower sensitivity at the 128 MHz bandpass edges) to cover the redshifted 
frequency of the CO(1--0) transition at 24.889~GHz. 
%We observed 3C48 for the flux calibration and J0010+1724 for the phase calibration every 6 minutes. 
The total on-source time was 9.0~hours. 

Except for adopting no Hanning smoothing, the raw data were processed in the 
same way as the C-band data described above. We then used the CLEAN procedure 
to create an image cube and the IMCONTSUB procedure to subtract continuum 
emission. The resulting map has a synthesized beam size of 
$0.90''\times 0.79''$ ($\rm PA = -53.7\degr$). 

The CO(1--0) line emission is reliably detected only in the counter-image C in 
two velocity channels, each $144.54~\rm km~s^{-1}$ wide. 
%at the expected frequency of the line. 
By summing over the two velocity channels using the IMMOMENTS procedure, we 
achieve a $4.7\,\sigma$ CO(1--0) detection in image C, and $2-3\,\sigma$ 
detections in the other images B, D, and E, as shown in Fig.~\ref{fig:contours}. 
The CO(1--0) line integrated flux of image C, as well as the total B+C+D+E 
CO(1--0) line integrated flux obtained by integrating the velocity-integrated 
map over the same apertures as those used for the CO(6--5) emission detected 
over all multiple images B, C, D, and E, can be found in 
Table~\ref{tab:photometry}. The latter perfectly matches the total integrated 
flux derived by propagating the CO(1--0)--{\it F814W} color properly measured 
in image C over the sum of multiple images B, C, D, and E (see 
Table~\ref{tab:photometry}).

%
%__________________________________________________________________

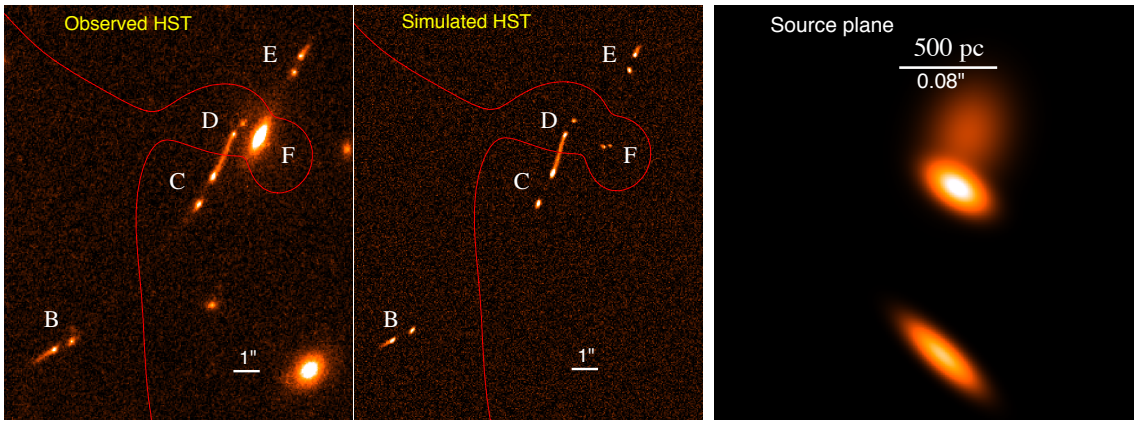
\begin{figure*}
\centering
\includegraphics[width=9.2cm,clip]{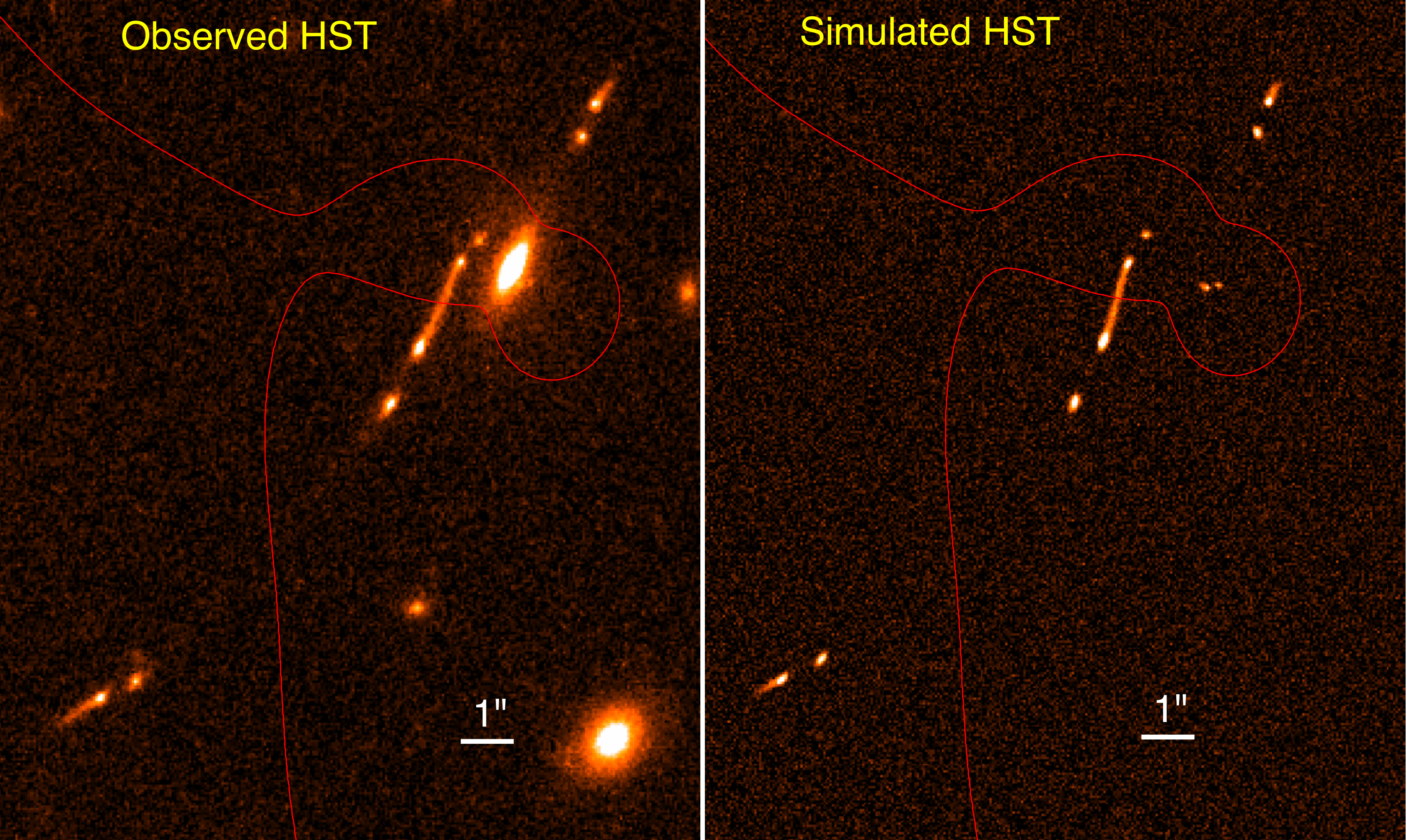}
\put(-247,35){\color{white}{\small B}}
\put(-200,87){\color{white}{\small C}}
\put(-188,110){\color{white}{\small D}}
\put(-165,135){\color{white}{\small E}}
\put(-158,97){\color{white}{\small F}}
%%\put(-200,100){\color{blue}{\small X}}
\raisebox{0pt}[0pt][0pt]{\hspace{0.05cm} \includegraphics[width=5.52cm,clip]{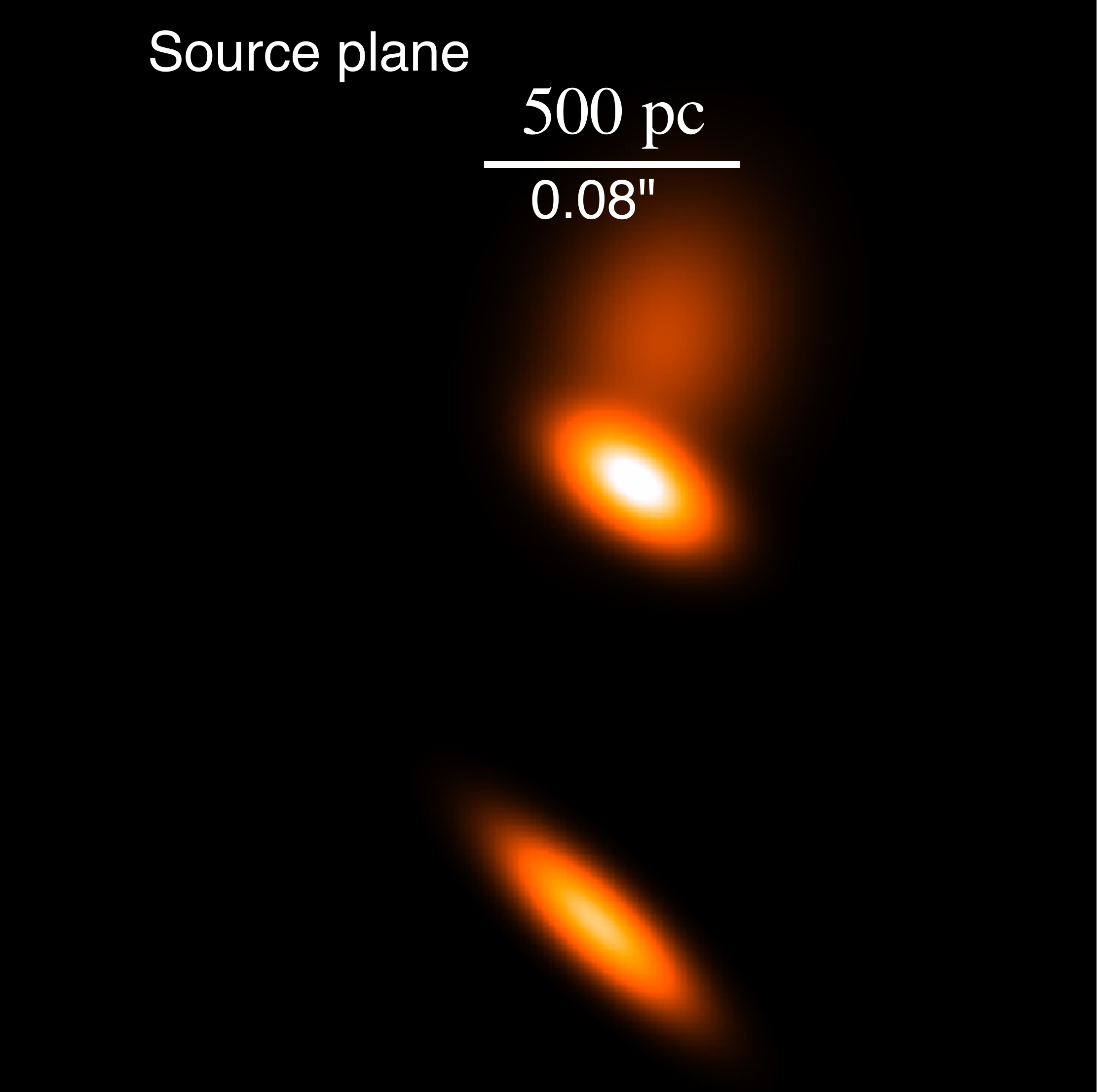}
\put(-281,35){\color{white}{\small B}}
\put(-232,87){\color{white}{\small C}}
\put(-222,110){\color{white}{\small D}}
\put(-199,135){\color{white}{\small E}}
\put(-191,97){\color{white}{\small F}}

}
\caption{Observed HST/ACS {\it F814W} optical image of the multiple images B, C, D, E, and F of MACSJ0032-arc with the critical line in red (left panel); simulated observation of MACSJ0032-arc resulting from the best lens model when accounting for the PSF of the HST image (middle panel); and source plane reconstruction of MACSJ0032-arc obtained by simulating the morphology of the source as the combination of three elliptical Gaussian light profiles (right panel). The two bright knots, resolved in the HST rest-frame UV images, are separated by $1.14\pm 0.28~\rm kpc$ in the source plane. The simulated HST observation includes a constant Gaussian noise background matched to the {\it F814W} noise level.} 
%In addition, we simulate, in the source plane, the CO(1--0) and CO(6--5) line emissions and the 2~mm and 5~GHz continuum emissions with a combination of one extended elliptical Gaussian light profile (labelled 1) placed between the two bright knots, and two elliptical Gaussian light profiles (labelled 2 and 3) co-spatial with the knot and its tail, respectively. The yellow contours in the right panel present the surface brightness contours of the three components of the simulated mm/radio emissions over a linear scale. The flux ratios are kept constant in the simulations. For each of the four observations (CO(1--0), CO(6--5), 2~mm, and 5~GHz) the absolute flux scaling is obtained by matching the corresponding beam-convolved map in the image plane with the observed flux at the peak of the emission in the counter-image C (Fig.~\ref{fig:simulatedimageplane}).}
%The corresponding beam-convolved emission maps in the image plane are shown in Fig.~\ref{fig:simulatedimageplane}.
\label{fig:lensmodel}
\end{figure*}
%
%__________________________________________________________________

\subsection{Flux anomalies}

A critical inspection of the fluxes measured for the individual counter-images
B, C, D, and E, found in Table~\ref{tab:photometry} shows that the 5~GHz 
continuum flux of the image E is a factor of 3 higher with respect to the 
expected flux from the color conservation that is supposed to hold over 
multiple images of a lensed object. We have carefully verified that image E is 
not affected by sidelobes from any bright cluster galaxy, and this is not the 
case.
%Performing various tests in the data reduction, by no means we manage to lower the 5~GHz continuum flux in E. 
A similar failure in the color conservation is also observed for the CO(6--5) 
emission in image E, although this is true to a smaller extent given the lower 
excess
%hk \LEt{ yes? also below. "Excess" suggests that it is too much, "increase" suggests that it is higher } 
of a factor of 1.6. 

We have no satisfying explanation for what can induce such an excess in both 
the 5~GHz JVLA and CO(6--5) PdBI data. We can only speculate on two phenomena.
%the relatively large uncertainties on the derived CO(6--5) line integrated fluxes of the individual counter-images B, C, and D smooth the observed colour variations from one image to the other. 
One possible phenomenon, which can possibly 
%hk \LEt{ eventually=it will definitely happen at a much later time, or do you mean "which can possibly"? } 
break the color conservation in a single counter-image, is the micro-lensing 
effect, which may very locally increase the flux of a specific region of the 
source. If this scenario is correct, it suggests that the radio continuum and 
the high-$J$ CO(6--5) transition have to be emitted from co-located regions of 
the MACSJ0032-arc galaxy. This is not true for either the CO(1--0) line emission 
or the 2~mm dust continuum; if it were true, their respective flux in image E 
would also be higher and this is not observed (see Table~\ref{tab:photometry}).
%If this is the right scenario, this suggests that the radio continuum and the high-$J$ CO(6--5) transition have to be emitted in a very close by region of the MACSJ0032-arc galaxy, on the contrary of the CO(1--0) line emission and the 2~mm dust continuum as their respective flux in image E would also be in excess and this is not the case (see Table~\ref{tab:photometry}).}
%Neither the CO(1--0) line emission nor the 2~mm dust continuum seem to be affected by such a micro-lensing effect, as their respective flux in image E would be higher; this would suggest that they are not exactly co-spatial with the 5~GHz and CO(6--5) emissions.
Another phenomenon that may also be considered is the lensing time-delay, 
where the light travel time or path length for each counter-image differs. 
If a variable AGN were present, it is possible that image E could be boosted 
at a given time in comparison to the other images. The continuum may also 
vary differently to the line emission, such that the resulting color varies 
in time.
%Whereas the light travel time is different for each counter-image, in presence of a variable AGN it could happen that the flux in the image E appears boosted at a given time, together with a chromaticism, since the continuum varies but not the CO(6--5) line emission, such that the resulting colour varies in time. 
Even if the AGN had an optical contribution, the variations at these wavelengths 
are not expected to be synchronous. However, 
%we do not have any evidence for an AGN in MACSJ0032-arc. Most importantly, 
the absence of Ly$\alpha$ and any other ISM lines in emission (especially 
C\,{\sc iv}) in the near-UV/optical Keck~I/LRIS spectrum (Richard et~al., in 
prep.) argues quite strongly against an AGN in MACSJ0032-arc.
%we see no sign of C\,{\sc iv} emission, which would be the strongest line in presence of an AGN, nor of Ly$\alpha$ emission in the near-UV/optical Keck~I/LRIS spectrum (Richard et~al., in prep.). Moreover, the presence of an AGN can also be excluded from the obtained FIR SED fits (Sect.~\ref{sect:SED}) and the FIR--radio relation (Sect~\ref{sect:COspatialdistr}).

To overcome the anomalous flux excess in the 5~GHz continuum and the CO(6--5) 
line emission in image E, we consider in subsequent analyses the expected 
total B+C+D+E flux derived by assuming the color conservation and the fluxes 
measured in counter-images unaffected by the excess (see 
Table~\ref{tab:physicalparameters}).

%
%__________________________________________________________________

\section{Analysis and results}
\label{sect:analysis-results}

\subsection{Gravitational lens model}
\label{sect:lensmodel}

%The HST optical images reveal six multiple images of this high-redshift galaxy, forming a giant arc.
We optimize a parametric mass model of the lensing cluster MACS\,J0032.1+1808 
based on strong-lensing constraints detected in the HST/ACS optical images. We 
use the positions of the five brightest multiple images A, B, C, D, and E 
forming the giant arc, including for each image the centroids of the two knots 
observed in the HST images (see Fig.~\ref{fig:thumbnail}). In addition, two 
triply imaged systems that we spectroscopically confirmed  (Richard et~al., in 
prep.) provide additional constraints on the large-scale mass distribution 
within the cluster lens. Our mass model considers components on two 
complementary scales: on large scales, we include two cluster-scale dark-matter 
haloes, parametrized by double pseudo-isothermal elliptical (dPIE) potentials 
\citep[e.g.,][]{limousin12}; and on galaxy scales, we model each member of the 
lensing cluster as a smaller-scale dPIE potential following a fixed scaling 
relation with its luminosity.  We use the Lenstool software \citep{jullo09} to 
optimize the lens parameters with a Bayesian Markov chain Monte Carlo (MCMC) 
sampler, which provides the best model by minimizing the distance between the 
predicted and observed locations of the counter-images, as well as a range of 
models sampling the parameter space of the potentials. Overall our best-fit 
models reproduce the location of all images with an RMS of 0.17''.

Deviating from the modeling prescription applied to all other cluster galaxies, 
we give special treatment to one bright cluster member, marked X in 
Fig.~\ref{fig:contours}. As its presence splits the giant arc into multiple 
images C to F, additional constraints ensue that allow us to model galaxy X 
separately from the scaling relation. Its effect on the shape of the critical 
line at $z\simeq 3.63$ is very noticeable in Fig.~\ref{fig:lensmodel} (left and 
middle panels). A counter-image F is predicted and indeed observed
%for the giant arc 
next to image D on the other side of the lensing galaxy X. By contrast, a 
seventh central counter-image is demagnified and, consequently, is not observed 
in our HST images. The combination of the potentials of the cluster and galaxy X 
leads to a very high magnification factor. To assess the value of the total 
magnification for the sum of the multiple images B, C, D, and E, which will be 
used in the rest of the paper (see Sect.~\ref{sect:photometry}), we need to 
properly simulate the merging pair of counter-images C and D, which experience 
the strongest magnification. 

We simulate the morphology of the source as the combination of three elliptical 
Gaussian light profiles (right panel of Fig.~\ref{fig:lensmodel}) and match 
their flux and shape to the photometry of the observed HST images 
(Table~\ref{tab:photometry}). This was performed by inverting the location of 
flux peaks in the image plane to the source plane, and producing for each 
component a grid of elliptical Gaussians with FWHM within 100~pc to 1~kpc in 
each direction and varying the orientation by steps of 10 degrees. We select for 
each component the source model that best matches the image plane morphology 
once lensed and convolved by the HST PSF. The morphology is well constrained 
along the shear direction, but is an upper limit in the direction perpendicular 
to it. The simulated observation of MACSJ0032-arc, corrected for the PSF of the 
HST/ACS {\it F814W} image, is shown in the middle panel of 
Fig.~\ref{fig:lensmodel} and globally reproduces the observed locations, shapes, 
and flux ratios of each counter-image, shown in the left panel of 
Fig.~\ref{fig:lensmodel}. We derive the total magnification of the sum of B, C, 
D, and E from the ratio between the total flux in the simulated image and the 
corresponding flux in the model image in the source plane (right panel of 
Fig.~\ref{fig:lensmodel}), and we get a value of $\mu = 62\pm 6$, where the 
quoted uncertainty represents the statistical error from the MCMC models.

%
%__________________________________________________________________

\begin{figure}
\centering
\includegraphics[width=8.9cm,clip]{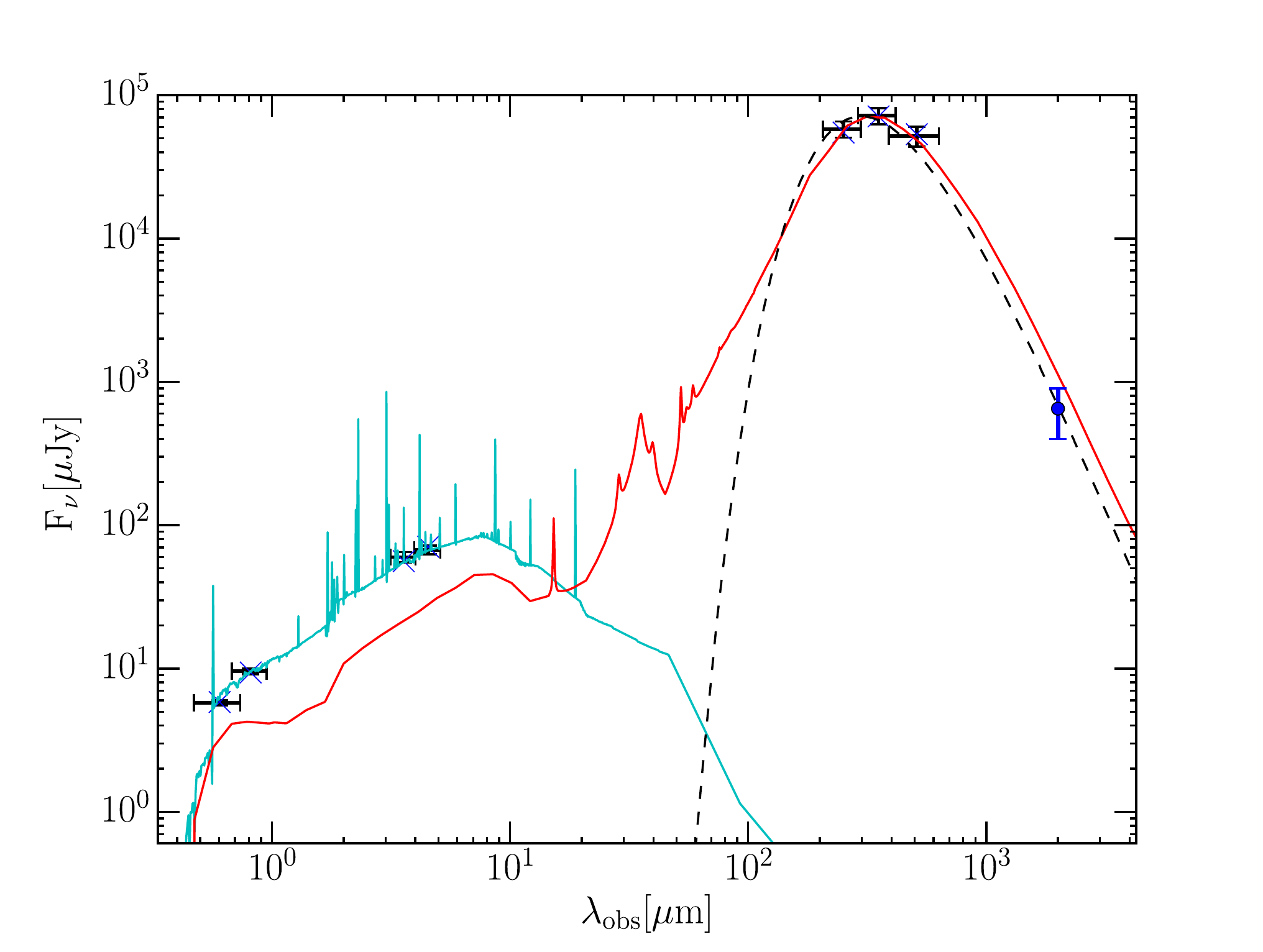}
\caption{Observed SED of MACSJ0032-arc at $z_{\rm CO} = 3.6314$. Overlaid are the best-fit stellar SED (solid cyan line) obtained with the energy-conserving model 
%with the extinction fixed to the observed $L_{\rm IR}/L_{\rm UV}$ ratio 
and constant $\mathit{SFR}$ SFH, as well as the best-fit FIR and mm SED obtained with the \citet{vega08} empirical template (solid red line) and the MBB with a $\beta$-slope fixed at 2 (dashed black line).}
\label{fig:SED}
\end{figure}
%
%__________________________________________________________________

Although the respective sizes of each of the three components in our source 
plane model are only constrained through upper limits because they are observed 
as compact sources after magnification, the separation between the two 
brighter knots is well-constrained.
%Since the three components of MACSJ0032-arc are observed as compact sources after magnification, their sizes in our model of the system in the source plane are only constrained from above (i.e., through upper limits). The physical separation of the two brighter knots, however, is well constrained at 
We derive a physical distance of $1.14\pm 0.28~\rm kpc$, taking into account 
both the statistical error from the MCMC models and systematic uncertainties 
assessed by comparing the distances for each of the counter-images B, C, D, and 
E in the source plane. MACSJ0032-arc is thus found to be a compact galaxy in the 
source plane with a global size $<2.5~\rm kpc$, after adding the three 
components used in our source model, each of which features a size of 
$<500~\rm pc$. This is a typical size for SFGs at $3<z<4$, 
%with $9.5 < \log(M_*/M_{\sun}) < 10.5$ and $L_{\rm UV} = 0.3-1~L^{\star}_{z=3}$, 
whose stellar effective radii range from 0.4~kpc to 3~kpc with a median at 
$1.3~\rm kpc$ \citep{shibuya15a}.
%The SFGs are all the more compact toward lower $M_*$.
%Just based on the morphology of the MACSJ0032-arc in the source plane, the two brighter knots may either testify from a merging scenario, or be two star-forming regions/clumps within a single, gravitationally bound galaxy, given their very small sub-kiloparsec sizes ($<500~\rm pc$ in diameter). Two scenarios further discussed in Sect.~\ref{sect:COspatialdistr}.}  

%
%__________________________________________________________________

\subsection{Stellar and dust content}
\label{sect:SED}

The combination of the available rest-frame UV, optical, and FIR photometry 
(see Table~\ref{tab:photometry}) enables constraints to be obtained on both the 
stellar and thermal dust spectral energy distributions (SEDs) of MACSJ0032-arc. 
We only consider the total photometry of the sum of the multiple images B, C, D, 
and E, 
%which is an indirect way to bring all the available bands to the same and scarce point spread function of the {\it Herschel}/SPIRE images, 
as motivated in Sect.~\ref{sect:photometry}. The fits to these SEDs determine 
the physical parameters of the galaxy, such as the IR luminosity ($L_{\rm IR}$), 
the extinction ($A_V$), $\mathit{SFR}$, $M_*$, and the age of the stellar 
population. We perform the fits by generating 1000 Monte Carlo realizations of 
the observed photometric data points, which are fit and used to determine the 
median values and the 68\% confidence intervals of the various physical 
parameters. The latter are then corrected for lensing effects by applying the 
total B+C+D+E magnification factor (Sect.~\ref{sect:lensmodel}).

We first model the dust-processed {\it Herschel} FIR SED using different 
empirical templates and the modified blackbody (MBB) function.
%and, in parallel, a modified black-body (MBB) function. 
The MBB provides the best fit to the FIR SED, with the PdBI 2~mm continuum
detection giving a strong constraint on the $\beta$-slope, which we fix at 2. 
The best empirical template reproducing the FIR SED is the one of 
\citet{vega08}, which, however, fails to reproduce the 2~mm dust continuum 
emission because the required $\beta$-slope is steeper than that of the 
template. The inferred respective total luminosities, $L_{\rm IR}$, of the 
multiple images B, C, D, and E taken together are in very good agreement and only  
differ by $0.01~\rm dex$. 
%(i.e., within the $1\,\sigma$ error) between the empirical template and MBB fits. 

%
%__________________________________________________________________

\begin{table}
\caption{Physical properties of MACSJ0032-arc.}             
\label{tab:physicalparameters}      
\centering          
\begin{tabular}{l l}     
\hline\hline   
Physical parameter & Value \\
\hline 
$z_{\rm CO}$                                & $3.6314\pm 0.0005$ \\
$\mu$~$^a$                                  & $62\pm 6$ \\
$L_{\rm UV}/\mu$ ($L_{\sun}$)~$^{\dag}$     & $1.1^{+0.2}_{-0.1}\times 10^{10}$ \\
$L_{\rm IR}/\mu$ ($L_{\sun}$)~$^{\dag b}$   & $4.8^{+1.2}_{-0.6}\times 10^{11}$ \\
$A_V$                                       & $1.29^{+0.12}_{-0.09}$ \\
$\mathit{SFR}_{\rm UV+IR}/\mu \simeq \mathit{SFR}_{\rm SED}/\mu$ ($M_{\sun}~{\rm yr}^{-1}$)~$^{\dag}$                        & $51^{+7}_{-10}$ \\
$M_*/\mu$ ($M_{\sun}$)~$^{\dag}$            & $4.8^{+0.5}_{-1.0}\times 10^9$ \\
$\mathit{sSFR} = \mathit{SFR}/M_*$ (Gyr$^{-1}$) & $10.6^{+2.1}_{-3.7}$ \\
$\mathit{SFR}_{\rm radio}/\mu$ ($M_{\sun}~{\rm yr}^{-1}$)~$^{\dag c}$ & $58\pm 11$ \\
$T_{\rm dust}$ (K)~$^d$                     & $43\pm 5$ \\
$M_{\rm dust}/\mu$ ($M_{\sun}$)~$^{\dag d}$ & $(1.9\pm 0.7)\times 10^7$ \\
$L'_{\rm CO(1-0)}/\mu$ (K~km~s$^{-1}$~pc$^2$)~$^{\dag e}$ & $(2.54\pm 0.73)\times 10^9$ \\
$L'_{\rm CO(4-3)}/\mu$ (K~km~s$^{-1}$~pc$^2$)~$^{\dag e}$ & $(1.53\pm 0.37)\times 10^9$ \\
$L'_{\rm CO(6-5)}/\mu$ (K~km~s$^{-1}$~pc$^2$)~$^{\dag e\,f}$ & $(0.70\pm 0.15)\times 10^9$ \\
r$_{4,1} = L'_{\rm CO(4-3)}/L'_{\rm CO(1-0)}$ & $0.60\pm 0.17$ \\
r$_{6,1} = L'_{\rm CO(6-5)}/L'_{\rm CO(1-0)}$ & $0.28\pm 0.08$ \\
$\alpha_{\rm CO}$ ($M_{\sun}/({\rm K~km~s^{-1}~pc^2})$)~$^g$ & $2.8-7.1$ \\
$M_{\rm molgas}/\mu$ ($M_{\sun}$)~$^{\dag h}$ & $(7.1-18)\times 10^9$ \\
$f_{\rm molgas} = M_{\rm molgas}/(M_{\rm molgas}+M_*)$ & $0.60-0.79$ \\
$t_{\rm depl} = M_{\rm molgas}/\mathit{SFR}$ (Gyr) & $0.14-0.35$ \\
\hline                  
\end{tabular}
\tablefoot{
\tablefoottext{$\dag$}{Lensing-corrected values divided by $\mu$.}
\tablefoottext{a}{Total magnification factor of the multiple images B+C+D+E.}
\tablefoottext{b}{IR luminosity obtained by integrating the [8,1000]~$\mu$m interval of the best-fit FIR SED.}
\tablefoottext{c}{Star formation rate as derived from the 5~GHz continuum and the $\mathit{SFR}$--1.4~GHz calibration from \citet{bell03}. The total 5~GHz flux (Table~\ref{tab:photometry}) is first corrected for the flux excess observed in image E by propagating the 5~GHz--F814W color properly measured in images B, C, and D over the counter-image E.}
\tablefoottext{d}{Dust temperature and dust mass obtained from the MBB fit to the FIR and 2~mm data (Table~\ref{tab:photometry}) with a $\beta$-slope fixed at 2 and Eq.~(\ref{eq:Mdust}).}
\tablefoottext{e}{CO luminosities inferred from the CO line integrated fluxes ($F_{\rm CO}$) in $\rm Jy~km~s^{-1}$ (Table~\ref{tab:photometry}) that have been beforehand corrected against the CMB (Sect.~\ref{sect:COSLED}), and derived using the \citet{solomon97} formula: $L'_{\rm CO}~({\rm K~km~s^{-1}~pc^2}) = 3.25\times 10^7 F_{\rm CO} \nu_{\rm obs}^{-2} D_L^2 (1+z)^{-3}$, where $\nu_{\rm obs}$ is the observed CO line frequency in GHz, and $D_L$ is the luminosity distance of the galaxy in Mpc.}
\tablefoottext{f}{The total CO(6--5) line integrated flux (Table~\ref{tab:photometry}) is first corrected for the flux excess observed in image E by propagating the CO(6--5)--F814W color properly measured in image C over the counter-image E.}
\tablefoottext{g}{CO-to-H$_2$ conversion factor as computed in Sect.~\ref{sect:alphaCO}.}
\tablefoottext{h}{Molecular gas mass obtained from $L'_{\rm CO(1-0)}$ and the $\alpha_{\rm CO}$ quoted above.}
}
\end{table}
%
%__________________________________________________________________

To model the stellar SED from the optical regime covered by the HST filters up 
to the {\it Spitzer}/IRAC IR bands, we use the energy-conserving models 
presented in \citet{sklias14}. The rationale behind this approach is to decrease 
the number of free parameters of the usual SED fitting by fixing the attenuation 
using $L_{\rm IR}$ and assuming that it is due to obscuration of the SED between 
0.912~$\mu$m and 3~$\mu$m. The observed $L_{\rm IR}/L_{\rm UV}$ ratio is then 
converted into $A_V$ using the calibration from \citet{schaerer13}. This 
attenuation is entered into the updated version of the {\it Hyperz} photometric 
redshift code \citep{schaerer09,schaerer10} to generate fits that conserve 
energy and alleviate the age-extinction degeneracy. 
%often encountered in obscured galaxies. 
Prior to this process, the stellar photometry is corrected for foreground 
extinction from the Galaxy, taken to be $E(B-V) = 0.11$ at the position of the 
galaxy cluster \citep{schlafly11}.
%, meaning that the stellar population model produced in this case will reproduce the actual observed $L_{\rm IR}$ without suffering from the eventual age-extinction degeneracy often encountered in obscured galaxies. These energy conserving models thus provide the most accurate physical parameters. 
When modeling SEDs, we consider star formation histories (SFHs) with both 
delayed exponential $\mathit{SFR}$ ($\propto t \exp(-t/\tau)$) and constant 
$\mathit{SFR}$ with a minimum age prior of $t_{\rm min} = 50~\rm Myr$. Either 
approach yields equally good best-fit SEDs with similar $\chi^2$ values and 
comparable outputs ($M_*$ and $\mathit{SFR}$) differing by about 10\%. We also 
explore the subsolar and solar metallicities; we adopt the subsolar 
metallicity, since it provides the best-fit SED with a $\chi^2$ of 0.40 against 
1.68 for the solar fit.
%the sub-solar ones which provide best-fit SEDs with much better $\chi^2$ values. 

%is fixed to $Z=0.2~Z_{\sun}$. 
%and the \citet{chabrier03} IMF is used. 

The resulting best-fit FIR and stellar SEDs with a constant $\mathit{SFR}$ SFH 
are shown in Fig.~\ref{fig:SED}. The inferred physical parameters for 
MACSJ0032-arc 
%resulting from the best-fit FIR and stellar SEDs shown in Fig.~\ref{fig:SED}, 
are summarized in Table~\ref{tab:physicalparameters} and reveal a low-mass, 
%hk IR luminous \LEt{ luminous infrared ? } 
luminous infrared galaxy (LIRG) with $M_*/\mu = 4.8^{+0.5}_{-1.0}\times 
10^9~M_{\sun}$. We discover that $\sim 90$\% of the total 
$\mathit{SFR}_{\rm UV+IR}$ of this high-redshift SFG is undetected at UV 
wavelengths, 
%because of dust obscuration, 
but is seen through the thermal FIR dust emission. We find our source to lie 
within about $-0.1$ to $+0.3~\rm dex$ of the main sequence at $z\sim 4$, 
meaning at most a factor of 2 above the MS depending on the adopted MS 
parametrization 
%(using, for example, the parametrizations from 
from \citet{tacconi13}, \citet{speagle14}, or \citet{tomczak16}. All these 
parametrizations still remain uncertain since the star formation main sequence 
and its scatter at $z\sim 4$ are not well defined yet, especially at low stellar 
masses comparable to MACSJ0032-arc.

Finally, the combination of the {\it Herschel} FIR photometry and the PdBI 2~mm 
continuum allows an estimate of 
%we consider the PdBI 2~mm continuum detected at $6\,\sigma$ in the most amplified multiple image C (no clear continuum is detected in any other multiple images as it can be appreciated in Fig.~\ref{fig:contours}). When integrated over the $1\,\sigma$ CO(6--5) detection contours of the multiple images B, C, D, and E (see Fig.~\ref{fig:contours}), we get an observed total integrated flux $F_{\rm 2~mm} = 0.65\pm 0.15~\rm mJy$ (not lensing corrected; see Table~\ref{tab:photometry}). We add it to the {\it Herschel} FIR photometry to estimate
%add the PdBI 2~mm continuum (see Table~\ref{tab:photometry}), detected at $6\,\sigma$ in the most amplified multiple image C (no continuum is detected in any other multiple images as appreciated in Fig.~\ref{fig:contours}), 
the dust mass ($M_{\rm dust}$) of MACSJ0032-arc to be obtained via the 
flux--$M_{\rm dust}$ calibration from \citet{kruegel03}
\begin{equation}\label{eq:Mdust}
M_{\rm dust} = \frac{S_{\nu}(\lambda_{\rm obs})D_L^2}{(1+z)\kappa(\lambda_{\rm rest})B_{\nu}(\lambda_{\rm rest},T_{\rm dust})}\,,
\end{equation}
where $S_{\nu}(\lambda_{\rm obs})$ is the flux at a given observed wavelength, 
$D_L$ the luminosity distance, $\kappa(\lambda_{\rm rest})$ the dust grain 
opacity per unit of dust mass, and $B_{\nu}(\lambda_{\rm rest},T_{\rm dust})$ 
the Planck function at a given rest-frame wavelength and dust temperature. For 
the opacities, $\kappa(\lambda_{\rm rest})$, we follow the \citet{li01} 
calibration.
%$\kappa(\lambda) \approx 2.29\times 10^5(\lambda/\mu{\rm m})^{-2}$ for $20 < \lambda < 700~\mu{\rm m}$, and $\kappa(\lambda) \approx 3.58\times 10^4(\lambda/\mu{\rm m})^{-1.68}$ for $700 < \lambda < 10^4~\mu{\rm m}$. 
We estimate $S_{\nu}$ from the best-fit MBB to the FIR and 2~mm data with the 
$\beta$-slope fixed at 2 (see Fig.~\ref{fig:SED}). 
%We consider the 2~mm continuum flux integrated over the $1\,\sigma$ CO(6--5) detection contours of the multiple images B, C, D, and E (see Fig.~\ref{fig:contours}) and get an observed total integrated flux of $F_{\rm 2~mm} = 0.65\pm 0.15~\rm mJy$ (not lensing corrected).
The resulting values for $M_{\rm dust}$ and $T_{\rm dust}$ are given in 
Table~\ref{tab:physicalparameters}. We propagate the error of the 2~mm continuum 
flux, carefully set in Sect.~\ref{sect:PdBI}, on $M_{\rm dust}$.

%
%__________________________________________________________________

\subsection{CO spectral line energy distribution}
\label{sect:COSLED}

At $z_{\rm CO} = 3.6314$ MACSJ0032-arc currently is the highest redshift typical 
LIRG with a measured CO(1--0) luminosity. 
%hk Taken all together  \LEt{ FYI: "altogether" is an adverb that means `completely', e.g., It stopped raining altogether.  }
The CO(1--0), CO(4--3), and CO(6--5) line detections within the arc offer the 
rare opportunity of characterizing the CO SLED and directly measure the CO 
luminosity correction factors, r$_{J,1} = 
L'_{{\rm CO}(J-(J-1))}/L'_{{\rm CO}(1-0)}$, for these high-$J$ CO transitions in 
a normal SFG at such a high redshift. This is of particular importance since the 
estimate of the molecular gas mass is based on the luminosity of the fundamental 
CO(1--0) line, which is rarely accessible at high redshift. In its absence, 
luminosity corrections have to be applied to the more readily observable 
rotationally excited lines.
%one needs to apply corrections for the ratio of the intrinsic Rayleigh-Jeans brightness temperatures in the 1--0 line to that in the available rotationally excited line \citep{papadopoulos12,lagos12,narayanan14}.

\begin{figure}
\centering
\includegraphics[width=8cm,clip]{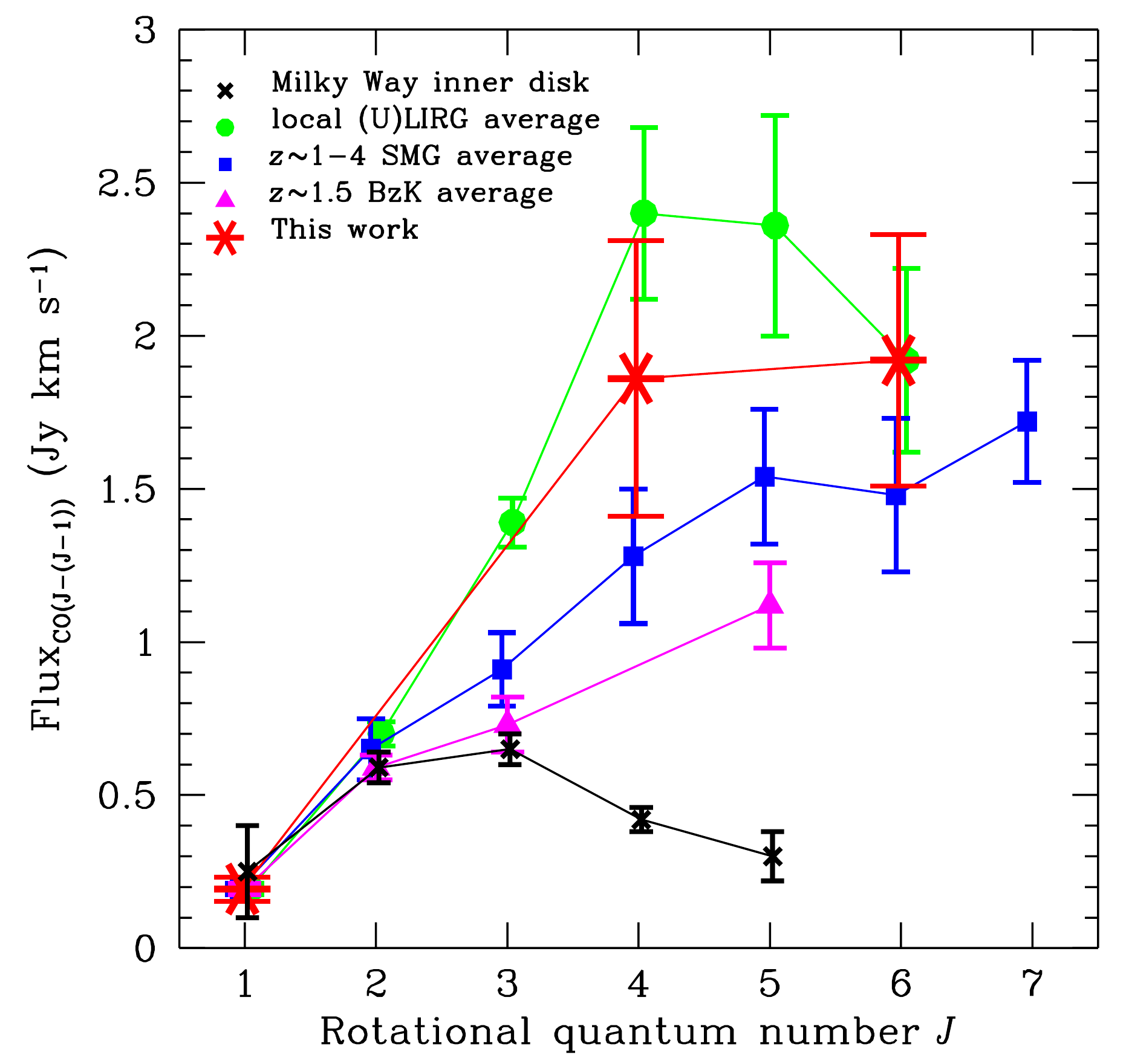}
\caption{CMB corrected CO SLED of MACSJ0032-arc at $z_{\rm CO} = 3.6314$ (red stars) compared to the average CO SLEDs of the Milky Way (black crosses), the local (U)LIRGs (green circles), the $z\sim 1-4$ SMGs (blue squares), and the $z\sim 1.5$ BzK galaxies (magenta triangles). All SLEDs are normalized to the CO(1--0) flux of the average SLED for BzK galaxies, except the Milky Way normalized using CO(2--1).}
\label{fig:COSLED}
\end{figure}

In Fig.~\ref{fig:COSLED} we show the CO SLED of MACSJ0032-arc that we have  
corrected for the cosmic microwave background (CMB) radiation, which is becoming 
non-negligible at $z\sim 3.6$, the redshift of the arc, with a temperature of 
$T_{\rm CMB} = 12.6~\rm K$. \citet{cunha13} computed the ratios between the line 
fluxes observed against the CMB and the intrinsic line fluxes for different CO 
transitions and redshifts in the local thermal equilibrium (LTE) case and in the 
non-LTE case. For the kinetic temperature of the gas in MACSJ0032-arc of 
$T_{\rm kin} \sim T_{\rm dust} = 43\pm 5~\rm K$ and the LTE case, the 
corresponding ratios are approximately 0.75 for CO(1--0), 0.85 for CO(4--3), and 
0.9 for CO(6--5). Moreover, the total CO(6--5) line integrated flux 
(Table~\ref{tab:photometry}) was first corrected for the flux excess observed in 
the counter-image E\footnote{This correction, derived by propagating the 
CO(6--5)--{\it F814W} color properly measured in image C over the 
counter-image E, represents a total CO(6--5) line integrated flux decrease of 
10\%.}. This yields the CO(1--0), CO(4--3), and CO(6--5) luminosities corrected 
against CMB listed in Table~\ref{tab:physicalparameters}.

%
%__________________________________________________________________

%
%__________________________________________________________________

The resulting CO SLED of MACSJ0032-arc is then compared to the average CO SLED 
of $z\sim 1.5$ BzK galaxies from \citet{daddi15}, $z\sim 1-4$ submm galaxies 
(SMGs) from \citet{bothwell13}, local (ultra-)LIRGs from \citet{papadopoulos12}, 
and the Milky Way inner disk from  \citet{fixsen99}. We find a clear CO SLED 
enhancement of MACSJ0032-arc over the Milky Way CO SLED and, more specifically, 
this galaxy demonstrates that high-$J$ transitions up to CO(6--5) may remain 
excited at least in some high-redshift normal SFGs. In fact, the CO SLED of this 
$z\sim 3.6$ galaxy appears to be similar to the SLED of high-redshift SMGs, with 
a turnover at $J>4$, even though its IR luminosity, 
$L_{\rm IR}/\mu = 4.8^{+1.2}_{-0.6}\times 10^{11}~L_{\sun}$, is 10 times lower 
than typically observed in SMGs. Although at first glance this seems unexpected, 
it is still in line with the CO SLED predictions for high-redshift SFGs from 
theoretical modeling. Indeed, the \citet{bournaud15} simulations predict a 
turnover at only $J=5$ for both disk and merger galaxies, and 
\citet{narayanan14} find that the overall CO excitation is regulated by the 
$\mathit{SFR}$ surface density, which also implies a turnover at $J=5$ for 
MACSJ0032-arc. In addition, \citet{papadopoulos12} propose a toy model with no 
turnover at $J<6$.

As a result, the CO(6--5) line detection in MACSJ0032-arc enables us to place a 
constraint on the CO SLED of high-redshift SFGs at a very high $J$ rotational 
level. We derive the following CO luminosity correction factors:
%The CO luminosity correction factors for MACSJ0032-arc are: 
r$_{4,1} = L'_{\rm CO(4-3)}/L'_{\rm CO(1-0)} = 0.60\pm 0.17$ and 
r$_{6,1} = L'_{\rm CO(6-5)}/L'_{\rm CO(1-0)} = 0.28\pm 0.08$\footnote{We find a 
slightly lower r$_{6,1} = 0.25$, but still within the $1\,\sigma$ uncertainty, 
when using the robust CO(1--0) and CO(6--5) line detections of the counter-image 
C alone (see Fig.~\ref{fig:contours}).} 
%rather than of the sum of the multiple images B, C, D, and E.} 
(Table~\ref{tab:physicalparameters}). These factors are comparable to those of 
high-redshift SMGs with r$_{4,1} = 0.41-0.60$ and r$_{6,5} = 0.21-0.46$ 
\citep{bothwell13,spilker14}. They are slightly higher than those of BzK 
galaxies at $z\sim 1.5$, with average r$_{3,1} = 0.42\pm 0.07$ and 
r$_{5,1} = 0.23\pm 0.04$ \citep{daddi15}, but formally still within the
uncertainties. The trend for a somewhat more excited CO SLED in SFGs at higher 
redshift needs to be further confirmed.
%When compared with the BzK galaxies at $z\sim 1.5$ with average r$_{3,1} = 0.42\pm 0.07$ and r$_{5,1} = 0.23\pm 0.04$ \citep{daddi15}, this $z\sim 3.6$ SFG suggests a trend for a somewhat more excited CO SLED at higher redshift. In fact, its CO SLED is comparable to that of high-redshift SMGs with r$_{4,1} = 0.41-0.60$ and r$_{6,5} = 0.21-0.46$ \citep{bothwell13,spilker14}. 
%The high excitation observed in the MACSJ0032-arc, while this SFG formally is not a starburst, ULIRG-type galaxy with its lensing-corrected $L_{\rm IR}/\mu = 4.8\times 10^{11}~L_{\sun}$ lower than $10^{12}~L_{\sun}$, but is a MS galaxy, may well be triggered by its compactness \citep{solomon97,weiss05,weiss07}
%(see Sect.~\ref{sect:lensmodel}) 
%rather than a merging process usually observed in local ULIRGs and high-redshift SMGs. This could reflect a general trend to be confirmed for $z\gtrsim 3.5$ SFGs, since galaxies are all the more compact toward higher redshifts \citep[e.g.][]{buitrago08,shibuya15}.

%
%__________________________________________________________________

\begin{figure*}[!]
\centering
\includegraphics[width=12.5cm,clip]{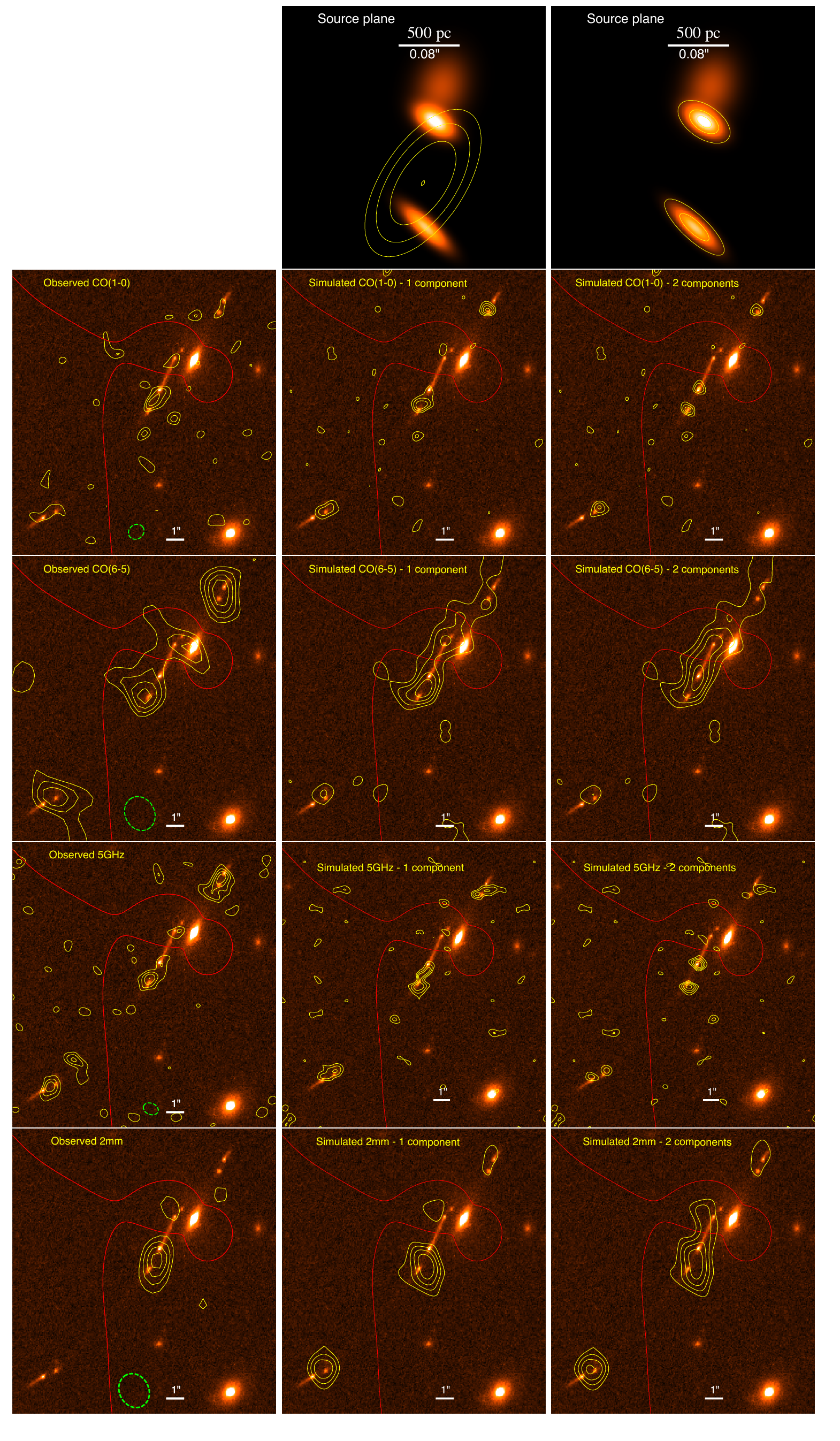}
\put(-340,30){\color{white}{\scriptsize B}}
\put(-300,75){\color{white}{\scriptsize C}}
\put(-293,94){\color{white}{\scriptsize D}}
\put(-270,115){\color{white}{\scriptsize E}}
\put(-266,87){\color{white}{\scriptsize F}}
\raisebox{0pt}[0pt][0pt]{
\put(-340,154){\color{white}{\scriptsize B}}
\put(-297,199){\color{white}{\scriptsize C}}
\put(-290,214){\color{white}{\scriptsize D}}
\put(-270,239){\color{white}{\scriptsize E}}
\put(-266,207){\color{white}{\scriptsize F}}

}
\raisebox{0pt}[0pt][0pt]{
\put(-343,275){\color{white}{\scriptsize B}}
\put(-303,320){\color{white}{\scriptsize C}}
\put(-296,339){\color{white}{\scriptsize D}}
\put(-273,360){\color{white}{\scriptsize E}}
\put(-269,332){\color{white}{\scriptsize F}}

}
\raisebox{0pt}[0pt][0pt]{
\put(-345,395){\color{white}{\scriptsize B}}
\put(-305,440){\color{white}{\scriptsize C}}
\put(-298,459){\color{white}{\scriptsize D}}
\put(-275,480){\color{white}{\scriptsize E}}
\put(-271,452){\color{white}{\scriptsize F}}

}
\caption{Comparison of the observed JVLA CO(1--0) line emission, PdBI CO(6--5) line emission, JVLA 5~GHz continuum, and PdBI 2~mm continuum contours, overlaid on the HST/ACS {\it F814W} multiple images B, C, D, and E of MACSJ0032-arc (left panels), with the respective simulated beam-convolved emission maps in the image plane (yellow contours in the middle and right panels) obtained for two different combinations of elliptical Gaussian light profiles in the source plane: one extended Gaussian component placed between the two HST UV-bright knots (yellow contours in the top middle panel) and two Gaussian components with locations and shapes following the two UV-bright knots (yellow contours in the top right panel). 
%The simulated maps have been derived with the combination of three elliptical Gaussian light profiles in the source plane shown in the right panel of Fig.~\ref{fig:lensmodel} (yellow contours).
Noise realizations based on the data are added to the simulated maps, and the absolute flux scaling is obtained by matching
%Only a global flux normalisation of the simulated images is adjusted to match
the observed flux at the peak of the emission in the counter-image C in each of the CO(1--0), CO(6--5), 5~GHz, and 2~mm simulated maps. 
%{\bf Note the slight offset in the predicted location of image B, due to the overall RMS of the lens model (0.17'', see Sect.~\ref{sect:lensmodel}).}
%once convolved by their respective beams. 
The critical line is in red. The size and orientation of the beams are indicated by the dashed green ellipses. Contour levels start at $2\,\sigma$ and are spaced in steps of $1\,\sigma$ (as in Fig.~\ref{fig:contours}).}
\label{fig:simulatedimageplane}
\end{figure*}
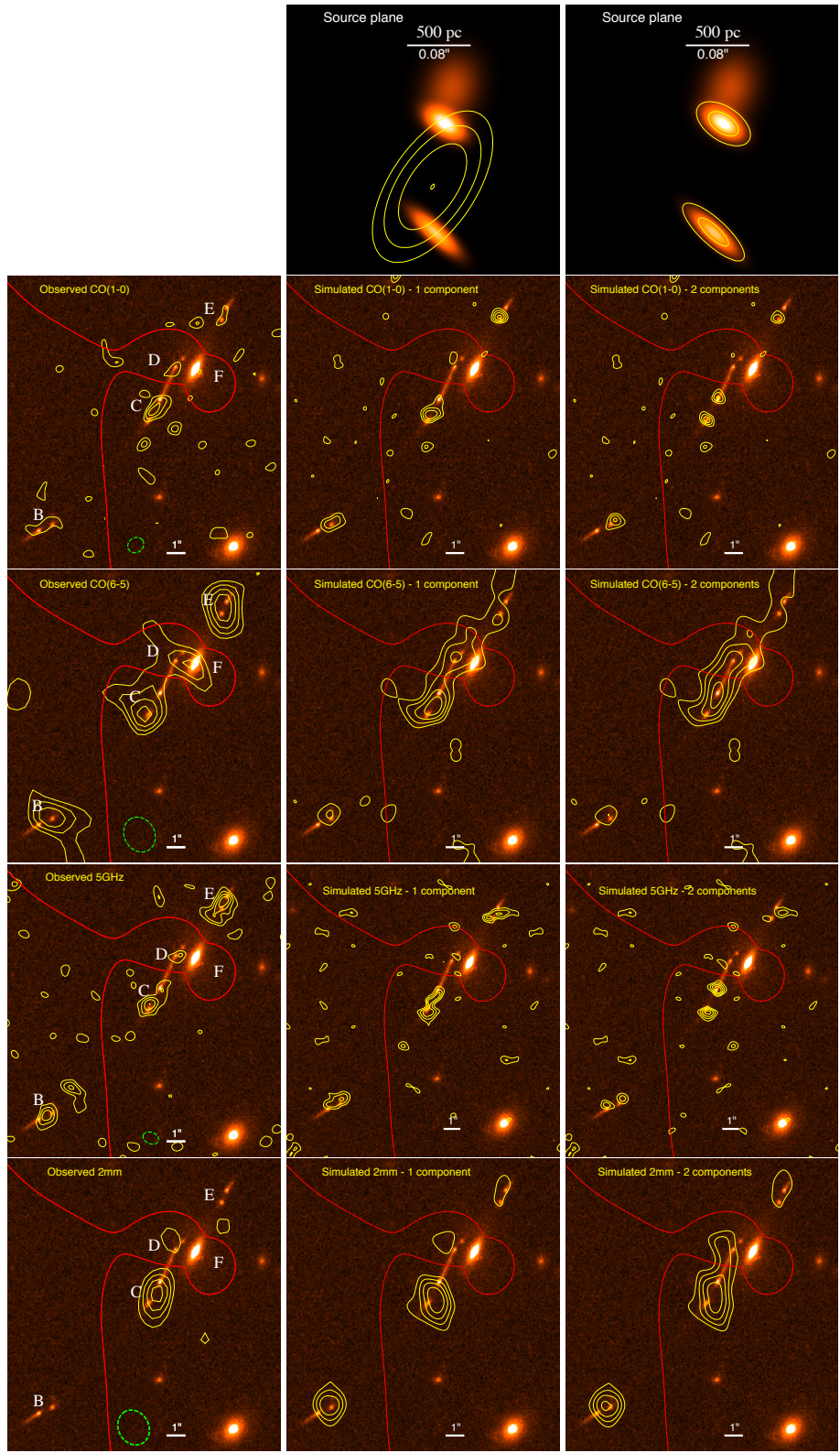
%
%__________________________________________________________________

\subsection{CO spatial distribution and kinematics}
\label{sect:COspatialdistr}

The CO(6--5) emission is successfully detected in the four most strongly 
amplified multiple images (B, C, D, and E) of MACSJ0032-arc (see 
Sect.~\ref{sect:PdBI}). As shown in Fig.~\ref{fig:contours}, where we overlay 
the velocity-integrated CO(6--5) contours on the HST/ACS {\it F814W} image, each 
individual counter-image is well resolved in CO(6--5), with the exception of 
some faint emission bridging the images C and D. However, the two knots that, at 
a separation of $\sim 0.8''$ ($1.14\pm 0.28~\rm kpc$ in the reconstructed source 
plane; see Sect.~\ref{sect:lensmodel}), are clearly resolved in the HST 
rest-frame UV data within each counter-image remain blended in the PdBI 
$1.94''\times 1.62''$ beam. They should be resolved in the JVLA CO(1--0) 
observations obtained with a beam size of $0.90''\times 0.79''$. However, a 
closer look at Fig.~\ref{fig:contours}, which also shows the velocity-integrated 
CO(1--0) contours overlaid on the HST/ACS {\it F814W} image, reveals that the 
peak of the CO(1--0) emission in the counter-image C falls between the two HST 
UV-bright knots. The JVLA radio continuum emission at 5~GHz, acquired at an even 
better angular resolution with a beam size of $0.87''\times 0.62''$, also  shown  
in Fig.~\ref{fig:contours}, exhibits a very similar spatial distribution pattern 
to that suggested by the CO(1--0) emission: detected in images B, C, D, and E 
(see Sect.~\ref{sect:JVLA}), it peaks between the two UV-bright knots, even in 
the most strongly amplified and stretched image C where it breaks into two 
centroids not co-spatial with the UV knots. Indeed, the observed centroid 
displacements between the UV and 5~GHz emissions is of $\sim 0.25''$ in image C, 
which cannot be accounted for either by the JVLA positional 
accuracy\footnote{Defined as $\rm \sigma_{position} \approx 
beam_{FWHM}/(2\times S/N)$ with the beam full width half maximum 
($\rm beam_{FWHM}$) in arcseconds and the signal-to-noise ratio (S/N) of the 
detection \citep{condon97}.} of $\sim 35$~mas at $5\,\sigma$ or by the absolute 
HST positional error of $\sim 45$~mas (Sect.~\ref{sect:photometry}).
%some weak signal seems also to be co-spatial with the knot in the tail. 
%as for the CO(6--5) emission (see the blue velocity component discussed below).

To ascertain the spatial origin of the CO(1--0) and CO(6--5) line emissions 
together with the 5~GHz and 2~mm continuum emissions 
%in the source plane 
with respect to the two UV-bright knots, we have searched for the best 
combination of elliptical Gaussian light profiles in the source plane 
%(yellow contours in the right panel of Fig.~\ref{fig:lensmodel}) 
that can reproduce all these emissions in the image plane.
%\footnote{{\color{red} It is evident that forcing all the emissions to arise from exactly the same location in the source model is a strong assumption,
%assuming all the emissions to be co-spatial is very simplistic; 
%especially for the CO(1--0) and CO(6--5) emissions that obviously trace gas with different densities and temperatures. However, this is the best we can do given the limited angular resolution and quality of the available CO(6--5) and 2~mm data, and also in order to prevent from hitting degeneracies in the model parameters.}}. 
%More precisely, the objective is to reproduce the best-resolution 5~GHz continuum data which provide the most stringent constraints, and get a consistent picture with the other data.
%with the most stringent constraints supplied by the best-resolution 5~GHz continuum data.
We are aware that forcing all the emissions to arise from exactly the same 
location in the source model is a strong assumption. Indeed, it is well known 
that the different CO lines of the rotational ladder require different physical
conditions to be excited: higher density and temperature at higher $J$. It is 
therefore expected that the CO(6--5) and CO(1--0) emissions are not exactly 
co-spatial, as is frequently observed in the nearby galaxies such as M82 
\citep{ward03} or NGC253 \citep{krips16}. Also, the continuum requires the 
heating of the dust, and does not have the same excitation requirements as the 
lines, thus is likely not to be exactly co-spatial. However, modeling this 
complex origin of CO and continuum emissions is not possible, first because of 
the limited angular resolution and quality of the available CO(6--5) and 2~mm 
data, and second because one would quickly be 
%also in order to prevent from 
hitting degeneracies in the model parameters.
  
%Moreover, our modelling already demands to handle several free parameters,
The free parameters used in our source plane simulations are the number of 
Gaussian light profiles, their locations, shapes, and relative fluxes. 
%that have not to hit degeneracies. 
To provide more realistic simulations of our observations, we have also 
constructed noise realizations based on the data, where all pixel values above 
$2.5\,\sigma$ were set to zero. We randomly attributed pixel values from these 
noise maps, convolved them with the corresponding beam, and scaled them to the 
same noise level as in the observations. We then follow the same procedure as in 
Sect.~\ref{sect:lensmodel} to model the source HST rest-frame UV emission using 
a grid of elliptical Gaussians. We first considered a grid of three Gaussians 
similar to the source plane reconstruction of the HST rest-frame UV emission, 
then a grid of two Gaussians coincident with the two HST UV-bright knots, and 
finally a grid of one single Gaussian. A global flux normalization of the 
simulated images was, in all explored combinations, adjusted to match 
%the brightest peak 
the observed flux at the peak of the emission in the most strongly amplified 
counter-image C in each of the CO(1--0), CO(6--5), 5~GHz, and 2~mm maps once 
convolved by their respective beams. 
%{\color{red} The resulting $2\,\sigma$ contours in our simulated noisy maps around the simulated sources show that the noise realisations are representative of the noise level in the observations. This can be appreciated in Fig.~\ref{fig:simulatedimageplane}, where we present two examples of combinations of elliptical Gaussian light profiles we have explored in the source plane:}

In Fig.~\ref{fig:simulatedimageplane} we show the combination of one and two 
elliptical Gaussian light profiles in the source plane that best reproduces
%hk \LEt{ i.e. the combination reproduces? } 
the observed CO/radio/mm emissions:
\begin{itemize}
\item[(i)] {\em One extended Gaussian placed between the two HST UV-bright 
knots (middle panels of Fig.~\ref{fig:simulatedimageplane})}: 
A compact elliptical Gaussian (PSF-dominated) fails to give the correct shape 
and orientation of the CO(1--0), CO(6--5), 5~GHz, and 2~mm emissions in the 
image plane, while an extended Gaussian component positioned between the two HST 
UV-bright knots in the source plane yields {the best} agreement between the 
simulated beam-convolved and observed maps in the image plane both in terms of 
spatial distribution (shape and orientation) and respective detection levels 
from one counter-image to the other.
%The CO(1--0) and 2~mm contours observed in the image plane are very well reproduced, but both the observed bridge between the counter-images C and D in CO(6--5) and the extended contours in the 5~GHz continuum over the knot in the tail in image C are absent in the simulated CO(6--5) and 5~GHz maps in the image plane. A compact, instead of an extended, elliptical Gaussian (PSF-dominated) even fails to give the correct shape and orientation in the image plane.
%
\item[(ii)] {\em Two Gaussians with locations and shapes following the two HST 
UV-bright knots (right panels of Fig.~\ref{fig:simulatedimageplane})}: 
The corresponding simulated contours of the unresolved CO(6--5) and 2~mm 
emissions show one single blended emission in the image plane located between 
the two UV-bright knots, as is seen in  the observations. 
%the fainter (less detected) images B, D, and E 
However, the simulated contours of the 5~GHz continuum and the CO(1--0) emission 
in the most strongly amplified image C are resolved and show two distinct 
emission peaks in the image plane
%a gap between the two knots, 
that fail to reproduce the observations. Bridging the two emission peaks is 
possible  
%This can adjust this 
by extending the size of each Gaussian in the direction of the other UV-bright 
knot in the source plane, but then the simulated contours in the image plane 
become too extended.
%
%\item[(iii)] {\em Three Gaussians with one extended placed between the two HST UV-bright knots (1), and two co-spatial with the knot (2) and its tail (3)}: 
%The 5~GHz contours in image C suggest two components (1) and (2), with the brightest and extended one (1) located between the UV-bright knots and a fainter and compact emission (2) centred on the knot in the tail. The bridge in CO(6--5) contours between images C and D suggests a third component (3) which follows the size and shape of the rest-frame UV emission in the tail. The flux ratio between (1) and (2) is fixed by the 5~GHz continuum, and the flux ratio between (1) and (3) is essentially fixed by the CO(6--5) emission. This combination of elliptical Gaussian light profiles (yellow contours in the right panel of Fig.~\ref{fig:lensmodel}) yields the best agreement between simulated beam-convolved and observed maps in the image plane, both in terms of spatial distribution and respective detection levels from one counter-image to the other, as shown in Fig.~\ref{fig:simulatedimageplane}.
\end{itemize}

%To further test the precise spatial origin of the CO(1--0) and radio continuum emissions in the source plane, we simulate both with an extended elliptical Gaussian placed between the two UV-bright knots (yellow contours in the right panel of Fig.~\ref{fig:lensmodel}). The corresponding beam-convolved emission map in the image plane, shown for the CO(1--0) line emission by the yellow contours in the middle panel of Fig.~\ref{fig:lensmodel}, resembles closely the data from the JVLA observations, both in terms of the spatial distribution and the respective detection levels from one counter-image to the other (see Fig.~\ref{fig:contours}). This good agreement is lost when the CO(1--0) emission is simulated with: (i)~a compact, instead of an extended, elliptical Gaussian (PSF-dominated) between the UV-bright knots in the source plane: the compact source remains compact in the image plane with neither the correct shape nor the orientation; and {\bf (ii)~two elliptical Gaussians in the source plane with locations and shapes following the two HST knots: in that case the contours along the less detected images B and E follow better the CO(1--0) observations in the image plane, but the most strongly amplified image C is resolved and shows a gap between the two knots, while this is not observed; one can adjust this by extending the size of each Gaussian in the direction of the other knot, but then the contours in the image plane are too extended}. 

%
%__________________________________________________________________

\begin{figure}[!t]
\centering
\includegraphics[width=8.0cm,clip]{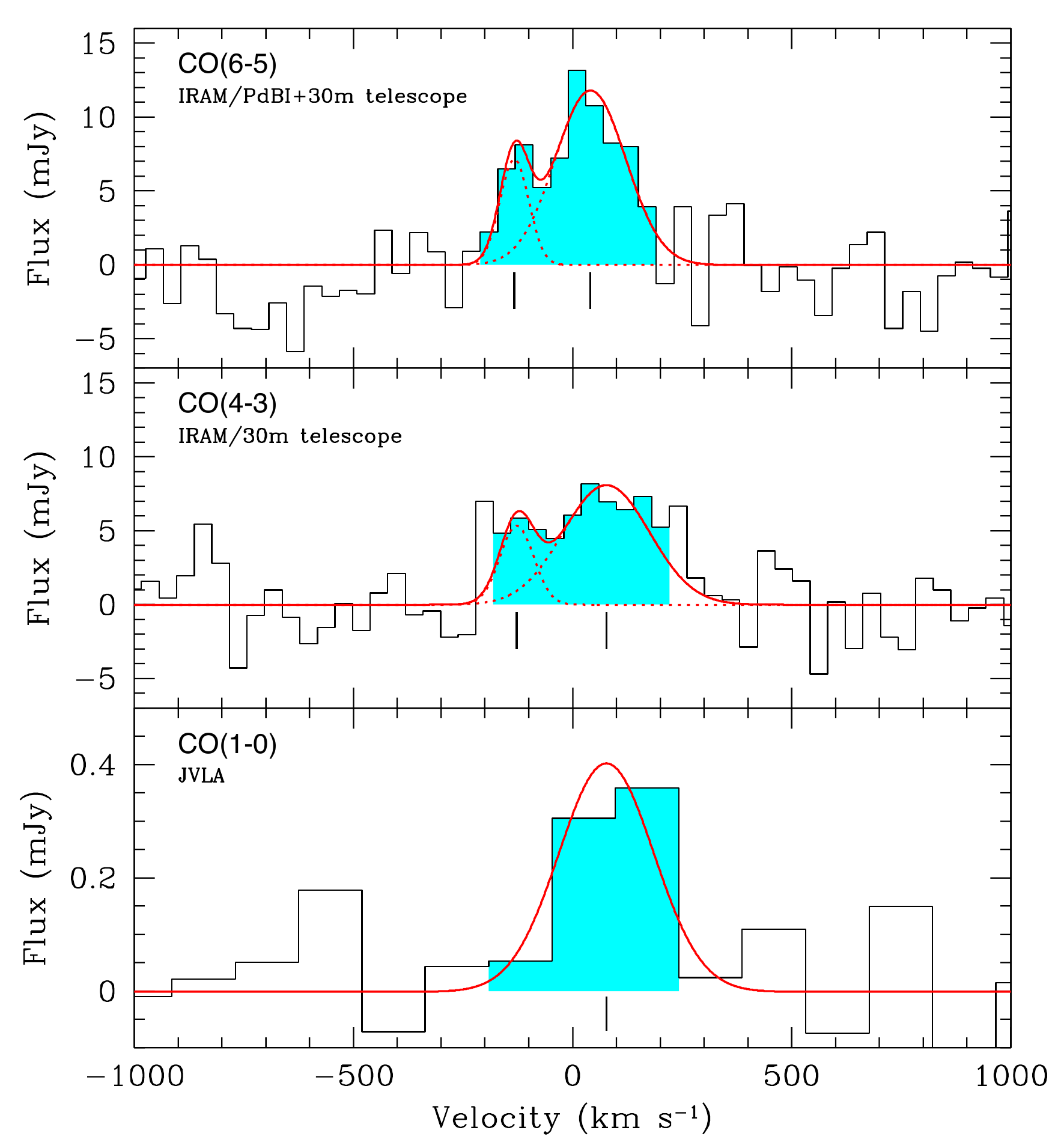}
\caption{Spectra of the CO(6--5), CO(4--3), and CO(1--0) emission lines detected in MACSJ0032-arc binned in steps of $40.16~\rm km~s^{-1}$, expect for the CO(1--0) line where channels have a resolution of $144.54~\rm km~s^{-1}$. 
%The CO(1--0) spectrum is also representative of the counter-image C only. 
The cyan-shaded regions indicate channels where positive emission is detected. These channels are used to derive the CO line integrated fluxes and the velocity-integrated and velocity-averaged maps shown in Figs.~\ref{fig:contours} and \ref{fig:CO65distr}, respectively. The solid red lines show the single or double Gaussian profiles that best fit the observed CO line profiles. The zero velocity is set to $z_{\rm CO} = 3.6314$.} 
\label{fig:COspectra}
\end{figure}
%
%__________________________________________________________________

\begin{figure*}[!t]
{\centering
\includegraphics[width=15.5cm,clip]{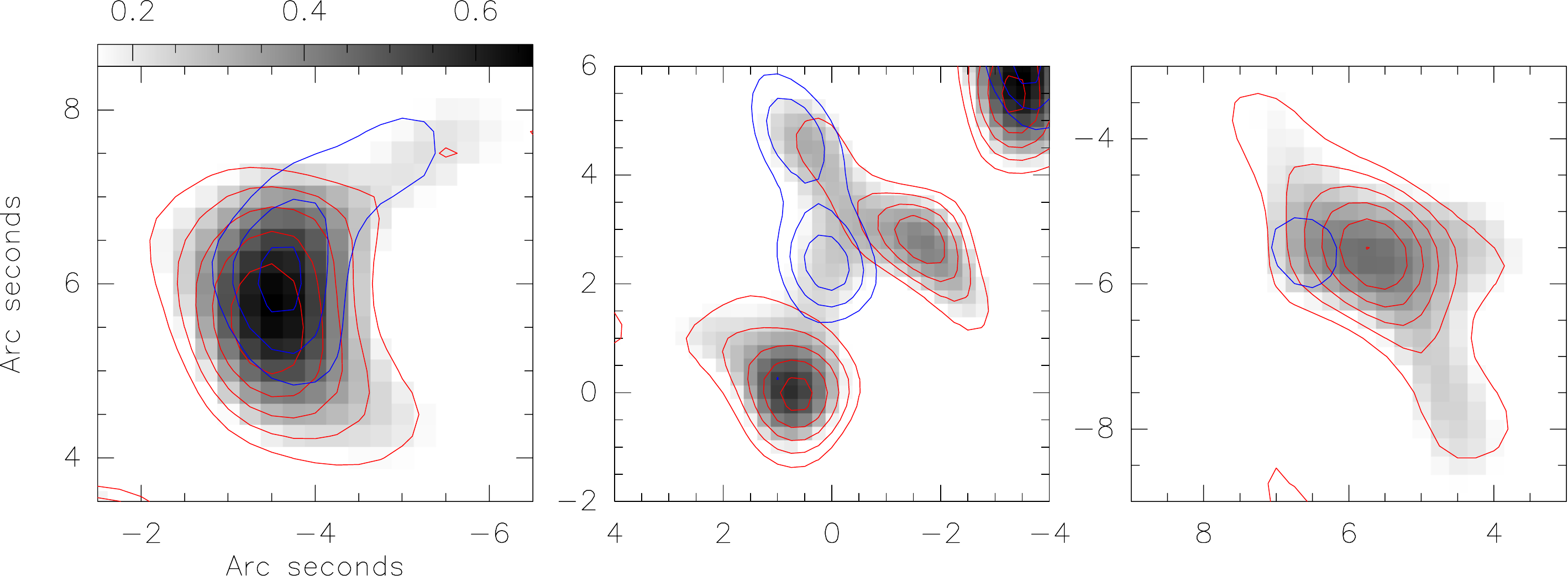}

}

\raisebox{1.5cm}[0pt][0pt]{\hspace{2.8cm} E \hspace{4.6cm} C+D \hspace{4.2cm} B}
\vspace{-0.3cm}
\caption{Velocity-averaged maps of the CO(6--5) emission of the counter-images E (left panel), C+D (middle panel), and B (right panel) of MACSJ0032-arc. Overlaid are the CO(6--5) contours integrated over, respectively, the channels that define the blue component (blue contours) and the red component (red contours) of the double-peaked CO(6--5) line profile shown in Fig.~\ref{fig:COspectra}. The color-coding is in units of integrated flux $\rm Jy~km~s^{-1}$ and starts at $2\,\sigma$. Contour levels also start at $2\,\sigma$ and are spaced in steps of $1\,\sigma$. A clear spatial offset between the blue and red contours is observed. The corresponding PdBI beam size of $1.94''\times 1.62''$ is about the size of the individual counter-images themselves.} 
%in the multiple images B and E, with the exception of the blue contours in the multiple images C and D which remain blended.
\label{fig:CO65distr}
\end{figure*} 
%
%__________________________________________________________________

%{\color{red} We have explored more complex combinations of Gaussian components in the source plane, {\bf but none of them yield} a better match with the observations in the image plane in comparison with the simple extended, one-component model.} 
In conclusion, the best simulated CO, 5~GHz, and 2~mm emissions are obtained 
with one extended Gaussian light profile placed between the HST UV-knots in 
the source plane, although some imperfections remain (see 
Fig.~\ref{fig:simulatedimageplane}) such as the slight offset in the predicted 
locations, seen in particular in image B, which is due to the overall RMS of the 
lens model (0.17'', see Sect.~\ref{sect:lensmodel}), the overestimated 2~mm 
continuum emission in images B and E that can result from our too optimistic 
noise realization (see also Sect.~\ref{sect:PdBI}), and the poor prediction of 
the CO(6--5) emission in image E. For this source plane reconstruction,
%Based on the best simulated CO, 5~GHz, and 2~mm emissions obtained with one extended Gaussian placed between the HST UV-knots, 
we estimate a total magnification factor of $\sim 65$ for the sum of the CO, 
5~GHz, and 2~mm emissions over the multiple images B, C, D, and E. Having 
derived a very similar magnification factor of $\mu = 62\pm 6$ for the total 
rest-frame UV emission of images B, C, D, and E (see 
Sect.~\ref{sect:lensmodel}), this indicates that differential magnification 
between the UV and CO/radio/mm emissions in MACSJ0032-arc is insignificant at 
the precision of our current measurements.

%The proposed spatial distribution and relative flux of the three elliptical Gaussian light profiles in the source plane (yellow contours in the right panel of Fig.~\ref{fig:lensmodel}), as the best model to reproduce the observed CO(1--0), CO(6--5), 5~GHz, and 2~mm emissions in the image plane, matched the CO kinematic signatures accessible in MACSJ0032-arc.
Searching for kinematic signatures, we observe that the CO(6--5) line, 
%spectrum extracted from the PdBI datacube 
extracted over all four multiple images B, C, D, and E, shows an asymmetric 
double-peaked emission line profile, which is similar to the CO(4--3) line, as shown in
%spectrum obtained with the IRAM 30~m telescope 
Fig.~\ref{fig:COspectra}. Adopting a double Gaussian model for the description 
of the observed CO(6--5) and CO(4--3) line profiles leads to a best-fit solution 
that has the two velocity components separated by $185\pm 15~\rm km~s^{-1}$, 
based on the nonlinear $\chi^2$ minimization and the Levenberg-Marquardt 
algorithm. 
%The resulting integrated flux of the CO(6--5) line is in perfect agreement with the value derived from the CO(6--5) velocity-integrated map. The observed CO(4--3) line integrated flux can be found in Table~\ref{tab:photometry}. 
In Fig.~\ref{fig:CO65distr} we plot, separately for the counter-images B, C+D, 
and E, the CO(6--5) contours integrated over, respectively, the velocity 
channels that define the red component and the blue component of the 
double-peaked CO(6--5) emission line profile. We find a clear spatial offset 
between the contours of the two velocity components in the counter-image C+D, 
while in images B and E only hints of some offset are observed. However, in 
all multiple images
%\footnote{The blue contours in the image C+D (Fig.~\ref{fig:CO65distr}, middle panel) show a single CO(6--5) emission peak, reflecting some residual blending between the counter-images C and D, as is also the case at a minor level in the HST images for the tails in C and D.}. Moreover,
the red and blue contours follow the same inversions from one counter-image to 
the other as observed in the HST images and as expected from the lens model 
(see Figs.~\ref{fig:thumbnail} and \ref{fig:lensmodel}). Overall, this  is 
suggestive of a possible signature of rotation.

The CO(1--0) line shows a single velocity component that is centered at the 
velocity of the strong red component of the CO(6--5) and CO(4--3) lines (see 
Fig.~\ref{fig:COspectra}). 
%This red velocity component observed in all CO line profiles, and containing most of the CO flux, hence very likely originates from between the two HST UV-bright knots, where most of CO(1--0) and CO(6--5) is observed to be emitted (see the right panel of Fig.~\ref{fig:lensmodel}). The $4\,\times$ fainter (in integrated flux) blue velocity component detected in the CO(6--5) and CO(4--3) line profiles is hence very likely associated with the knot and the tail, in agreement with the location of the CO(6--5) blue velocity component in Fig.~\ref{fig:CO65distr}. 
The non-detection of the blue, $4\,\times$ fainter (in integrated flux) CO(1--0) 
component is easily explained by both the low S/N of the CO(1--0) detection and 
the low spectral resolution of the CO(1--0) line of $144.54~\rm km~s^{-1}$, 
which implies that the blue component with $\rm FWHM = 83\pm 8~km~s^{-1}$, as 
derived by the Gaussian best-fitting solution of the CO(6--5) and CO(4--3) line 
profiles, is detected over one single velocity channel. 

We conclude 
%from the spatial distribution of the CO emission 
that the bulk of the molecular gas reservoir and cold dust seems, 
at the resolution of our data, to peak
%looks co-spatial and located 
between the two UV-bright knots, while cold gas and dust still extend over the 
whole galaxy (see the top middle panel of Fig.~\ref{fig:simulatedimageplane}).
%, confined to a region of at most 1.14~kpc in size. 
The fact that the radio continuum, which mostly traces the unobscured star 
formation through the synchrotron radiation from supernovae remnants, seems to 
arise from the same location indicates that most of the star formation is also taking place between the two UV-bright knots, a region
%taking place in this region
%this dusty star-forming region 
which is dark at the HST rest-frame UV wavelength\footnote{Not detected down to 
the $5\,\sigma$ magnitude limit of 27.4 as measured in a $0.2''$ diameter 
aperture in the HST/ACS {\it F814W} filter.}
%probed by our HST images 
because it is too dusty. This is the location where $\sim 90$\% of the star 
formation of MACSJ0032-arc occurs since $\mathit{SFR}_{\rm UV}$ represents only 
10\% of the total $\mathit{SFR}_{\rm UV+IR}$ of the galaxy 
(Sect.~\ref{sect:SED}). This agrees with the lensing-corrected 
$\mathit{SFR}_{\rm radio}/\mu = 58\pm 11~M_{\sun}~\rm yr^{-1}$ (see 
Table~\ref{tab:physicalparameters}), as determined from the radio 5~GHz 
continuum flux and the $\mathit{SFR}$--1.4~GHz calibration from \citet[][see 
also \citealt{rieke09}]{bell03}. The 5~GHz flux was first corrected for the flux 
excess observed in the counter-image E (see Sect.~\ref{sect:JVLA}) by 
propagating the 5~GHz--F814W color properly measured in images B, C, and D over 
the counter-image E. The spatial offset between UV clumps and the bulk of star 
formation has been found in many $z\sim 2$ SFGs, with only 5--10\% of the star 
formation emerging in the rest-frame UV \citep{rujopakarn16,dunlop17}.

%
%__________________________________________________________________

\subsection{Morphology: single galaxy versus merger}
\label{sect:morphology}

Gathering all information available on MACSJ0032-arc, we now attempt to arrive 
at a self-consistent interpretation of its observed morphological properties 
summarized in the following. The arc is shown to be a fairly low-mass, normal 
SFG with a physical size of $\sim 2.5$~kpc or smaller, comparable to the MS 
galaxy population at $3<z<4$. It is composed of two very compact rest-frame 
UV-bright knots with sizes $<500~\rm pc$ and separated by $1.14\pm 0.28$~kpc in 
the source plane (Sect.~\ref{sect:lensmodel}). The bulk of the molecular gas, 
dust content, and star formation, found to be dust obscured, is spatially offset 
from the rest-frame UV emission and is located between these two UV-emission 
regions (Sect.~\ref{sect:COspatialdistr}). Neither the CO(1--0) emission nor the 
radio continuum at 5~GHz is fully resolved in the JVLA observations and may 
hence originate from a region 
%in between the two UV-bright knots 
with a physical size as large as the separation between the UV-bright knots or 
smaller ($\lesssim 1.14~\rm kpc$). Our gravitational lens model suggests an 
extended rather than a compact CO, dust, and radio emission 
(Fig.~\ref{fig:simulatedimageplane}). 

Two scenarios may explain these observational facts. First, the two UV-bright 
knots may be two separate entities in the process of merging, creating a 
compression zone in their middle where most of the molecular gas mass is found, %and along a tidal stream or bridge of molecular gas that is connecting the two objects 
and triggering new, still dust obscured, star-bursting regions, similar to 
what is observed in some local galaxies (e.g., the Antennae, 
\citealt{whitmore99}, and NGC\,6240, \citealt{rieke85}). This scenario would 
most easily explain the more diffuse tail seen extending from one of the 
UV-bright knots.
%However, this scenario would all but require MACSJ0032-arc to be not a MS galaxy, but a starburst galaxy observed above the MS, as are the local ultra-luminous IR galaxies (ULIRGs) and the high-redshift submm galaxies (SMGs) generally identified as mergers. 
On the other hand, the two very small knots with sub-kpc sizes may be two 
star-forming regions, or clumps, within a single gravitationally bound galaxy, 
whose molecular gas and star formation resides and occurs primarily in its 
center, similarly to the numerous high-redshift clumpy galaxies observed in HST 
images \citep[e.g.,][]{elmegreen05,guo15}. This scenario is supported by the 
arc's main sequence and LIRG nature, and could explain both the observed 
double-peaked CO(4--3) and CO(6--5) emission line profiles 
(Fig.~\ref{fig:COspectra}) and the spatial offset between the red and blue 
velocity components in the CO(6--5) velocity-averaged map 
%of the counter-images B, C+D, and E 
(Fig.~\ref{fig:CO65distr}) as signatures of rotation. The measured velocity 
separation of $185~\rm km~s^{-1}$ between the red and the blue velocity 
components is also compatible with this scenario.

%The highly excited molecular gas observed in the MACSJ0032-arc, with a CO SLED comparable to that of high-redshift SMGs (Sect.~\ref{sect:COSLED} and Fig.~\ref{fig:COSLED}), should a priori bring to a conclusion between these two scenarios by favouring the merging scenario, since SMGs are known mergers. However, the high CO excitation in the arc cannot be assigned to the ULIRG/SMG type nature given its lensing-corrected $L_{\rm IR}/\mu = 4.8\times 10^{11}~L_{\sun}$ found below $10^{12}~L_{\sun}$, the lower $L_{\rm IR}$ limit defining the ULIRGs. It is thus
Given the observed excited state of the CO molecular gas in MACSJ0032-arc, 
comparable to that of high-redshift SMGs (Sect.~\ref{sect:COSLED} and 
Fig.~\ref{fig:COSLED}), we might be tempted to favor the merging scenario. 
%though not a major one which would be at odds with the arc's main sequence nature. 
However, this high excitation is more fundamentally related to the 
compactness of the object, 
%More likely the cause of the high excitation lies in the system's compactness, 
irrespective of its origin (merger or not). 
%This is because the compactness induces, on average, a higher molecular gas density 
Indeed, a compact object will, on average, have a higher molecular gas density 
that will lead to more excitation of CO by collisions with H$_2$ 
\citep[see][]{solomon97,weiss05,weiss07}. 
%This could reflect a general trend at $z\gtrsim 3.5$ (to be confirmed) of 
%galaxies becoming more compact at ever higher redshifts 
This compactness could reflect the general fact that galaxies are more compact 
at higher redshifts ($z\gtrsim 3.5$) because of their smaller sizes 
\citep[e.g.,][]{buitrago08,shibuya15a}. Therefore, we interpret this system as 
a single galaxy with two UV-bright star-forming regions, but this configuration 
may still be the result of a recent merger or accretion event, albeit not one 
big enough to completely disrupt the kinematic state of the galaxy.
%Overall, we therefore prefer the single-galaxy with two star-forming regions over the merging scenario for MACSJ0032-arc. Such a configuration may be the result of a recent merger or accretion event, albeit not one big enough to completely disrupt the kinematic state of the galaxy.
%In the global balance, the single galaxy scenario is hence preferred over the merging scenario for the MACSJ0032-arc.

%
%__________________________________________________________________

\section{Discussion}
\label{sect:discussion}

%At $z_{\rm CO} = 3.6314$ MACSJ0032-arc currently is the highest redshift normal star-forming galaxy (LIRG-type) with a measured CO(1--0) luminosity. To convert $L'_{\rm CO(1-0)}$ to a molecular gas mass, we need to assume a CO-to-H$_2$ conversion factor. 
%The CO-to-H$_2$ conversion factor is often assumed to be the same as that measured in the Milky Way. However, 
%It has been shown, both theoretically and empirically, that this conversion factor can vary with physical conditions, such as the metallicity, the temperature and density, and the dynamical state of a galaxy \citep{narayanan12,bournaud15,solomon97,leroy11,bolatto13}. 
In this section, we make use of the wealth of data we have on MACSJ0032-arc to 
explore the CO-to-H$_2$ conversion factor in this galaxy. We then add our new 
CO(1--0) measure to the small sample of five prior CO detections in SFGs at
$z>2.5$ \citep{riechers10,johansson12,tan13,saintonge13}\footnote{We do not 
include the very tentative CO measurement from \citet{livermore12} obtained in a 
strongly lensed galaxy at $z\sim 4.9$.} to study the cosmic evolution of the 
molecular gas depletion timescale and that of the molecular gas fraction. To 
place our object in the general context, we consider the compilation by 
\citet[][and references therein]{dessauges15} of local spirals and main sequence 
SFGs at both $z<0.4$ and $z>1$ with CO measurements from the literature.

%
%__________________________________________________________________

\subsection{CO-to-H$_2$ conversion factor}
\label{sect:alphaCO}

%To derive the molecular gas mass ($M_{\rm molgas}$) of the MACSJ0032-arc and ofgalaxies in general, the major difficulty resides in evaluating 
A major difficulty in measuring the molecular gas mass ($M_{\rm molgas}$) of 
individual galaxies is to determine the CO-to-H$_2$ conversion factor 
($\alpha_{\rm CO}$) that relates $M_{\rm molgas}$ to the CO(1--0) luminosity as 
$M_{\rm molgas} = \alpha_{\rm CO}\times L'_{{\rm CO(1-0)}}$, shown to vary with 
physical conditions (metallicity, temperature, density, dynamical state of the 
galaxy, etc.). There is a growing consensus that $\alpha_{\rm CO}$ and 
metallicity are inversely correlated for galaxies with metallicities from 
$Z\sim 0.2~Z_{\sun}$ to $Z\sim 2~Z_{\sun}$, because of the increasing fraction 
of CO that is photo-dissociated at low metallicity due to the underabundance of 
dust that yields more intense radiation fields, resulting in molecular gas that 
is deficient (dark) in CO \citep{maloney88,wilson95,israel00,wolfire10,leroy11,
bolatto13,sternberg14}. 
%For the sake of uniformity, we determine their molecular gas masses ($M_{\rm molgas}$) by assuming 

\begin{figure}
\centering
\includegraphics[width=9cm,clip]{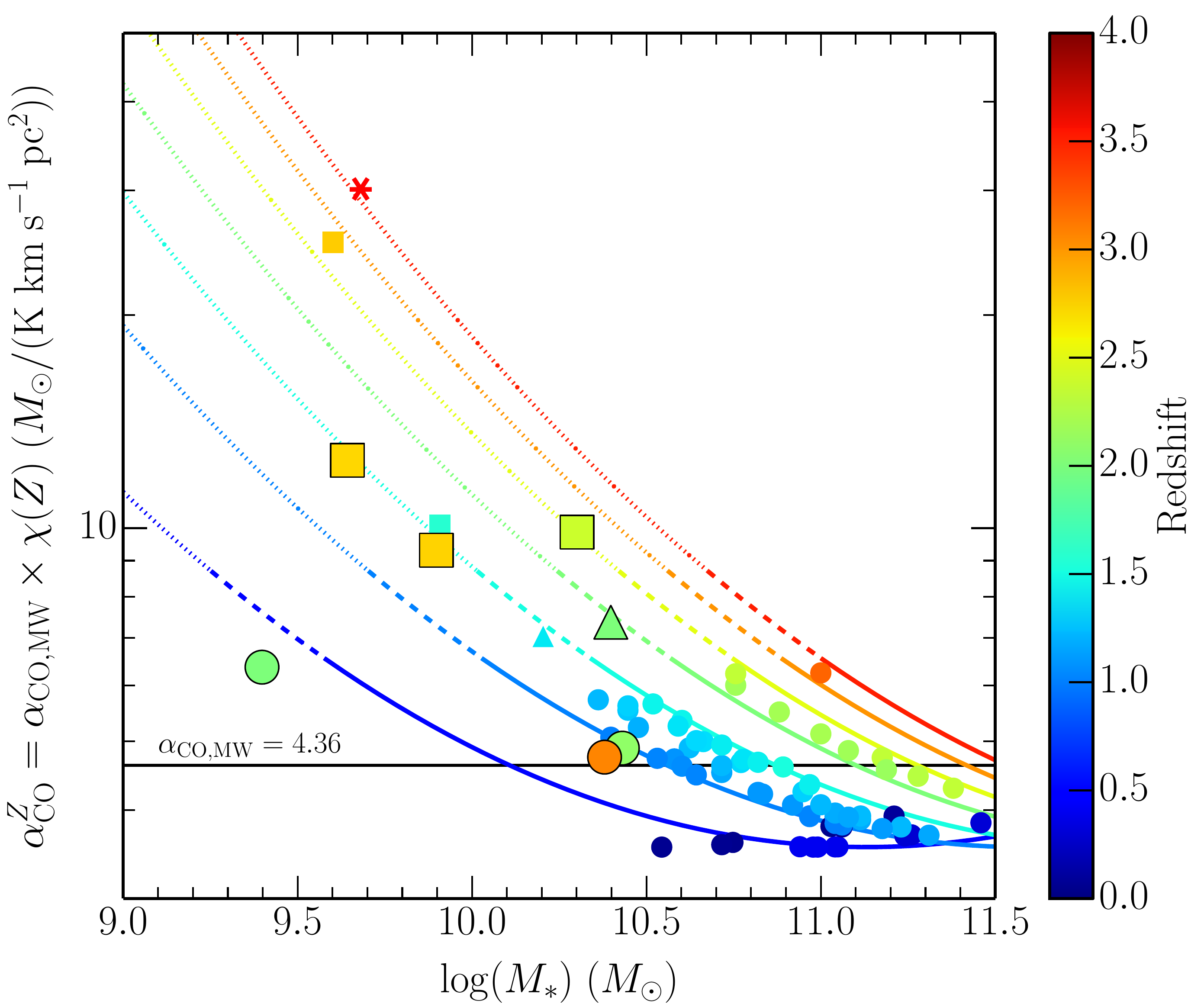}
\caption{Dependence of the metallicity-dependent CO-to-H$_2$ conversion function, computed from Eqs.~(\ref{eq:alphaCO},\,\ref{eq:Xsi},\,\ref{eq:metallicity}), on stellar mass and redshift. The lines are the iso-redshift curves, color-coded as a function of redshift, for $z=0.5$ to $z=3.5$ in steps of $\Delta z = 0.5$: the solid sections correspond to $\alpha_{\rm CO}^Z/\alpha_{\rm CO,MW} < 1.5$, the dashed segments to $1.5 < \alpha_{\rm CO}^Z/\alpha_{\rm CO,MW} < 2$, and the dotted portions to $\alpha_{\rm CO}^Z/\alpha_{\rm CO,MW} > 2$. The horizontal black line marks the Milky Way CO-to-H$_2$ conversion factor. Small symbols represent the computed $\alpha_{\rm CO}^Z$ values for MACSJ0032-arc (red star) as well as the compilation of MS SFGs with CO measurements from the literature \citep{dessauges15}. Large symbols correspond to the few MS SFGs for which gas-phase metallicities have been measured from rest-frame optical nebular lines. The symbols are color-coded according to galaxy redshift and their shapes indicate $\alpha_{\rm CO}^Z/\alpha_{\rm CO,MW}$ ratios ($<1.5$: circles, $1.5-2$: triangles, and $>2$: squares and star).}
\label{fig:alphaCO}
\end{figure}

Following \citet[][see their Eqs.~(7,\,8,\,12a) and their respective 
justifications]{genzel15}, we consider in what follows the metallicity-dependent 
CO-to-H$_2$ conversion function and the mass-metallicity relation calibrated to 
the \citet[][PP04]{pettini04} metallicity scale given by
%to estimate the metallicities of galaxies 
\begin{equation}\label{eq:alphaCO}
\alpha_{\rm CO}^Z = \alpha_{\rm CO,MW} \times \chi(Z)
\end{equation}
with 
\begin{equation}\label{eq:Xsi}
\chi(Z) = 10^{-1.27 (12+\log({\rm O/H})_{\rm PP04}-8.67)}
\end{equation}
and 
\begin{equation}\label{eq:metallicity}
12+\log({\rm O/H})_{\rm PP04} = a-0.087(\log(M_*) -b)^2
,\end{equation}
\begin{eqnarray}
{\rm where}~a&=&8.74~{\rm and} \nonumber \\
b&=&10.4+4.46\log(1+z)-1.78(\log(1+z))^2\,. \nonumber
\end{eqnarray}
We assume the Milky Way CO-to-H$_2$ conversion factor to be $\alpha_{\rm CO,MW} 
= 4.36~M_{\sun}/({\rm K~km~s^{-1}~pc^2})$\footnote{Equivalent to 
$X_{\rm CO,MW} = 2\times 10^{20}~\rm cm^{-2}/(K~km~s^{-1})$.}, which includes a 
correction factor of 1.36 for helium \citep{strong96}, and the solar abundance 
of $12+\log({\rm O/H})_{\sun} = 8.67$ from \citet{asplund04}. The resulting 
metallicity-dependent CO-to-H$_2$ conversion function is essentially a function 
of two physical parameters, $M_*$ and redshift, as illustrated in 
Fig.~\ref{fig:alphaCO} by the iso-redshift curves. The uncertainties typically 
are $\pm 0.25$~dex on the inferred $\alpha_{\rm CO}^Z$, which is basically the 
scatter of the mass-metallicity relation, and $\pm 0.15$~dex on the stellar 
masses. The value of 
%hk \LEt{ ok? to avoid the symbol at the beginning of the sentence  } 
$\alpha_{\rm CO}^Z(M_*,z)$ increases with redshift for any given $M_*$, and at 
any given redshift increases with decreasing $M_*$. 
%These trends are particularly pronounced at $\log(M_*/M_{\sun}) \lesssim 10$ and $z > 1.5$. 
Adopting these prescriptions, we find most of the CO-detected MS SFGs from our 
literature compilation \citep{dessauges15} to feature $\alpha_{\rm CO}^Z$ values 
comparable to $\alpha_{\rm CO,MW}$, with 
$0.8 < \alpha_{\rm CO}^Z/\alpha_{\rm CO,MW} < 1.5$, owing to their high stellar 
masses (small circles in Fig.~\ref{fig:alphaCO}). 
%Only for MS SFGs at $z\gtrsim 1.5$
%where the $\alpha_{\rm CO}(M_*,z)$ function becomes steeper at a given $M_*$, 
%does $\alpha_{\rm CO}^Z$ exceed $\alpha_{\rm CO,MW}$ by a factor of few (small triangles and squares in Fig.~\ref{fig:alphaCO}).

%
%__________________________________________________________________

%
%__________________________________________________________________

Using the derived $M_*$ (Table~\ref{tab:physicalparameters}), 
$z_{\rm CO} = 3.6314$, and $12+\log({\rm O/H})_{\rm PP04} = 8.0\pm 0.2$ 
($Z=0.21~Z_{\sun}$) estimated from Eq.~(\ref{eq:metallicity}), we obtain a 
high value of $\alpha_{\rm CO}^Z = 30\pm 9~M_{\sun}/({\rm K~km~s^{-1}~pc^2})$ 
for MACSJ0032-arc, as expected for a high-redshift galaxy with a low stellar 
mass (star in Fig.~\ref{fig:alphaCO}). Such a high $\alpha_{\rm CO}^Z$ value 
translates into a very important gas mass of $M_{\rm molgas}/\mu = 
(7.6\pm 3.0)\times 10^{10}~M_{\sun}$ (lensing-corrected). Since most of this 
molecular gas is concentrated within a $1-2~\rm kpc$ region (see 
Sect.~\ref{sect:COspatialdistr}), this also implies a very high
%an unphysically high 
H$_2$ surface density of $\Sigma_{\rm H_2}/\mu \gtrsim 
2.4\times 10^4~M_{\sun}~\rm pc^{-2}$, 
%is inferred, 
which is about two orders of magnitude higher than the typical value of 
$\sim 200~M_{\sun}~\rm pc^{-2}$ found for giant molecular clouds (GMCs) in the 
Milky Way, and even higher than that of GMCs in local ULIRGs of 
$\sim 10^4~M_{\sun}~\rm pc^{-2}$ \citep{downes98}. In ULIRGs, however, the 
CO-to-H$_2$ conversion factor is significantly lower, typically 
$0.8~M_{\sun}/({\rm K~km~s^{-1}~pc^2})$ (see below). Although there is evidence 
for $\alpha_{\rm CO}^{\rm Z}$ values of 
$20-100~M_{\sun}/({\rm K~km~s^{-1}~pc^2})$ in dwarf galaxies like the Magellanic 
Clouds \citep{leroy11}, we argue against such a high CO-to-H$_2$ conversion 
factor for MACSJ0032-arc based on the fact that its morphology, star formation 
rate, and CO SLED, among other factors, represent compelling evidence of a much 
denser and more active environment than found in local dwarf galaxies. 
%(see Sect.~\ref{sect:morphology}).

%Our findings thus suggest a possible limitation to the metallicity-dependent CO-to-H$_2$ conversion function.
%that the adopted metallicity-dependent CO-to-H$_2$ conversion factor function does not hold for MACSJ0032-arc. Indeed, said function suffers from several shortcomings. First, it is overly simplistic, 
%One thing {\bf that the metallicity-dependent CO-to-H$_2$ conversion function} does not take into account, is the gas excitation state linked to the physical conditions of the galaxy (temperature and density).
%is the CO excitation, which depends on the galaxy temperature and density. 
%Moreover, it 
The metallicity-dependent CO-to-H$_2$ conversion function strongly relies on the 
adopted mass-metallicity calibration, which has known caveats, in particular for 
objects at $z>2.5$ and with low stellar masses. With the mass-metallicity 
relation specifically calibrated for SFGs at $z\sim 3-5$ and valid for galaxies 
with $\log(M_*/M_{\sun}) = 9-11$ \citep{troncoso14}, we find a similarly low 
metallicity 
%of $12+\log({\rm O/H}) = 8.1\pm 0.2$ and a similarly high $\alpha_{\rm CO}^Z = 23\pm 10~M_{\sun}/({\rm K~km~s^{-1}~pc^2})$ 
for MACSJ0032-arc. On the other hand, the fundamental metallicity relation (FMR) 
calibrated for low-$M_*$ galaxies \citep{mannucci10,mannucci11,christensen12}, 
characterized by $\mu_{0.32} = \log(M_*) -0.32\log(\mathit{SFR}) < 9.5$ (a 
criterion satisfied by MACSJ0032-arc with $\mu_{0.32} = 9.1$), yields a higher 
metallicity of $12+\log({\rm O/H})_{\rm FMR} = 8.5\pm 0.2$ ($Z=0.68~Z_{\sun}$). 
This higher metallicity is also supported by the low 
$R3 =~$[O\,{\sc iii}]\,$\lambda$5007/H$\beta$ line flux ratio of $0.63\pm 0.07$, 
measured in our recently acquired near-IR LBT/LUCI spectrum of MACSJ0032-arc 
(Walth et~al., in prep.). According to the latest $R3$--metallicity calibration 
from \citet{curti17}\footnote{http://www.arcetri.astro.it/metallicity/
calibrazioni.pl}, although known to be very approximative, this ratio points to 
a metallicity of $12+\log({\rm O/H})_{\rm neb} = 8.6-8.7$. At this higher 
metallicity, we obtain a much smaller $\alpha_{\rm CO}^Z$ value of 
$7.1\pm 2.2~M_{\sun}/({\rm K~km~s^{-1}~pc^2})$. Thus, the uncertainty of the 
metallicity of MACSJ0032-arc induces an error of at least a factor of 4 into the 
$\alpha_{\rm CO}^Z$ determination. 

Similar problems are encountered for the few MS galaxies at high redshift for 
which direct measurements of their gas-phase metallicities (large symbols in 
Fig.~\ref{fig:alphaCO}) have been obtained from the 
[N\,{\sc ii}]\,$\lambda 6583$/H$\alpha$\,$\lambda 6563$ line flux ratio and 
the PP04 metallicity calibration\footnote{Except for the Cosmic Eye at 
$z=3.074$, where the gas-phase metallicity is determined from the $R23$ 
metallicity calibration.} \citep{teplitz00,hainline09,law09,richard11,
dessauges11,genzel13}. For most of these systems, the $\alpha_{\rm CO}^Z$ values 
computed from their measured metallicities, which are higher than the 
metallicities estimated from Eq.~(\ref{eq:metallicity}), fall significantly 
below the iso-redshift curves at their $M_*$ and $z$ values (as illustrated in 
Fig.~\ref{fig:alphaCO} by a redshift color-code that does not match that of the 
corresponding iso-redshift curve). 

\citet{magdis11} proposed another way of determining the CO-to-H$_2$ conversion 
factor at high redshift by using the dust mass and the CO(1-0) luminosity 
through the relation
\begin{equation}\label{eq:alphaCOdust}
\alpha_{\rm CO}^{\rm dust} = \frac{1}{\delta_{\rm DGR}}\times \frac{M_{\rm dust}}{L'_{\rm CO(1-0)}}\,.
\end{equation}
%Apart from the uncertainties linked to estimating $M_{\rm dust}$ (see Eq.~(\ref{eq:Mdust})), 
This method requires the knowledge of another physical parameter that is the 
dust-to-gas mass ratio, defined as $\delta_{\rm DGR} = 
M_{\rm dust}/M_{\rm molgas}$\footnote{At high redshift, it is generally 
assumed that $M_{\rm H_2} \gg M_{\rm H\,I}$, or equivalently 
$M_{\rm H_2+H\,I} \simeq M_{\rm molgas} = M_{\rm dust}/\delta_{\rm DGR}$.}. The 
value of $\delta_{\rm DGR}$ is assumed to vary with metallicity,
%is taken to vary only with metallicity (but not with redshift nor galaxy mass), 
following the prescription derived by \citet{leroy11} for local galaxies
\begin{equation}\label{eq:DGR}
\delta_{\rm DGR} = 10^{(-2+0.85(12+\log({\rm O/H})_{\rm PP04}-8.67))}\,.
\end{equation}
From the $M_{\rm dust}$ and $L'_{\rm CO(1-0)}$ values listed in 
Table~\ref{tab:physicalparameters} and the metallicity estimate of 
$12+\log({\rm O/H})_{\rm PP04} = 8.0\pm 0.2$ computed using
Eq.~(\ref{eq:metallicity}), we find $\alpha_{\rm CO}^{\rm dust} = 
2.8\pm 0.9~M_{\sun}/({\rm K~km~s^{-1}~pc^2})$ for MACSJ0032-arc. This in turn 
yields much more reasonable $M_{\rm molgas}$ and H$_2$ surface density.
%compared to the values observed in local GMCs. 
%The corresponding $\alpha_{\rm CO}^{\rm dust}$ yield $M_{\rm molgas} = (3.2-8.3)\times 10^9~M_{\sun}$, which implies a much more reasonable H$_2$ surface density in this high-redshift MS SFG compared to what is observed in local GMCs.
%The derived $\alpha_{\rm CO}^{\rm dust}$ is in good agreement with the average$\alpha_{\rm CO}^{\rm dust} = 5.5\pm 0.4~M_{\sun}/({\rm K~km~s^{-1}~pc^2})$ computed by \citet{magdis12} in 6 MS BzK galaxies at $z\sim 1.5$, using the same method with the same $\delta_{\rm DGR}$ (Eq.~(\ref{eq:DGR})). For the identical BzK galaxy sample, based on the dynamical masses, \citep{daddi10} argued for an average $\alpha_{\rm CO}^{\rm dyn} = 3.6\pm 0.8~M_{\sun}/({\rm K~km~s^{-1}~pc^2})$, also in perfect agreement. 

This $\alpha_{\rm CO}^{\rm dust}$ value, however, differs by a factor of 11 from 
the value $\alpha_{\rm CO}^Z = 30\pm 9~M_{\sun}/({\rm K~km~s^{-1}~pc^2})$ 
computed before from Eqs.~(\ref{eq:alphaCO},\,\ref{eq:Xsi}), while the two 
derivations of the CO-to-H$_2$ conversion factor, if correct, should yield the 
same result for any given galaxy. 
%For the $12+\log({\rm O/H}) = 8.4\pm 0.2$ metallicity obtained with the FMR calibrated for low-$M_*$ galaxies, we get $\alpha_{\rm CO}^{\rm dust} = 1.3\pm 0.6~M_{\sun}/({\rm K~km~s^{-1}~pc^2})$, still differing by a factor of $\sim 7$ with the corresponding $\alpha_{\rm CO}$. 
%Hence, apart from the uncertainty on the MACSJ0032-arc metallicity,
The cause of the discrepancy does not lie in erroneous determinations of either
%might lie in erroneous determinations of 
$L'_{\rm CO(1-0)}$ or $M_{\rm dust}$ (Table~\ref{tab:physicalparameters}), 
which both enter in Eq.~(\ref{eq:alphaCOdust}). Indeed, we detect the CO(1--0) 
emission, and therefore know the $L'_{\rm CO(1-0)}$ within the given 
uncertainty. Accounting for the CMB radiation leads to a higher intrinsic 
$L'_{\rm CO(1-0)}$ by approximately 25\% (Sect.~\ref{sect:COSLED}). 
%It is unlikely though that the CO(1--0) luminosity is overestimated, which could have led to a too low value of $\alpha_{\rm CO}^{\rm dust}$, and while an underestimated 2~mm continuum emission would have the same effect, this could increase $M_{\rm dust}$ by no more than a factor of 2.
%However, to increase the derived $\alpha_{\rm CO}^{\rm dust}$ nohow the CO(1--0) luminosity could be overestimated, and whereas the 2~mm continuum flux might be underestimated, but by no more than a factor of 3 and would hence result in an increase of the $M_{\rm dust}$ by a factor of 2 at most. 
The value of $M_{\rm dust}$ is notoriously more difficult to determine 
(Eq.~(\ref{eq:Mdust})), and a factor of 2 uncertainty in the dust mass is 
commonly accepted. An underestimation of the 2~mm continuum emission could, in 
addition, increase the $M_{\rm dust}$ estimate, but by no more than another 
factor of 1.4 according to the data (see Sect.~\ref{sect:PdBI})
%It is, however, unlikely that $M_{\rm dust}$ could absorb the entire factor of 11 discrepancy. Furthermore, 
%Our $M_{\rm dust}$ measurement is, on the other hand, in 
and given the good agreement with the constant dust-to-stellar mass ratio of 
$\log(M_{\rm dust}/M_*) \approx -2.6$ 
%predicted by the well defined $M_{\rm dust}$--$M_*$ relation 
\citep{santini10,smith12,sklias14}. The value of $\delta_{\rm DGR}$
%, and last parameter in Eq.~(\ref{eq:alphaCOdust}), 
could also vary by a factor 2 given the dispersion in the 
$\delta_{\rm DGR}$--metallicity relation. However, we are still far from the 
factor 11. The situation is less dire if we assume the higher metallicity 
for MACSJ0032-arc (at $12+\log({\rm O/H})_{\rm FMR} = 8.5\pm 0.2$ the 
discrepancy falls to a factor of 7), but we are nevertheless left, {\em in 
fine}, with 
%We are thus left with the puzzling 
the realization that Eqs.~(\ref{eq:alphaCO},\,\ref{eq:Xsi}) and 
Eqs.~(\ref{eq:alphaCOdust},\,\ref{eq:DGR}) cannot produce consistent results for 
$\alpha_{\rm CO}^Z$ and $\alpha_{\rm CO}^{\rm dust}$. 
%unless unrealistic metallicity values of $Z\gg Z_{\sun}$ are assumed.

There is another fundamental physical parameter that should be taken into 
account in the estimation of 
%physical effect that affects 
the CO-to-H$_2$ conversion factor, namely 
%and might allow us to reconcile this discrepancy: 
the gas excitation state, which depends on the galaxy temperature and density. 
%Consequently, another way to reconcile $\alpha_{\rm CO}$ with $\alpha_{\rm CO}^{\rm dust}$ in the MACSJ0032-arc is still the excitation. 
The CO SLED analysis presented in Sect.~\ref{sect:COSLED} shows that the CO 
excitation in MACSJ0032-arc is much higher than that of the Milky Way, and is in 
fact comparable to that of high-redshift SMGs. Given this similarity in the 
excitation state of their molecular gas, it appears appropriate to replace, for 
MACSJ0032-arc, $\alpha_{\rm CO,MW}$ in Eq.~(\ref{eq:alphaCO}) by
%It follows that our target's $\alpha_{\rm CO}^Z$ value might be $\sim 5\,\times$ lower than standard when, in Eq.~(\ref{eq:alphaCO}), we adopt 
$\alpha_{\rm CO,ULIRG}\simeq 0.8~M_{\sun}/({\rm K~km~s^{-1}~pc^2})$ 
%instead of $\alpha_{\rm CO,MW}$, 
as observed in local ULIRGs 
%and high-redshift SMGs
\citep{solomon97,bolatto13,carilli13}. It follows that our target's
$\alpha_{\rm CO}^Z$ value is likely to be about $5\times$ lower. 
%(this prescription giving an $\alpha_{\rm CO}^Z \approx 7~M_{\sun}/({\rm K~km~s^{-1}~pc^2})$). 
With this alternative $\alpha_{\rm CO}^Z$ of reference, it becomes possible 
%This is precisely the factor needed. 
%Applying this excitation correction enables us 
to reconcile $\alpha_{\rm CO}^Z$ with $\alpha_{\rm CO}^{\rm dust}$ for 
MACSJ0032-arc down to a small factor between 1.3 and 2 depending on the adopted 
metallicity ($12+\log({\rm O/H})_{\rm FMR} = 8.5\pm 0.2$ and 
$12+\log({\rm O/H})_{\rm PP04} = 8.0\pm 0.2$, respectively).

Since we have, at present, no way to constrain the CO-to-H$_2$ conversion 
factor more tightly in MACSJ0032-arc (and in general), in the following we 
consider the two extreme but still reasonable $\alpha_{\rm CO}$ values for the 
arc (see Table~\ref{tab:physicalparameters}): 
\begin{itemize}
\item $\alpha_{\rm CO}^Z \simeq 7.1~M_{\sun}/({\rm K~km~s^{-1}~pc^2})$, 
corresponding to an $\alpha_{\rm CO}^Z$ 
%based on $\alpha_{\rm CO,ULIRG}$ and motivated by the high observed CO excitation, 
along with a metallicity of $12+\log({\rm O/H})_{\rm FMR} = 8.5\pm 0.2$ 
computed using the FMR calibrated for low-$M_*$ galaxies;
\item $\alpha_{\rm CO}^{\rm dust} \simeq 2.8~M_{\sun}/({\rm K~km~s^{-1}~pc^2})$, 
corresponding to an $\alpha^{\rm dust}_{\rm CO}$ along with a metallicity of 
$12+\log({\rm O/H})_{\rm PP04} = 8.0\pm 0.2$ derived from 
Eq.~(\ref{eq:metallicity}).
%derived by requiring $\alpha_{\rm CO}^{\rm dust}$ to equal the $\alpha_{\rm CO}^Z$ based on $\alpha_{\rm CO,ULIRG}$. This solution implies that $Z\simeq Z_{\sun}$ in the arc.
\end{itemize}
These two $\alpha_{\rm CO}$ values yield a range of lensing-corrected molecular 
gas masses in MACSJ0032-arc of $M_{\rm molgas}/\mu = 
(7.1-18)\times 10^9~M_{\sun}$ (see Table~\ref{tab:physicalparameters}) and 
lensing-corrected H$_2$ surfaces densities of $\Sigma_{\rm H_2}/\mu \gtrsim 
2200-5800~M_{\sun}~{\rm pc^2}$. 
%in comparison with the values observed in local GMCs.

Although it is clearly ill-constrained, we note that this range of 
$\alpha_{\rm CO}$ values for MACSJ0032-arc is consistent with results from 
studies of $z\sim 1-2$ BzK and MS SFGs in which $\alpha_{\rm CO}$ was found to 
vary between 2 and $20~M_{\sun}/({\rm K~km~s^{-1}~pc^2})$.
%In comparison with the adopted $\alpha_{\rm CO}$ values for the MACSJ0032-arc, 
Indeed, \citet{magdis12} obtained an average $\alpha_{\rm CO}^{\rm dust} = 
5.5\pm 0.4~M_{\sun}/({\rm K~km~s^{-1}~pc^2})$ computed for 6 BzK galaxies at 
$z\sim 1.5$ using the same Eqs.~(\ref{eq:alphaCOdust},\,\ref{eq:DGR}), 
%and the same metallicity-dependent $\delta_{\rm DGR}$ (Eq.~(\ref{eq:DGR})), 
whereas for the same BzK galaxy sample \citet{daddi10} argued for 
$\alpha_{\rm CO}^{\rm dyn} = 3.6\pm 0.8~M_{\sun}/({\rm K~km~s^{-1}~pc^2})$ in a 
work based on the dynamical masses. \citet{magnelli12} obtained, on average, a 
much higher value of $\alpha_{\rm CO}^{\rm dust} = 
14\pm 4~M_{\sun}/({\rm K~km~s^{-1}~pc^2})$ for seven MS SFGs at $z\sim 1$. 
%, but with similar metallicities. 
In another study, \citet{genzel12} used the Kennicutt-Schmidt (KS) relation 
in reverse to estimate $M_{\rm molgas}$ from the $\mathit{SFR}$ and reported, 
for 44 SFGs at two median redshifts of $\langle z\rangle = 1.2$ and 
$\langle z\rangle = 2.2$, a range of $\alpha_{\rm CO}^{\rm KS}$ values of 
$2-20~M_{\sun}/({\rm K~km~s^{-1}~pc^2})$ at metallicities from 
$12+\log({\rm O/H}) = 8.4$ to 8.9. 
%The inferred $\alpha_{\rm CO}^{\rm KS}$ values follow the $z=0$ metallicity-dependent CO-to-H$_2$ conversion function from \citet{leroy11} used in the $\alpha_{\rm CO}^Z$ parametrisation.  
%\citet{aravena16} derived $\alpha_{\rm CO}^{\rm dust}$, using a constant $\delta_{\rm DGR}$, for dusty, starburst, ULIRG-type galaxies at $2.5<z<5.7$ and found lower values varying between $0.4-1.8~M_{\sun}/({\rm K~km~s^{-1}~pc^2})$.
Finally, our adopted $\alpha_{\rm CO}$ values do not agree as closely with 
those of $\alpha_{\rm CO}^{\rm dust} = 
0.4-1.8~M_{\sun}/({\rm K~km~s^{-1}~pc^2})$ estimated for dusty starburst 
galaxies at $2.5<z<5.7$ by \citet{aravena16} using a constant 
$\delta_{\rm DGR}$.

For our compilation of CO-detected MS SFGs from the literature 
\citep{dessauges15}, in the following we adopt $\alpha_{\rm CO}^Z$, as computed 
from Eqs.~(\ref{eq:alphaCO},\,\ref{eq:Xsi},\,\ref{eq:metallicity}) and, when 
available, we consider the measured gas phase metallicities. Special attention 
was paid to galaxies at $z>2.5$ and/or with $\log(M_*/M_{\sun})<10$ because of 
the expected larger uncertainty in their metallicities when derived from the 
assumed mass-metallicity relation (Eq.~(\ref{eq:metallicity})). We therefore 
retain these galaxies only when direct measurements of their gas phase 
metallicities are available (two galaxies at $z=1.5859$ and $z=2.7793$, 
respectively, were thus removed from our compilation).

%
%__________________________________________________________________

\begin{figure}
\centering
\includegraphics[width=8cm,clip]{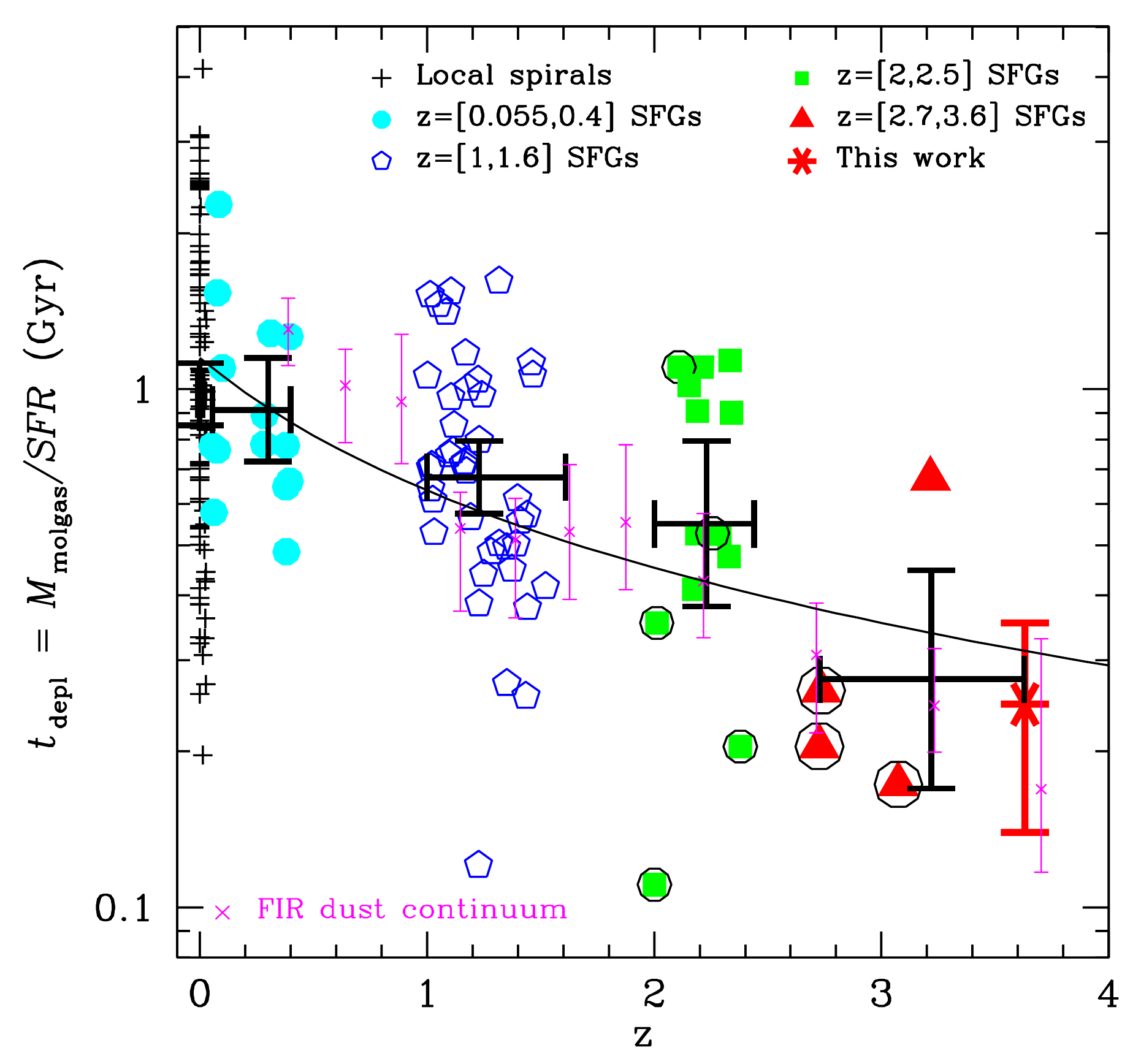}
\caption{Molecular gas depletion timescale as a function of redshift for our compilation of CO-detected MS SFGs from the literature \citep{dessauges15} and MACSJ0032-arc at $z_{\rm CO} = 3.6314$ (star). The encircled data points correspond to MS SFGs with measured gas-phase metallicities. Five redshift bins are considered: $z\sim 0$, $z=[0.055,0.4]$, $z=[1,1.6]$, $z=[2,2.5]$, and $z=[2.7,3.6]$. The corresponding mean values of $t_{\rm depl}$ are shown as large black crosses with the associated $1\,\sigma$ dispersion as standard deviation. Magenta crosses represent $t_{\rm depl}$ values from \citet{bethermin15} derived from the FIR dust continuum and by adopting the same mass-metallicity relation (Eq.~(\ref{eq:metallicity})). The observed evolution of $t_{\rm depl}$ with redshift is parametrized by a power law; the best fit, $1.15 (1+z)^{-0.85}$, is shown as a solid line.}
\label{fig:tdepl-z}
\end{figure}
%
%__________________________________________________________________

\subsection{Molecular gas depletion timescale}
\label{sect:tdepl}

The depletion timescale, defined as 
$t_{\rm depl} = M_{\rm molgas}/\mathit{SFR}$, describes how long each galaxy 
could sustain star formation at the current rate before running out of fuel, 
assuming that the gas reservoir is not replenished. Models predict the depletion 
of the molecular gas in galaxies over cosmic time
%how the molecular gas in MS galaxies gets depleted over cosmic time 
\citep[e.g.,][]{dave11,dave12}. More precisely, $t_{\rm depl}$ has been shown 
to scale as $(1+z)^{-1.5}$ for the canonical disk model 
\citep{mo98,bouche10,genel10}. The decrease in $t_{\rm depl}$ with redshift is 
amply supported by CO observations of SFGs \citep[e.g.,][]{combes13,tacconi13,
saintonge13}, but with growing evidence of a more modest redshift evolution of 
$(1+z)^\gamma$, with $\gamma$ possibly as low as $-0.16$ \citep{genzel15}. 
\citet{genzel15}, \citet{bethermin15}, and \citet{schinnerer16} also find 
$t_{\rm depl}(z)$ to be only slowly evolving when estimating $t_{\rm depl}$ from 
the FIR dust continuum. 
%The decrease tends to be steepen, however, at $z\gtrsim 2.5$ \citep{bethermin15}.

In Fig.~\ref{fig:tdepl-z} we show $t_{\rm depl}$ as a function of redshift for 
MACSJ0032-arc, and for our compilation of CO-detected galaxies from the 
literature \citep{dessauges15}, which we restrict to SFGs lying within the 
accepted thickness of the MS ($0.3 < \mathit{sSFR/sSFR}_{\rm MS} < 3$), a 
condition also satisfied by MACSJ0032-arc (see Sect.~\ref{sect:SED}).
%for the adopted the MS parametrization from \citet{tacconi13}.
%in Eq.~(\ref{eq:MS}). 
The underlying CO-to-H$_2$ conversion factors are uniformly derived for the 
sample of SFGs as explained in Sect.~\ref{sect:alphaCO}. We extend the redshift 
range explored by \citet{genzel15} by adding the measured depletion times from 
CO of 5 MS SFGs in the $z=[2.7,3.6]$ bin to the analysis, including that of 
MACSJ0032-arc.  
%we add a measurement in the $z=[2.7,3.6]$ bin, obtained by averaging over 5 MS SFGs. 
%We confirm the shallow %(almost non-existent) $t_{\rm depl}$ redshift-evolution and observe that it is mostly sustained by this new higher redshift data point. 
The new best-fit power law description of $t_{\rm depl}(z)$ is then given by the 
$(1+z)^{-0.85}$ scaling and is on average steeper than that determined by 
\citet{genzel15} only for galaxies at $z\lesssim 2.5$. With $t_{\rm depl}$ 
declining from $\sim 1.0~\rm Gyr$ at $z\sim 0$ to $\sim 270~\rm Myr$ at 
$z\sim 3.2$, the observed $t_{\rm depl}$ evolution is very similar to that 
inferred from the FIR dust continuum, which yields only moderate redshift 
evolution up to $z\sim 2.5$ followed by more rapid evolution at higher 
redshift \citep{bethermin15}. We almost invoke a two-step power law for the 
redshift evolution of $t_{\rm depl}$, with a turnover around $z\sim 2.8$, but 
this still needs to be confirmed with more $t_{\rm depl}$ measurements at 
$z>2.5$.

Overall, $t_{\rm depl}$ shows a more modest redshift evolution than predicted 
by galaxy evolution models.
%Thus, $t_{\rm depl}$ evolves with cosmic time, but we confirm that this evolution seems to happen at a much more moderate scale than predicted by models. 
%according to current observations. 
%We stress in this context that this finding 
This may be partly due to the metallicity-dependent CO-to-H$_2$ conversion  
function we have adopted ($\alpha_{\rm CO}^Z$ computed using 
Eqs.~(\ref{eq:alphaCO},\,\ref{eq:Xsi},\,\ref{eq:metallicity})).
%, if it does not provide a correct estimate of the real $\alpha_{\rm CO}$ of high-redshift MS SFGs (see Sect.~\ref{sect:alphadust}). 
As discussed in Sect.~\ref{sect:alphaCO} and shown in Fig.~\ref{fig:alphaCO}, 
$\alpha_{\rm CO}^Z$ increases with redshift for all $M_*$.
%and increases with decreasing $M_*$ at all redshifts. 
%The increase of $\alpha_{\rm CO}$ is particularly dramatic when $\log(M_*/M_{\sun})\lesssim 10$ and $z>1.5$. This precisely concerns the MACSJ0032-arc and explains its high $t_{\rm depl}$ obtained with a high $\alpha_{\rm CO} = 39~M_{\sun}/({\rm K~km~s^{-1}~pc^2})$ resulting from its expected sub-solar $\sim 0.2~Z_{\sun}$ metallicity. 
Adopting this $\alpha_{\rm CO}^Z(M_*,z)$ may thus artificially flatten the 
redshift evolution of $t_{\rm depl}$. A steeper evolution is indeed observed when a simple $\alpha_{\rm CO,MW}$ conversion factor is applied to MS SFGs. In this case, the power law index $\gamma$ is found to vary between $-1.5$ and $-1$
\citep[see][]{tacconi13,dessauges15}. 
%corresponding to $(1+z)^{\gamma}$ with $\gamma$ varying between $-1.5$ and $-1$ is obtained when a uniform $\alpha_{\rm CO,MW}$ conversion factor is applied to MS SFGs \citep[see][]{tacconi13,dessauges15}. 
%Moreover, we also observe that the few objects with direct gas-phase metallicity measurements (the encircled data points in Fig.~\ref{fig:tdepl-z}) have among the lower $t_{\rm depl}$ per redshift bin, because their $t_{\rm depl}$ are computed with lower $\alpha_{\rm CO}$ with respect to the $\alpha_{\rm CO}$ that would be inferred from the mass-metallicity relation (Eq.~(\ref{eq:metallicity})) as discussed above and illustrated in Fig.~\ref{fig:alphaCO}. 
%Unfortunate as the dependence of $t_{\rm depl}(z)$ on $\alpha_{\rm CO}^Z(M_*,z)$ is in general, it means that 
In any event, the lower $t_{\rm depl}$ average observed at the highest 
redshifts, i.e., in the $z=[2.7,3.6]$ bin, as well as the resulting evolution 
of $t_{\rm depl}$ with cosmic time only become the more solid results. 
%it induces (even if shallow) are all the more solid results. 
%However, it urges to get more constraints on $\alpha_{\rm CO}$ in high-redshift MS SFGs and, in particular, its dependence on redshift, since among all the $M_{\rm molgas}$ available at $z>1$ only 6 are derived for MS SFGs with $\log(M_*/M_{\sun}) < 10$ so far.

%
%__________________________________________________________________

\subsection{Molecular gas fraction}
\label{sect:fgas}

%The cosmic evolution of the molecular gas fraction with Hubble time is a direct result of the expansion of the Universe. New matter being accreted by galaxies along surrounding filaments comes from farther out as time progresses, adding material with higher angular momentum in increasingly outer parts of galaxies. This leads to the well known growth in size of galaxies with cosmic time \citep{trujillo06,buitrago08,stewart13}, which in turn has a direct impact on the H$_2$/H\,{\sc i} evolution and thus on the H$_2$ evolution with cosmic time \citep[e.g.,][]{obreschkow09,lagos11,lagos14}.

The molecular gas fraction is defined as
\begin{equation}\label{eq:fmolgas}
f_{\rm molgas} = \frac{M_{\rm molgas}}{M_{\rm molgas}+M_*} \equiv \frac{1}{1+(\mathit{sSFR} \times t_{\rm depl})^{-1}}\,.
\end{equation}
Here, 
%hk \LEt{ Hence=Therefore? "Thence"=from that place } 
$f_{\rm molgas}$ depends on both $\mathit{sSFR}$
%the specific star formation rate, $\mathit{sSFR} = \mathit{SFR}/M_*$, 
and $t_{\rm depl}$. The monotonic rise of $\mathit{sSFR}$ with redshift, 
commonly parametrized as $(1+z)^\gamma$, where $\gamma = 1.5-3$, is now well 
established observationally and theoretically 
\citep[e.g.,][]{schaerer09,schreiber15,tacchella16,faisst16}. Combined with the 
modest redshift evolution of $t_{\rm depl}$ discussed in Sect.~\ref{sect:tdepl}, 
an increase in $f_{\rm molgas}$ with redshift is similarly expected. This
$f_{\rm molgas}$ evolution with cosmic time is in line with various model 
predictions \citep[e.g.,][]{obreschkow09,lagos11,lagos14}, and could partly be 
at the origin of the steep decline in the star formation rate density after the 
peak of activity around $z = 1.5-2$ \citep{madau98,hopkins06} as the molecular 
gas is the fuel for star formation.

\begin{figure}
\centering
\includegraphics[width=8cm,clip]{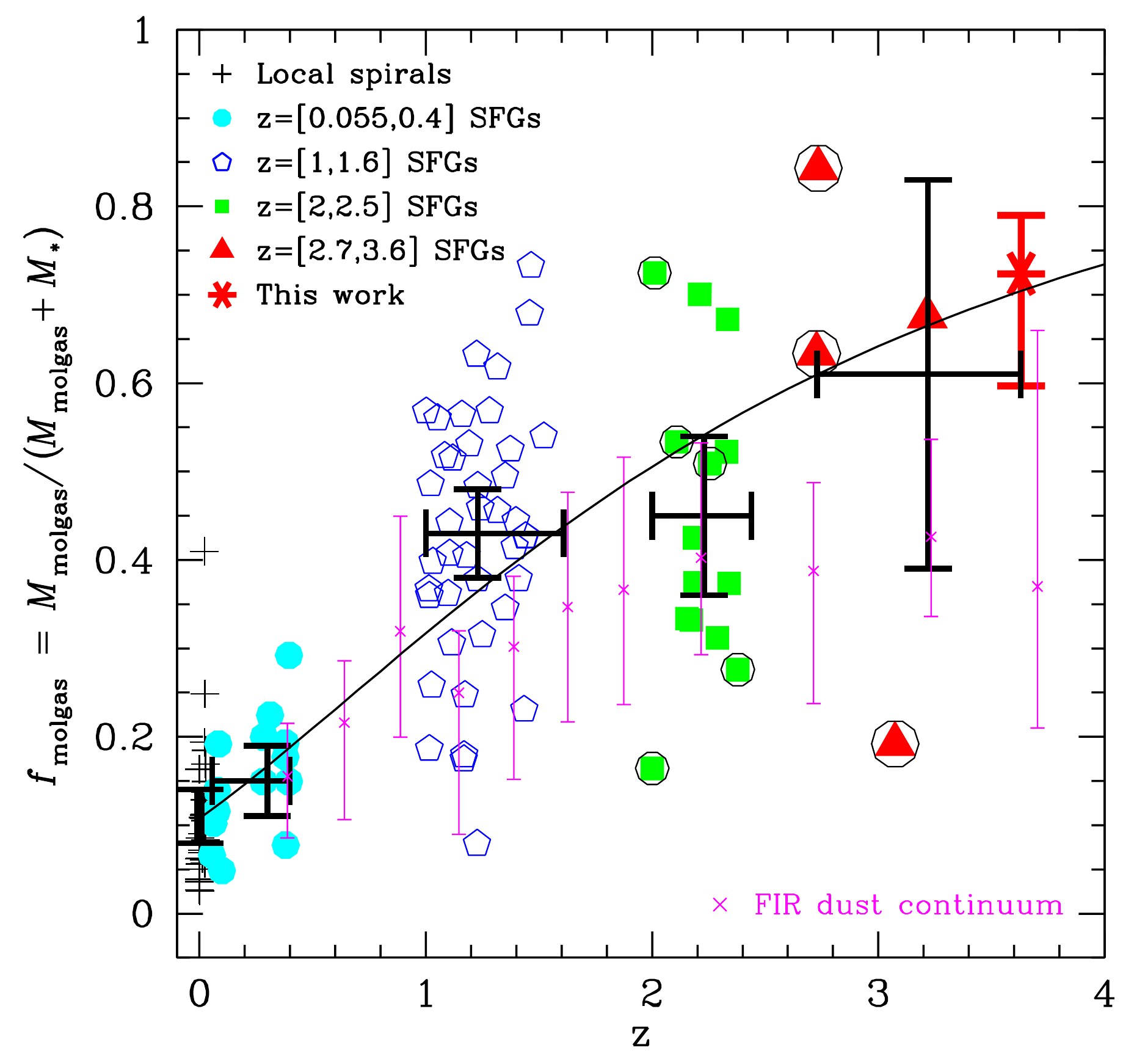}
\caption{Same as Fig.~\ref{fig:tdepl-z}, but for the molecular gas fraction. A clear evolution of $f_{\rm molgas}$ with redshift is observed, in good agreement with $1/(1+(0.12(1+z)^{1.95})^{-1})$ (solid line).}
%$1/(1+(0.11 (1+z)^{2.17})^{-1}) \propto (1+z)^{2.8}\times (1+z)^{-0.63}$ 
\label{fig:fgas-z}
\end{figure}

In Fig.~\ref{fig:fgas-z} we show $f_{\rm molgas}$ as a function of redshift for 
MACSJ0032-arc, and for our compilation of CO-detected galaxies from the 
literature \citep{dessauges15}, the latter having again been restricted to SFGs 
lying within the accepted thickness of the MS. We assume for the SFGs from the 
literature the same underlying CO-to-H$_2$ conversion factors as for 
$t_{\rm depl}$ measurements  (see Sect.~\ref{sect:alphaCO}). Extending the 
compilation of CO measurements to higher redshifts, the observed 
%increase of $f_{\rm molgas}$ with redshift 
evolution of $f_{\rm molgas}$ with redshift is well reproduced by the scaling 
law $1/(1+(0.12(1+z)^{2.8}\times (1+z)^{-0.85})^{-1})$ (Eq.~(\ref{eq:fmolgas})), 
composed of the $(1+z)^\gamma$ best-fit descriptions of, respectively, 
$\mathit{sSFR}(z)$ \citep[e.g.,][]{tacconi13}
%(Eq.~(\ref{eq:MS})) 
and $t_{\rm depl}(z)$ (Sect.~\ref{sect:tdepl}). This scaling law predicts an 
increase of $f_{\rm molgas}$ with redshift, which is at odds with 
%Extending earlier work to higher redshifts, our compilation of CO measurements does not support 
the flattening of $f_{\rm molgas}(z)$ beyond $z\gtrsim 2$ previously claimed 
for CO-detected MS SFGs \citep{saintonge13,dessauges15}, 
%as expected given the shallower $t_{\rm depl}$ redshift evolution now found with the adopted metallicity-dependent $\alpha_{\rm CO}$ function.
and 
%in galaxies for which the FIR dust continuum was used to derive $f_{\rm molgas}$
for FIR dust continuum $f_{\rm molgas}$ estimates
%when $f_{\rm molgas}$ was derived from the FIR dust continuum 
\citep{troncoso14,bethermin15,schinnerer16}.
%Two effects cause this change. First, 
%In fact the larger sample of CO measurements in the highest redshift bin 
%$z=[2.7,3.6]$ now favours a 
Nevertheless, currently, taking the few available CO measurements at 
$z\gtrsim 2$ at face value, we cannot exclude either the increase in or the 
flattening of $f_{\rm molgas}$ at high redshifts, since the two molecular gas 
fraction means, $\langle f_{\rm molgas}\rangle = 0.44\pm 0.08$ at $z=[2,2.5]$ 
and $\langle f_{\rm molgas}\rangle = 0.61\pm 0.22$ at $z=[2.7,3.6]$, agree 
within their $1\,\sigma$ errors defined as the standard deviation of the 
respective $f_{\rm molgas}$ dispersions.
%The data are, however, still consistent with an intermediate plateau around the mean value of $\langle f_{\rm molgas}\rangle = 0.44\pm 0.08$within the redshift interval from $z\sim 1.5$ to $z\sim 2.5$, followed by an increase to the mean $\langle f_{\rm molgas}\rangle = 0.61\pm 0.22$ in the highest $z=[2.7,3.6]$ bin.
% and is then followed by a $f_{\rm molgas}$ rise toward higher redshift, reaching a $\langle f_{\rm molgas}\rangle = 0.62\pm 0.22$ mean in the $z=[2.7,3.6]$ bin.

%
%__________________________________________________________________

%
%__________________________________________________________________

Moreover, the adopted metallicity-dependent $\alpha_{\rm CO}^Z$ function may 
artificially boost the redshift evolution of $f_{\rm molgas}$ for the same 
reasons that it may lessen the redshift evolution of $t_{\rm depl}$ (see 
Fig.~\ref{fig:alphaCO}). Indeed, the aforementioned plateau in $f_{\rm molgas}$ 
at $z\gtrsim 2$ is more clearly observed when instead a simple 
$\alpha_{\rm CO,MW}$ conversion factor is applied to MS SFGs 
\citep{dessauges15}. 
%We can further test the robustness of the $f_{\rm molgas}$ evolution now observed and eliminate
We can eliminate, in our estimate of $\alpha_{\rm CO}^Z$, at least the 
uncertainty in the mass-metallicity relation (Eq.~(\ref{eq:metallicity})) by 
considering only the few objects with direct gas-phase metallicity measurements 
(encircled data points in Fig.~\ref{fig:fgas-z}). 
%This results in lower $\alpha_{\rm CO}^Z$, on average (see Sect.~\ref{sect:alphaCO} and Fig.~\ref{fig:alphaCO}). 
Nevertheless, the mean $f_{\rm molgas}$ at $z=[2.7,3.6]$, in that case, still 
remains above that of the $z=[1,1.6]$ and $z=[2,2.5]$ bins, but with the same 
large standard deviation.
%We find that, at the highest redshifts $z=[2.7,3.6]$, the corresponding mean $f_{\rm molgas}$, computed with the now lower $\alpha_{\rm CO}^Z$ (see the discussion in Sect.~\ref{sect:alphaCO} and Fig.~\ref{fig:alphaCO}), still remains higher than the mean values obtained for the $z=[1,1.6]$ and $z=[2,2.5]$ bins. 
%A steady increase of $f_{\rm molgas}$ with redshift is also supported when we further include 
%including, in the highest redshift bin, 
Interestingly, the new $f_{\rm molgas}$ measurement as high as 60--79\% 
obtained in MACSJ0032-arc, which is the highest redshift object with a CO 
measure, favors a steady increase in $f_{\rm molgas}$ with redshift irrespective 
of the $\alpha_{\rm CO}$ employed.
%hk \LEt{ not at all sure I have understood here. Have I interpreted correctly? }
%even though the accepted range of $\alpha_{\rm CO}$ values yields very different $f_{\rm molgas}$ values in this galaxy.
%at a redshift of $z=3.6314$ and estimated with two distinct $\alpha_{\rm CO}$ values yielding possibly very different $f_{\rm molgas}$ in this MS SFG. 
%yielding possibly very different $f_{\rm molgas}$ in this MS SFG.
%at itself strongly supports the steady $f_{\rm molgas}$ increase with cosmic time, 
%Given the large $f_{\rm molgas}$ dispersion observed per redshift bin, the number of CO measurements in MS SFGs at $z>3$ obviously needs to be increased. 
%observed in the MACSJ0032-arc still strongly advocates a continuous redshift-rise of $f_{\rm molgas}$. 
%The MACSJ0032-arc may well be at the high end of the $f_{\rm molgas}$ distribution given the large $f_{\rm molgas}$ dispersion observed per redshift bin, therefore the number of CO measurements in MS SFGs at $z>3$ needs to be increased. 

Finally, we would like to note that all the CO measurements at $z>2.5$ (except 
one, the non-encircled triangle in Fig.~\ref{fig:fgas-z}) come from lensed 
galaxies with low stellar masses ($\log (M_*/M_{\sun}) < 10.4$), whereas CO is 
most commonly measured in massive galaxies at $z<2.5$. Similarly, all the 
molecular gas mass estimates from the FIR dust continuum are derived for massive 
galaxies. Given the expected increase in $f_{\rm molgas}$ with decreasing $M_*$ 
\citep{bouche10,dave11}, the possible trend for an increase
%the observed increase 
in $f_{\rm molgas}$ beyond $z\sim 2.5$ may reflect the higher $f_{\rm molgas}$ 
expected to be found in low-mass galaxies \citep[see][]{dessauges15} if the mass 
dependence of $f_{\rm molgas}$ is stronger than its dependence on redshift.

%
%__________________________________________________________________

\section{Summary and conclusions}
\label{sect:conclusions}

%We report on a newly discovered galaxy at $z_{\rm CO} = 3.6314$, MACSJ0032-arc, strongly lensed by the galaxy cluster MACS\,J0032.1+1808 with 
%The lensing configuration of the MACSJ0032-arc has favoured the 
%successful detections of its rest-frame UV, optical, FIR, mm, and radio continua as well as of CO emission in various transitions, including the important CO(1--0) line. 
For the first time, we can simultaneously characterize the stellar, dust, and 
molecular gas properties in a normal star-forming galaxy at a redshift as high 
as $z_{\rm CO} = 3.6314$.
%at such a high redshift. 
MACSJ0032-arc has a fairly low stellar mass $M_*/\mu = 4.8^{+0.5}_{-1.0}\times 
10^9~M_{\sun}$ and a star formation rate $\mathit{SFR}/\mu = 
51^{+7}_{-10}~M_{\sun}~{\rm yr^{-1}}$ when corrected for the lensing 
magnification $\mu = 62\pm 6$, 
%MACSJ0032-arc falls above less than a factor of $\sim 2$ of typical main sequence parametrisations of the $\mathit{SFR}$--$M_*$ distribution of SFGs at $z\sim 3.6$ and at this $M_*$ \citep{speagle14,tomczak16}. 
placing it relatively close to the main sequence at $z\sim 4$ with an offset 
$\lesssim +0.3~\rm dex$ depending on the MS definition \citep{tacconi13,
speagle14,tomczak16}. Moreover, it follows the tight correlation between IR and 
CO(1--0) luminosities \citep{dessauges15,scoville16}.
%Unlike massive MS SFGs at $z\gtrsim 3$ studied by \citet{tan13}, 
MACSJ0032-arc shows no weakening of its $L'_{\rm CO(1-0)}$ \citep[see][]{tan13} 
compared to the expectations based on the $L_{\rm IR}$--$L'_{\rm CO(1-0)}$ 
relation satisfied by MS SFGs at $z\lesssim 2.5$. 
%Thanks to the high lensing amplification, this system provides the first anchor for some of the physical properties of main sequence star-forming galaxies at a look-back time of $\sim 12~\rm Gyr$, many more should be routinely analysed within the next decade. 
Our findings 
%regarding the physical properties we derive 
for MACSJ0032-arc can be summarized as follows:

\begin{enumerate}
\item About 90\% of the total $\mathit{SFR}_{\rm UV+IR}$ of MACSJ0032-arc is 
inaccessible at rest-frame UV wavelengths, but is detected through thermal dust 
emission in the FIR regime.
%originates from thermal dust emission in the FIR regime that is inaccessible at rest-frame UV wavelengths.

\item In HST images, MACSJ0032-arc is resolved into two UV-bright knots, 
separated by $1.14\pm 0.28~\rm kpc$. The bulk of the molecular gas mass and star 
formation of the lensed galaxy is, however, decoupled from the rest-frame UV 
emission and comes from between these two UV knots, i.e., from a region too 
dusty to be detected at rest-frame UV wavelengths, shown to also host the 2~mm 
cold dust emission. 
%We note that this observation was made possible only thanks to the stretching power of lensing combined with the high angular resolution provided by our JVLA imaging observations of the CO(1--0) emission and the radio continuum at 5~GHz.

\item With the detection of the CO(1--0), CO(4--3), and CO(6--5) emission lines, 
whose fluxes have been corrected against the CMB, we can constrain the CO 
luminosity correction factors for the high $J=4$ and $J=6$ CO transitions in a 
normal SFG at $z\sim 3.6$. The derived values of r$_{4,1} = 0.60\pm 0.17$ and 
r$_{6,1} = 0.28\pm 0.08$ describe a CO SLED featuring a slightly more highly 
excited CO than found in lower-redshift SFGs, but in line with CO excitation 
levels observed in high-redshift SMGs even though the IR luminosity, 
$L_{\rm IR}/\mu = 4.8^{+1.2}_{-0.6}\times 10^{11}~L_{\sun}$, of MACSJ0032-arc is 
10 times lower than typically observed in SMGs. The high CO excitation is likely 
due to the compactness of MACSJ0032-arc, that could result from 
%, but we argue that the compactness might not mandatorily be due to a merger. It could simply be the reflect of 
the general fact that galaxies are more compact at higher redshifts because of 
their smaller sizes. These results are of particular interest in the context of 
high-redshift normal star-forming galaxies and their molecular gas mass 
estimates as only high CO rotational transitions ($J\geq 4$) are accessible 
beyond $z\sim 4$ 
%the expected detections of high CO rotational transitions 
with the NOrthern Extended Millimeter Array (NOEMA) and ALMA.
%($J=4$ up to $z=4.5$, $J=6$ up to $z=7.2$). 
%in normal star-forming galaxies at high redshift 
%that will have to be used to determine these objects' molecular gas masses.

\item We interpret the morphology of MACSJ0032-arc as indicative of it being a 
single system whose molecular gas and intense dusty star formation resides and 
occurs primarily in its center,
%with a centre of highly excited molecular gas and intense dusty star formation 
surrounded by two UV-bright star-forming regions, possibly in rotation. This 
configuration may also be the result of a recent merger or accretion event, 
albeit not one big enough to completely disrupt the kinematic state of the 
galaxy, as we still observe double-peaked CO(4--3) and CO(6--5) line profiles. 
%for the CO(4--3) and CO(6--5) emissions.
%In any case, the highly concentrated on-going star formation within a central region of at most 1.14~kpc in size could testify of the bulge formation in MACSJ0032-arc.

\item Measurements of both the dust mass and the CO(1--0) luminosity of 
MACSJ0032-arc enable us to estimate its CO-to-H$_2$ conversion factor, 
$\alpha_{\rm CO}$, using several independent methods. First, we consider the 
redshift-dependent mass-metallicity relation and metallicity-dependent 
$\alpha_{\rm CO}^Z$ function adopted by \citet{genzel15}. This method yields a 
metallicity of $12+\log({\rm O/H})_{\rm PP04} = 8.0\pm 0.2$ and a value of 
$\alpha_{\rm CO}^Z = 30\pm 9~M_{\sun}/({\rm K~km~s^{-1}~pc^2})$.
%It requires, however, a large extrapolation of the above relations given the low-mass and high-redshift our object.
Alternatively, assuming the metallicity-dependent dust-to-gas mass ratio from 
\citet{leroy11}, we obtain the markedly different value of 
$\alpha_{\rm CO}^{\rm dust} = 2.8\pm 0.9~M_{\sun}/({\rm K~km~s^{-1}~pc^2})$. 
Part of the discrepancy may be due to the uncertainty on the metallicity 
--~we get a metallicity as high as $12+\log({\rm O/H})_{\rm FMR} = 8.5\pm 0.2$ 
when using the FMR calibrated for low-$M_*$ galaxies --~and/or the high CO 
excitation observed in the arc, which may imply an additional correction factor 
to $\alpha_{\rm CO}^Z$ on the order of $\alpha_{\rm CO,MW}/\alpha_{\rm CO,ULIRG} 
\sim 5$. 
%resulting in $\alpha_{\rm CO}^Z \simeq 5.5~M_{\sun}/({\rm K~km~s^{-1}~pc^2})$. 
When this high CO excitation state is accounted for, the discrepancy between 
$\alpha_{\rm CO}^Z$ and $\alpha_{\rm CO}^{\rm dust}$ falls to a factor between 
1.3 and 2 depending on the adopted metallicity.
%($12+\log({\rm O/H})_{\rm FMR} = 8.5\pm 0.2$ as derived from the FMR calibrated for low-$M_*$ galaxies or $12+\log({\rm O/H})_{\rm PP04} = 8.0\pm 0.2$, respectively).
%Finally, by requiring that $\alpha_{\rm CO}^{\rm dust} = \alpha_{\rm CO}^Z$, we can solve for the metallicity, we derive $Z\sim Z_{\sun}$ and $\alpha_{\rm CO} \sim 0.7~M_{\sun}/({\rm K~km~s^{-1}~pc^2})$. A solar metallicity is not incompatible with the arc when the FMR calibrated for low-$M_*$ galaxies is applied and an error of a factor of $2-3$ on the measured $M_*$ is deemed acceptable. 
We thus assume the CO-to-H$_2$ conversion factor for MACSJ0032-arc to lie 
between $\alpha_{\rm CO}^{\rm dust} \simeq 
2.8~M_{\sun}/({\rm K~km~s^{-1}~pc^2})$ along with the 
$12+\log({\rm O/H})_{\rm PP04}$ metallicity and $\alpha_{\rm CO}^Z \simeq 
7.1~M_{\sun}/({\rm K~km~s^{-1}~pc^2})$ along with the 
$12+\log({\rm O/H})_{\rm FMR}$ metallicity. 
%in agreement with studies of BzK and MS SF galaxies at $z \sim 1-2$, in which $\alpha_{\rm CO}$ was found to vary from $\sim 2$ to $20~M_{\sun}/({\rm K~km~s^{-1}~pc^2})$.

\item Adopting the above $\alpha_{\rm CO}$ interval, we derive a depletion 
time for the molecular gas in this $z\sim 3.6$ star-forming galaxy of 
$t_{\rm depl} = 0.14-0.35~\rm Gyr$. Along with that of other high-redshift MS 
SFGs, generally limited to $z\lesssim 2.5$ because of the scarce number of CO 
measurements existing beyond this redshift, our analysis confirms the decrease 
in $t_{\rm depl}$ with cosmic time, although to a lesser degree than predicted 
by galaxy evolution models.

\item For the same $\alpha_{\rm CO}$ interval, we find the molecular gas 
fraction of MACSJ0032-arc to lie in the range $f_{\rm molgas} = 0.60-0.79$. When 
combined with the results obtained previously for the few $z\gtrsim 2.7$ MS SFGs 
with available CO measurements, the corresponding $f_{\rm molgas}$ mean supports 
the continued increase in $f_{\rm molgas}$ with redshift;
%beyond $z=2.5$, 
%despite the $f_{\rm molgas}(z)$ plateau observed between $z\sim 1.5$ and $z\sim 2.5$. 
however, its large standard deviation does not enable us to exclude the possible 
flattening of $f_{\rm molgas}$ beyond $z\sim 2$. More CO measurements in MS SFGs 
at $z>2.5$ are clearly needed to confirm the redshift evolution of 
$f_{\rm molgas}$.
%both $t_{\rm depl}$ and $f_{\rm molgas}$.
\end{enumerate}

%
%__________________________________________________________________

\begin{acknowledgements}
We thank the IRAM staff of both the Plateau de Bure Interferometer and the 30~m 
telescope for the high-quality data acquired and for their support during 
observations and data reduction. We are, in particular, grateful to M\'elanie 
Krips, Jan Martin Winters, and Nicolas Billot. This work was supported by the 
Swiss National Science Foundation.
\end{acknowledgements}

%
%__________________________________________________________________

\end{document}